%% file: dub.tex
\documentclass[usenatbib]{mn2e} 
\usepackage{ads_refs}

\usepackage{subfigure}

\usepackage{amsmath, amssymb}
\usepackage[english]{babel}
\usepackage{graphicx}
\usepackage{verbatim}

\begin{document}

\title[Nature and evolution of FR~IIs]{Nature and evolution of powerful radio galaxies at $z\sim1$ and their link with the quasar luminosity function}

\author[van Velzen, Falcke, and K\"ording]{Sjoert van Velzen$^{1,2}$\thanks{Hubble Fellow; e-mail: sjoert@jhu.edu}, Heino Falcke$^{1,3,4}$, and Elmar K\"ording$^{1}$ \\
$^{1}$Department of Astrophysics/IMAPP, Radboud University, PO Box 9010, 6500 GL Nijmegen, The Netherlands\\
$^{2}$Department of Physics and Astronomy, The Johns Hopkins University, Baltimore, MD 21218, USA \\
$^{3}$ASTRON, Dwingeloo, The Netherlands \\
$^4$Max-Planck-Institut f\"ur Radioastronomie, Bonn, Germany}
\date{}
\pubyear{2014}
\maketitle

\begin{abstract}
Current wide-area radio surveys are dominated by active galactic nuclei, yet many of these sources have no identified optical counterparts. Here we investigate whether one can constrain the nature and properties of these sources, using Fanaroff-Riley type II (FR~II) radio galaxies as probes. 
These sources are easy to identify since the angular separation of their lobes remains almost constant at some tens of arcseconds for $z>1$. Using a simple algorithm applied to the FIRST survey, we obtain the largest FR~II sample to date, containing over $10^4$ double-lobed sources. A subset of 459 sources is matched to SDSS quasars. This sample yields a statistically meaningful description of the fraction of quasars with lobes as a function of redshift and luminosity. This relation is combined with the bolometric quasar luminosity function, as derived from surveys at IR to hard X-ray frequencies, and a disc-lobe correlation to obtain a robust prediction for the density of FR~IIs on the radio sky. We find that the observed density can be explained by the population of known quasars, implying that the majority of powerful jets originate from a radiatively efficient accretion flow with a linear jet-disc coupling. Finally, we show that high-redshift jets are more often quenched within 100 kpc, suggesting a higher efficiency of jet-induced feedback into their  host galaxies.
\end{abstract}

\begin{keywords}
accretion, accretion discs -- galaxies: jets ---  quasars: general --- catalogues.
\end{keywords}
\section{Introduction}
This year marks the fortieth anniversary of the famous publication by \citet{Fanaroff74}, reporting that the morphology of extra-galactic radio sources is correlated with their luminosity.  The powerful FR~II radio galaxies are edge brightened; their radio output is dominated by two lobes with a hotspot at a typical separation of a few hundred kpc \citep{Mullin08}. The fainter FR~I sources, on the other hand, show a radial decrease of radio intensity and often have disrupted, non-symmetric morphologies. 

With radio lobes we can probe powerful active nuclei that are obscured at higher frequencies. The lobes of FR~II radio sources can be used as calorimeters of the kinetic jet luminosity \citep{Scheuer74,Falle91,Kaiser97,Willott99}, which allows us to test theories of black hole jet formation \citep[e.g.,][]{Rawlings91}. 
FR~II radio sources are also interesting since they single out galaxies at an important epoch of their evolution. Feedback during the quasar phase \citep{Kauffmann00, Churazov05, Hopkins06} most likely explains why the most massive galaxies are ``red and dead'' \citep{Bower06, Croton06}.  This active galactic nucleus (AGN) feedback \citep[for reviews see][]{Cattaneo09, Fabian12, McNamara12,Alexander12} can be exercised by winds from the accretion disc or by the relativistic jet \citep[e.g.,][]{Morganti13}. Jets from quasars are particularly interesting since they can carry large amounts of energy up to a distance of $\sim 1$~Mpc, thus expanding the black hole's influence to the scale of galaxy groups \citep{Fabian12}.

Since stellar mass black holes consistently show powerful jets during a state change of the accretion disc at high luminosity \citep{Fender04}, a scale-invariant view of black hole accretion \citep{Falcke99, McHardy06} tells us that all quasars will host powerful jets for some part of their lifetime \citep{Nipoti05, Koerding08}. The radio lobes of these FR~II quasars are bright; the sensitivity of current radio surveys is sufficient to detect the most powerful sources (such as Cygnus~A) at the earliest possible cosmological epoch ($z\sim 10$). To tap this potential of radio surveys, we need a robust method of source identification (e.g., we wish to avoid mixing up low-redshift starforming galaxies with high-redshift blazars). This selection can be done using the radio spectrum, information from higher frequency (e.g., optical/IR) observations, or the radio morphology.

In this work, we apply a simple morphological selection method to the catalogue obtained from Faint Images of the Radio Sky at Twenty-cm \citep*[FIRST;][]{Becker95} to collect double-lobed radio sources. We obtain a relatively clean sample of over ten thousand supermassive black holes with powerful jets. 
To derive the areal density of these sources (i.e., the number per unit area on the sky), we postulate that they must be coupled to the quasar luminosity function (QLF) ---after all, jets are ultimately powered by accretion. Indeed there exists a well-known (near) linear correlation between the accretion disc luminosity and the power of the compact radio core \citep{Falcke95I} or the radio luminosity of the lobes \citep{Rawlings91, Willott99}. This correlation is the beating heart of our method; we will assume that all double-lobed radio sources obey a single relation between disc power and lobe radio luminosity and we demonstrate that our observations support this assumption. 

Our approach is different from previous work on the radio source population, which has been focused on: $(i)$ the luminosity function of radio galaxies and radio-loud quasars \citep[e.g.,][]{Dunlop90,Willott01}, $(ii)$ estimating the kinetic jet luminosity function using a model for the conversion of jet power to radio emission \citep{Koerding08,Kapinska12,Mocz13}, or $(iii)$ a parametric redshift-dependent model of the source counts \citep{Condon84,Massardi10}.

\subsection{FR~II radio sources}
The division in radio luminosity between FR~I and FR~II sources is nominally placed at $L_{\rm 1.4GHz} \sim 10^{25}~{\rm W}\,{\rm Hz}^{-1}$ \citep{Fanaroff74}, although recent observations have  found FR~IIs at lower luminosity \citep[e.g.,][]{Antognini12}, implyng that the transition from FR~I to FR~II happens gradually with luminosity \citep[e.g.,][]{Singal14}. The typical FR~I/II transition luminosity increases with host galaxy mass \citep{Ledlow96}, which provides strong evidence that the Fanaroff-Riley classification separates jets based on their ability to drill through the gas in their environment: FR~I jets are not powerful enough and get disrupted \citep[see][and references therein]{Kaiser07}.

Contrary to radio emission from the compact core of the jet, which is subject to relativistic Doppler boosting, the observed lobe luminosity is independent of the jet inclination. Radio lobes can be used as calorimeters of jet power \citep{Godfrey13}, which makes them useful to test theories of jet formation. For example, we recently used the correlation between the optical disc luminosity and lobe radio luminosity to constrain the importance of the \citet{BlandfordZnajek77} mechanism for powering quasar jets \citep{vanVelzenFalcke13}.

\begin{figure}
\begin{center}
\includegraphics[trim=6mm 0mm 0mm 6mm, clip, width=0.5\textwidth]{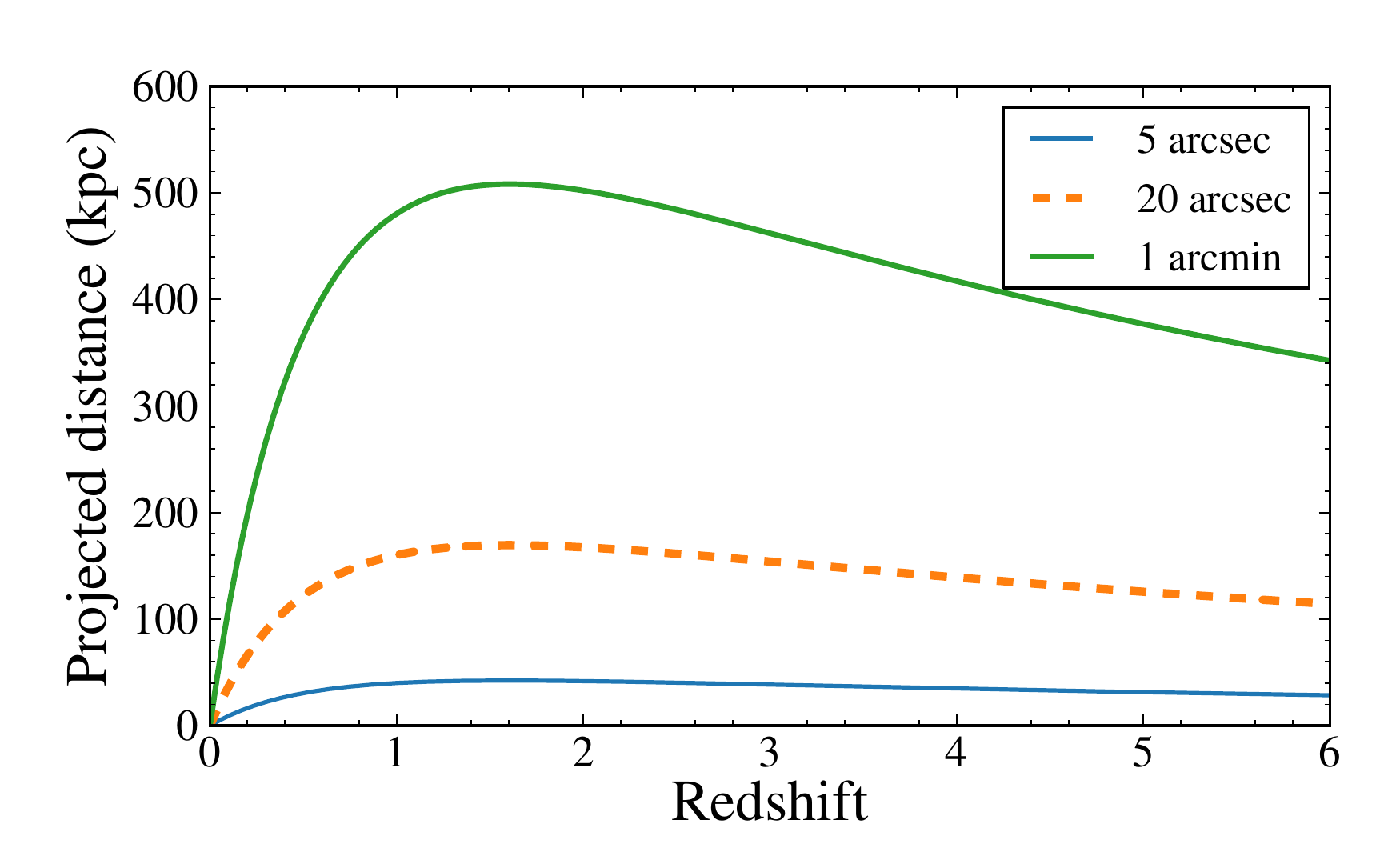}
\caption{Physical separation for a fixed angular distance as a function of redshift. A survey with a resolution better than 5~arcsec can resolve sources with a projected separation $>50$~kpc throughout the entire universe. In our search for double-lobed radio sources we used a maximum angular distance of 1~arcmin, corresponding to a maximum projected physical size of $\approx 500$~kpc.}\label{fig:sep-redshift}
\end{center}
\end{figure}

The observation that radio morphology and power are correlated was derived from images of the 3C survey \citep{MacDonald68,Mackay71}. Since then, the number of identified FR~II sources has grown with subsequent revisions and extensions of the 3C catalogue \citep[][]{Laing83, Blundell99} and other radio surveys \citep[e.g.,][]{Schoenmakers01}. Other methods of finding FR~II quasars or radio galaxies include (targeted) radio observations of galaxy clusters \citep{Owen93,Croft07, Antognini12}, and more recently, cross-matching of large area optical and radio surveys \citep{Ivezic02, Best05a, deVries06, Kimball08, Hodge09, SingalPetrosian13, Condon13}. Most of these authors used FIRST and the Sloan Digital Sky Survey \citep[SDSS;][]{york02}. 

To identify a pair of astronomical sources with a projected separation of 100~kpc in a survey with a resolution of 1~arcmin, one is limited to $z<0.1$. With a resolution of 10~arcsec, however, the pair can be resolved throughout the entire universe (Fig. \ref{fig:sep-redshift}). 
The FIRST survey covers $\sim 10^4$~deg$^2$ with a resolution of $5$~arcsec full width at half maximum (FWHM) to a limiting peak flux of 1~mJy. This combination of depth, large area, and high resolution implies that the images of the FIRST survey are fruitful fields to search for powerful jets at moderate to high redshift.

The difference between the existing FR~II research, as summarized above, and the work presented here is that we shall not rely on information from other wavelengths for the  construction of our FR~II sample, instead we employ a simple, fully automated and purely morphological criterion to compile the largest possible FR~II sample.  We note that  \citet{Kimball11} have manually classified SDSS quasars that match to FIRST sources (finding 387 `Lobe' and 619 `Triple' sources), but this advanced classification is not available for the fast majority of FIRST sources that have no matching SDSS quasars. \citet{Proctor11} has applied a combination of human identification and supervised pattern recognition software to the  FIRST images to provide morphological annotations, but this method was optimized for identifying different bent types (e.g., wide-angle tail) and does not provide a measurement of the flux of the core or the lobes (nor does it provide an estimate for the background due to chance associations). 

Finally, we emphasize that in this work we focus on a subset of radio-loud quasars, namely those with large ($>100$~kpc) radio lobes. This frees us from the controversy around the radio-loudness distribution, which has been claimed to be bi-modal \citep{Strittmatter80, Kellermann89, Falcke96, Xu99} or continuous \citep{Condon81, White00, Ivezic02, SingalPetrosian13}. This discrepancy is mostly due to the treatment of radio emission on galaxy scales ($<10$~kpc), which can be due to star formation \citep[e.g.,][]{Condon13},  winds from the accretion disc \citep{Stocke92, Zakamska14}, or compact/extended jets \citep[e.g.,][]{Falcke96}. In this work we remove the emission on galactic scales, allowing us to measure only the jet-induced radio output.

\subsection{Short guide to this paper}
We can summarize the method and the scientific aim of this work in one sentence: {\it obtain a large sample of FR~II radio sources from the FIRST survey, study their properties, and reproduce their areal density by combining the bolometric quasar luminosity function with linear jet-disc coupling}. 

Readers with very limited time can skip straight to Fig.~\ref{fig:logSlogN}, which displays the main result of this work. To obtain a good overview of the data selection, read Table~\ref{tab:cuts}. The summary of our method is given in Section~\ref{sec:sum_dub}. 

Throughout this paper, we work with the following cosmology, $h=0.70$, $\Omega_{\rm m}=0.3$, and $\Omega_\Lambda=0.7$.

\section{Source selection}\label{sec:algo}
In this section we first present the details of our FR~II selection algorithm\footnote{The development of this algorithm was motivated by the large density of small doubles that appeared in the first wide-field, high-resolution images obtained with the Low-Frequency Array (LOFAR) at 150~MHz (Orr\`u et al., in prep).}. The method has a few parameters which need to be adjusted to the properties of the survey and the desired purity and completeness of the resulting candidate FR~II sample. Here we discuss the optimal choice of these parameters in the context of the FIRST survey, but we stress that our algorithm can be applied to other radio catalogues as well. We used the 12FEB16 version of the FIRST catalog, containing 946,464 sources in $1.05\times 10^4$~deg$^2$.

\subsection{Automatic identification of double-lobed radio sources}\label{sec:init}
An FR~II radio source at $z\sim 1$ observed with 5~arcsec resolution will often simply appear as two point sources. They are readily spotted by eye, but given the large area of the FIRST survey and the even larger size of upcoming Square Kilometre Array (SKA) surveys, we want to be able to identify them automatically. The essence of our approach is to simply search for pairs of sources. We need to account for two exceptions: ({\it i}) images of FR~IIs at GHz frequencies can also contain unresolved emission from the compact core of the jet and ({\it ii}) some lobes are resolved into separate components (i.e., multiple elliptical Gaussians are fitted to the observed brightness). The aim of our lobe-finding algorithm is to retrieve all catalogued  components that are part of a double-lobed source and separate the two lobes from the core. We proceed in five steps as given below. 

\begin{enumerate}
\item Match all entries in the catalogue to each other, with a maximum radius $d_{\rm max}$. For this computationally expensive step we use \verb k3match , a very efficient coordinate matching algorithm \citep{vanVelzen12,Schellart13}. In the following steps we refer to the collection of all matches as a group. 
\item\label{step:sym} Find the symmetry axis of the group by fitting a straight line to the coordinates of the sources, weighted by their flux.
\item\label{step:reject} Reject members of the group that are more than $d_{\rm rej}$ from the best-fitting line and redo the fit for the symmetry axis. 
\item \label{step:onelast} Use the centre of the line that describes the symmetry axis to define the centre of the group. We thus assume that the origin of the twin jet system is near this geometrical center. The core flux is given by the sum of the flux of all group members within $d_{\rm core}$ of the center. 
\item\label{step:last} Project the (remaining) group members on to the symmetry axis. The sources that fall north of the centre make the northern lobe and vice versa. The lobe-lobe separation ($d$) is given by the largest angular distance between the group members.
\end{enumerate}
This FR~II detection algorithm thus has three parameters, $d_{\rm max}$, $d_{\rm rej}$, and $d_{\rm core}$, we discuss these below. 

\begin{figure}
\centering
		\includegraphics[trim=6mm 0mm 0mm 4mm, clip, width=0.5\textwidth]{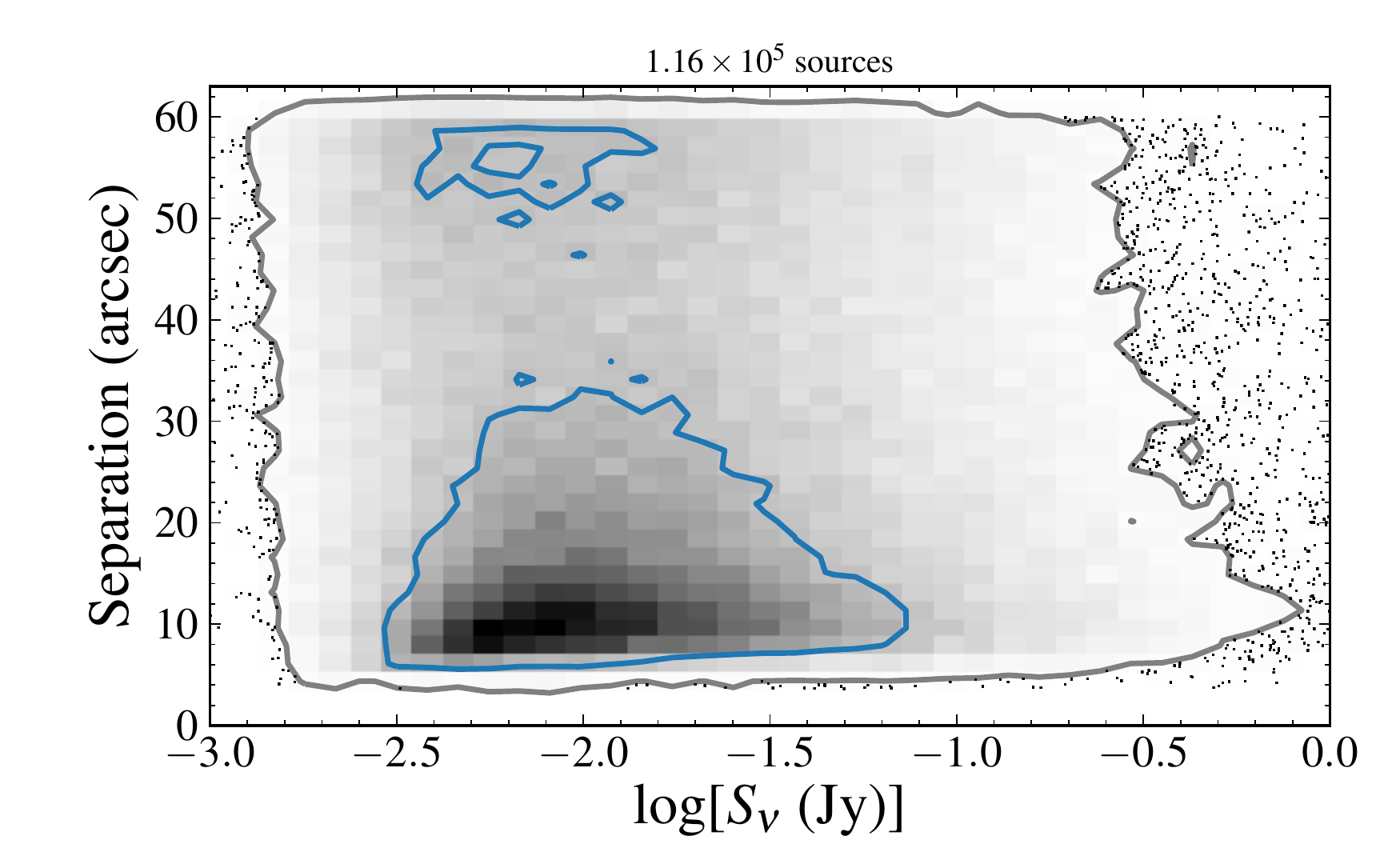}
		\includegraphics[trim=6mm 0mm 0mm 6mm, clip, width=0.5\textwidth]{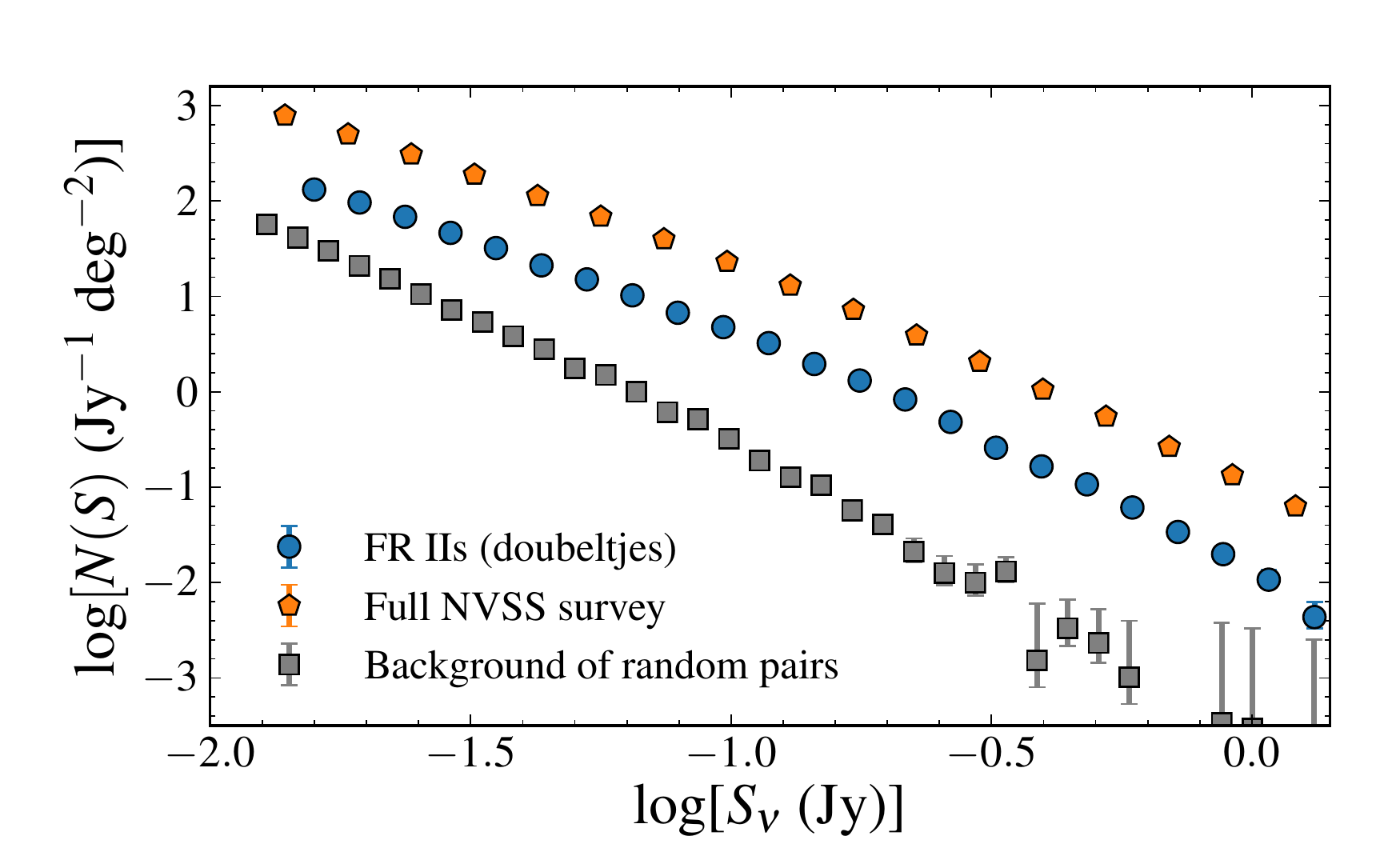}
\caption{Top: 115,889 candidate FR~II sources selected from the FIRST survey. We show the sum of the integrated flux of the lobes ($S_\nu$) and their angular separation. Contours encompass 50\% and 99\% of the population. As the separation approaches 1~arcmin, the background of random matches starts to dominate over the population of true FR~IIs. Bottom: the source count using 24,973 \mbox{doubeltjes} that remain after applying the quality cuts (Table~\ref{tab:cuts}).  We also show the counts for a sample of random \mbox{doubeltjes}, obtained by applying the same selection criteria to a uniform distribution of source coordinates, and for the full catalogue of the NVSS survey.
}\label{fig:sep-flux}
\end{figure}

The maximum separation of the two lobes ($d_{\rm max}$) needs to be smaller than the typical separation of single sources. Otherwise too many true FR~IIs are rejected because the initial group contains an unrelated bright source which results in a wrong measurement of the symmetry axis, leading to rejection in step~\ref{step:reject}. Furthermore, a larger value of $d_{\rm max}$ implies a larger background of random matches between unrated sources (see Fig.~\ref{fig:sep-flux}). We found that for the FIRST survey, the background due to random matches starts to become important for $d>1$~arcmin (Fig.~\ref{fig:sep-flux}). We therefore adopted $d_{\rm max}=1$~arcmin. This cut implies we can identify lobes with a maximum separation of $500$~kpc at $z=1$ (Fig.~\ref{fig:sep-redshift}). 

Of the FIRST sources that pass our maximum distance criterion, 10\% consisted of more than two Gaussian components and are thus processed further ---steps \ref{step:sym} to \ref{step:onelast}--- to reject unrelated sources and separate the core flux from the lobe flux. The maximum angular distance from the symmetry axis ($d_{\rm rej}$) needs to be larger than the resolution of the survey because extended lobes can be resolved into multiple Gaussian components. The angular distance that is used to identify the core emission ($d_{\rm core}$) should be similar to the resolution of the survey because the radio core is unresolved. We reject unrelated components using $d_{\rm rej}=10$~arcsec and identified core emission using $d_{\rm core}=5$~arcsec. After applying these criteria we are left with 115,889 candidate FR~IIs, the lobe flux and separation is plotted in Fig.~\ref{fig:sep-flux} (left panel). 

\subsection{Flux limit and quality cuts}\label{sec:qcut}
In the previous section we obtained a long list of double and triple radio sources. We now apply a series of quality criteria to this sample to obtain a well-defined, flux-limited sample of FR~IIs.

The lower limit for the peak flux ($F_{\rm p}$) of sources in the FIRST catalogue is $1$~mJy \citep{White97}. Yet a turnover in the number counts as a function of integrated flux ($F_{\rm i}$) can be seen at $F_{\rm i }\approx3$~mJy. This is due to extended sources that are not detected because their peak flux falls below the detection threshold. 
Double-lobed sources can also be missed because the flux of one of the lobes is below the flux limit. We thus obtain the following flux limit for the sum of the integrated flux of the two lobes: 
\begin{equation}\label{eq:flim}
S_{\nu,\rm lim} = 1~{\rm mJy} \times F_{\rm i}/F_{\rm p} (1+ f_{ l/l})\quad. 
\end{equation}
The median lobe-lobe flux ratio ($f_{l/l}$) is unity and the median of $F_{\rm i}/F_{\rm p}$ is 1.4 (see Fig.~\ref{fig:prop}), yielding a typical flux limit of $S_{\nu,\rm lim} \approx 3$~mJy. Based on the distribution of $F_{\rm i}/F_{\rm p}$ and $f_{ l/l}$ we find that for $S_{\nu}>11.9$~mJy, less than 1\% of sources are missing because one of the lobes is not detected. We therefore adopt $S_{\nu,\rm lim}=12$~mJy as our flux limit. To suppress the background due to random matches we remove sources with extreme flux ratio's: $F_{\rm i}/F_{\rm p}>5$ and $f_{l/l}>10$.

\begin{table}
\centering
	\begin{tabular}{l r p{125pt}}
	\hline
	Cut & \# left & Explanation \\
	\hline\hline
		$d_{\rm max}=1'$ & 115,889 & Maximum angular separation of the lobes. \\ 
		$S_\nu>12$~mJy & 59,192 & Flux limit for a complete sample (applied to the sum of the integrated flux of the lobes). \\ \
		$d_{\rm min}=18"$ &  35,851 & Minimum angular separation. \\ 
	$f_{ l/l}<10$ & 30,021 & Upper/lower limit on the ratio of the integrated flux of the lobes.  \\ 
	$F_{\rm i}/F_{\rm p}<5$ & 24,973 & Integrated flux over the core flux (applied to each lobe).  \\
	\hline
	\end{tabular}
\caption{Cuts for a well-defined sample of FR~II radio sources.}\label{tab:cuts}
\end{table}

Finally, we have to apply one more important quality cut. 
While the resolution of the FIRST survey implies that double-lobed sources can be resolved for $d>8$~arcsec (Fig.~\ref{fig:sep-flux}), we found that sources with a core require $d>18$~arcsec (for lower separations the core and lobes start to blend). We wish to avoid a bias for FR~IIs without a core and thus require a lower limit to the lobe-lobe separation of $d_{\rm min}=18$~arcsec.
 
Applying all quality cuts (Table~\ref{tab:cuts}) and the flux limit leaves 24,973 FR~II sources.  We will sometimes refer to these as {\it doubeltjes}, as derived from the Dutch word for small double\footnote{The first high-resolution images of LOFAR showed that the low-frequency radio sky is teeming with small doubles. We called them doubeltjes to emphasizes the difference with `Classic Doubles', which usually refers to well-resolved lobes with a separation of several arcminutes.}. The areal density of the doubeltjes at $S_\nu>12$~mJy is 2.4~deg$^{-2}$. In the right panel of Fig. \ref{fig:sep-flux} we show the areal density as a function of flux. 

\subsection{Correction for missing flux}\label{sec:missingflux}
The UV coverage that was used for the FIRST survey (VLA B-configuration) implies that some of the extended lobe flux will be ``resolved out'' (i.e., the radio interferometer acts as a high-pass filter). We compared the flux of our doubeltjes to the flux measured from the lower resolution images of NVSS \citep[NRAO VLA Sky Survey;][obtained with VLA D- and DnC-configurations]{Condon98}. Since we selected relatively compact sources, the median flux difference between the NVSS flux and the FIRST flux is only 0.05~dex. We measured this flux ratio as a function of the FIRST flux of the lobes and found $S_{\nu, \rm corrected} = 0.983\, S_\nu^{0.951}$ (with $S_\nu$ in Jy). Hereafter we use the lobe flux that is corrected for missing flux using this function.  

\subsection{Background subtraction}\label{sec:background}
The last step  before we can analyze our sample of \mbox{doubeltjes} is to estimate and subtract the background of random, unrelated matches. To this end we generate a homogeneous source distribution with fluxes drawn from the FIRST catalogue and repeat our lobe-finding algorithm (Sec.~\ref{sec:init}) and apply our quality cuts (Sec.~\ref{sec:qcut}). Because sources with high flux are rare, the fractional contribution of the background is very small for $S_\nu>100$~mJy (see Fig.~\ref{fig:sep-flux}). The areal density of background matches for $S_\nu >12$~mJy is 0.5~deg$^{-2}$. Our method of identifying random pairs slightly underestimates the true background since the sources are clustered. By counting how many SDSS quasars that match to a single FIRST source (see Section~\ref{sec:sdss_dub})  are identified as doubles by our algorithm we obtain an upper limit to the mean background of 4\%. 

After correcting for random matches, we find that, about 20\% of NVSS sources with $S_\nu \sim 10^2$~mJy are identified as doubeltjes. In the interval $S_\nu =[10, 100]$~mJy, the background-corrected fraction of NVSS sources that are doubeltjes increases with increasing flux.

\subsection{Comparison to the CONFIG sample}
The Combined NVSS-FIRST Galaxies \citep[CoNFIG;][]{Gendre08,Gendre10} sample is constructed by manual classification of subsets of NVSS sources. The most relevant subset for this work is the so-called CONFIG-4 sample which is complete to $S_\nu=50$~mJy and covers 64~deg$^2$. This sample contains 92 identified FR~IIs. Of these manually identified doubles, 90\% are retrieved by our lobe-finding algorithm. Of the nine sources that we miss, six have $d>1$~arcmin and three have $d<10$~arcsec. As explained in Section~\ref{sec:qcut} we applied a cut on the minimum lobe-lobe separation ($d_{\rm min}=18$~arcsec) to avoid a selection bias for FR~IIs without a core. If we apply our quality cuts to the CONFIG-4 sample, we recover 100\% of their manually identified FR~II sources.

\begin{figure*}
\centering
\includegraphics[trim=10mm 0mm 0mm 6mm, clip, width=0.44\textwidth]{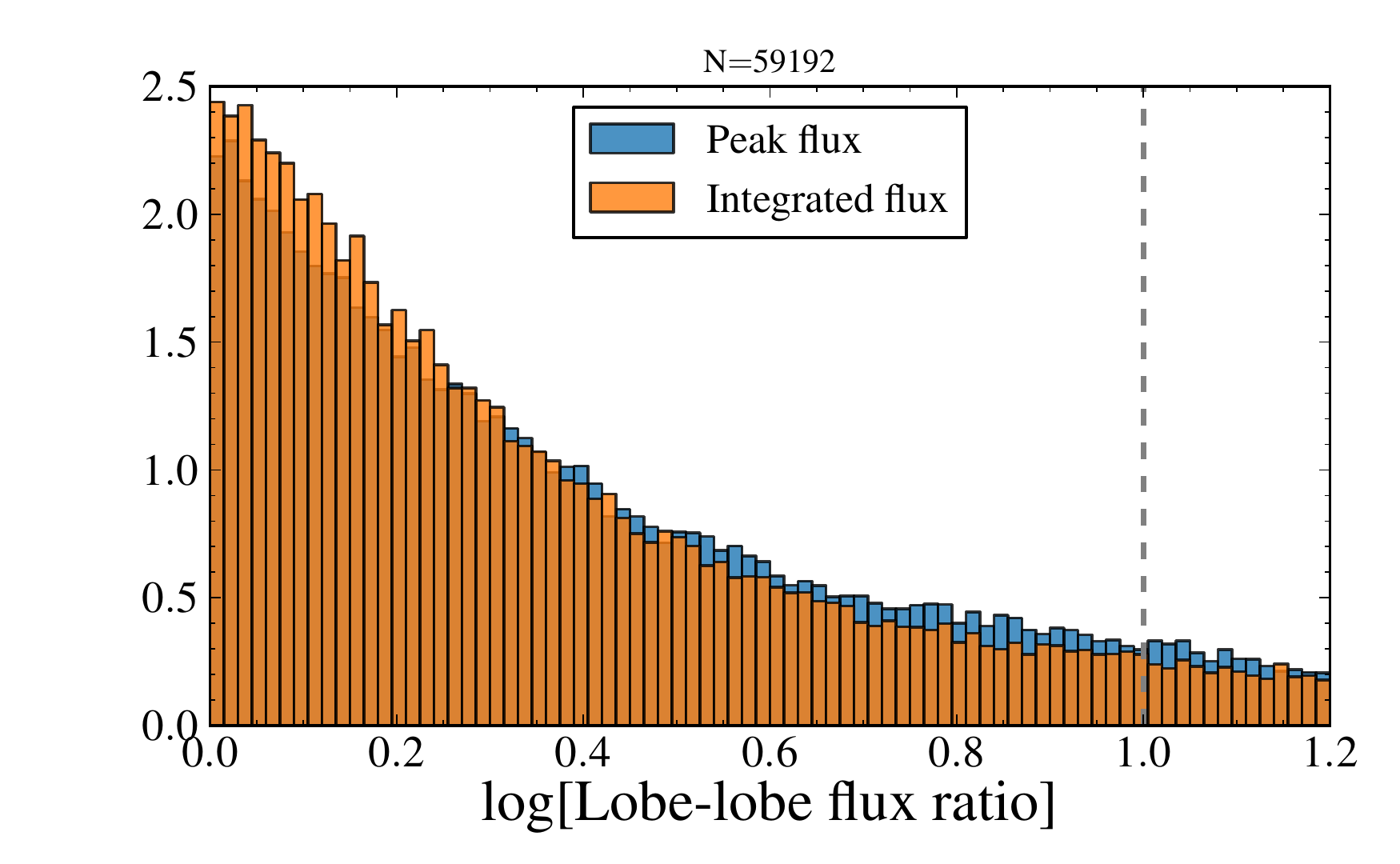} \quad 
\includegraphics[trim=10mm 0mm 0mm 6mm, clip, width=0.44\textwidth]{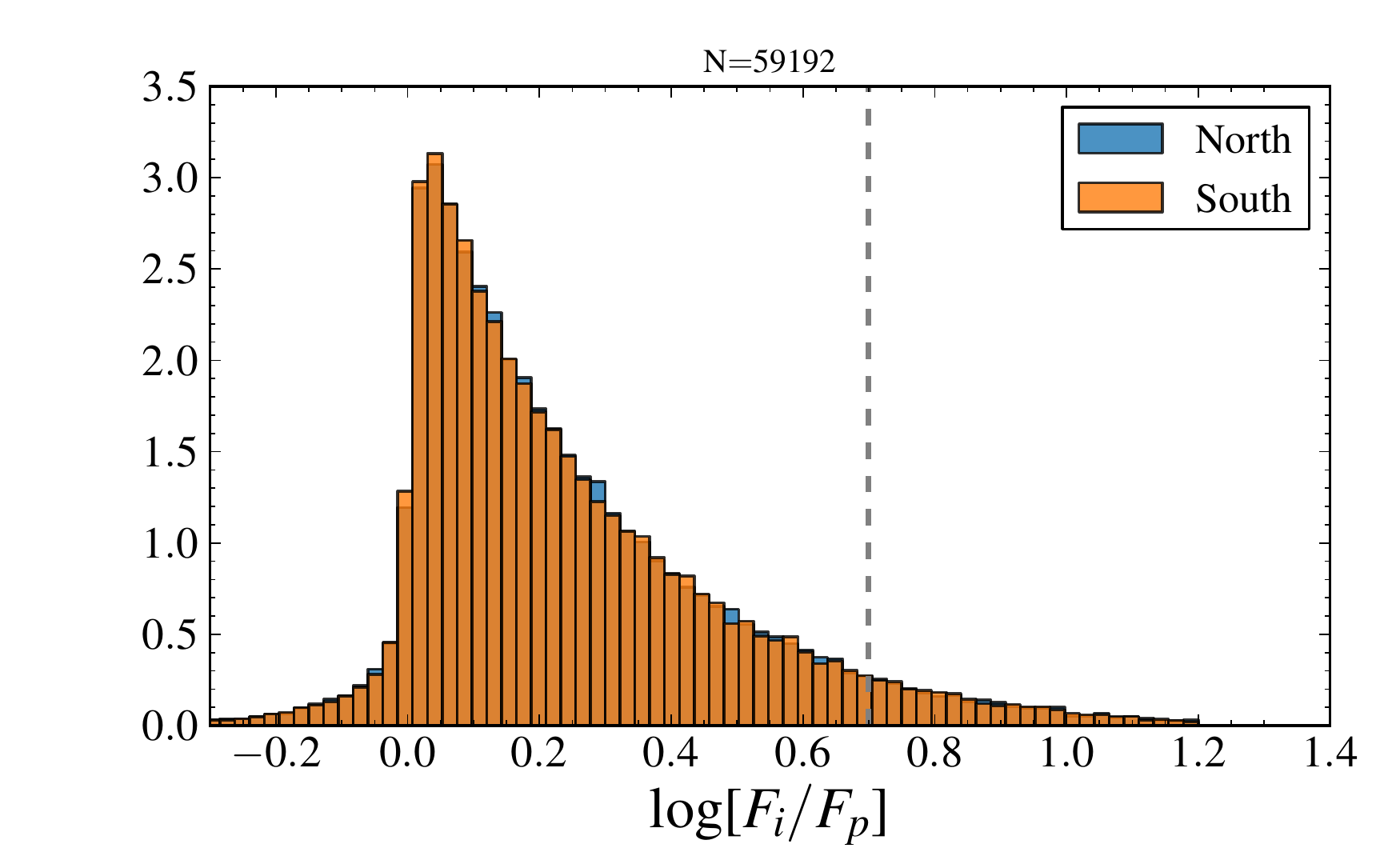} \\
\includegraphics[trim=10mm 0mm 0mm 6mm, clip, width=0.44\textwidth]{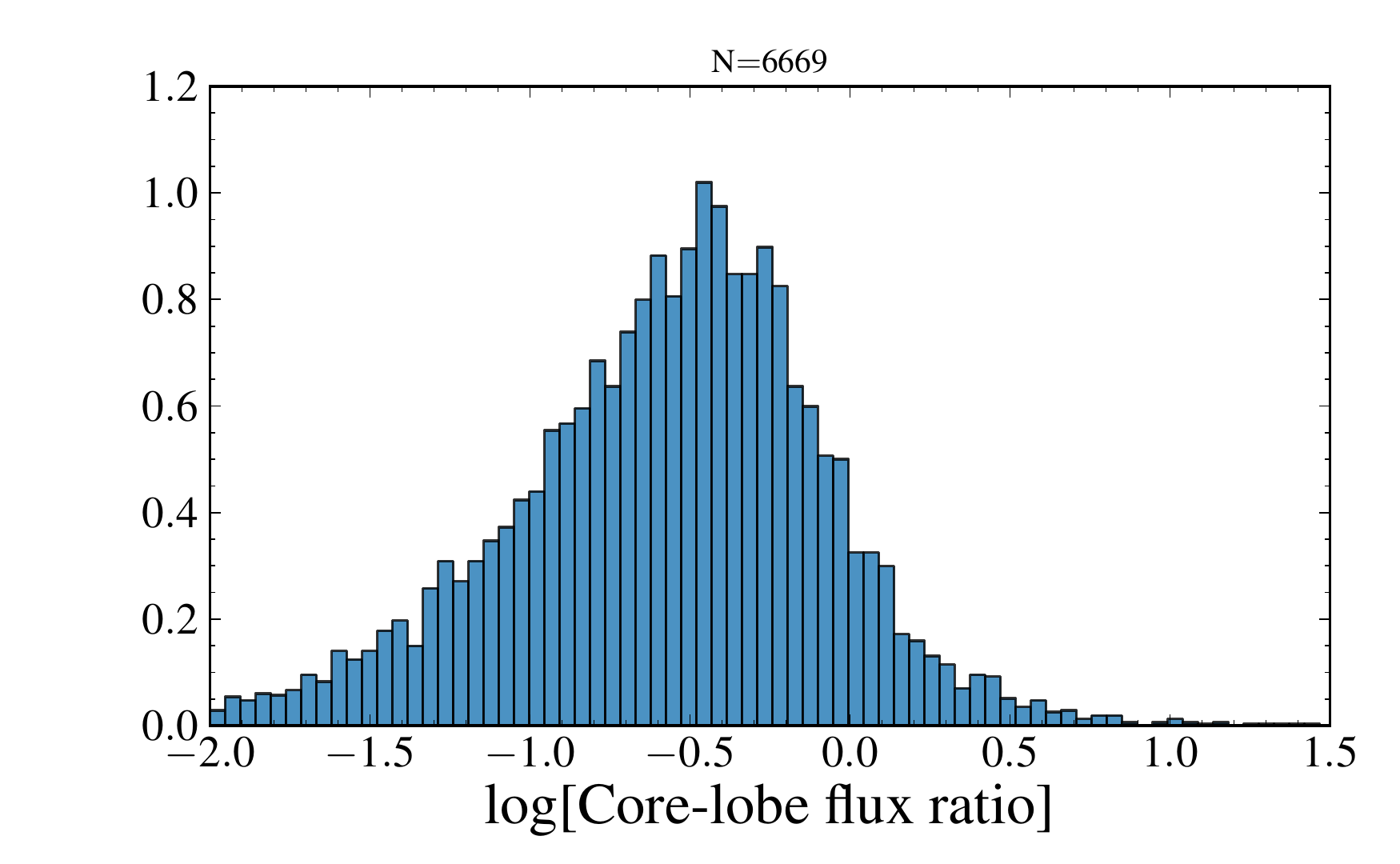} \quad
\includegraphics[trim=10mm 0mm 0mm 6mm, clip, width=0.44\textwidth]{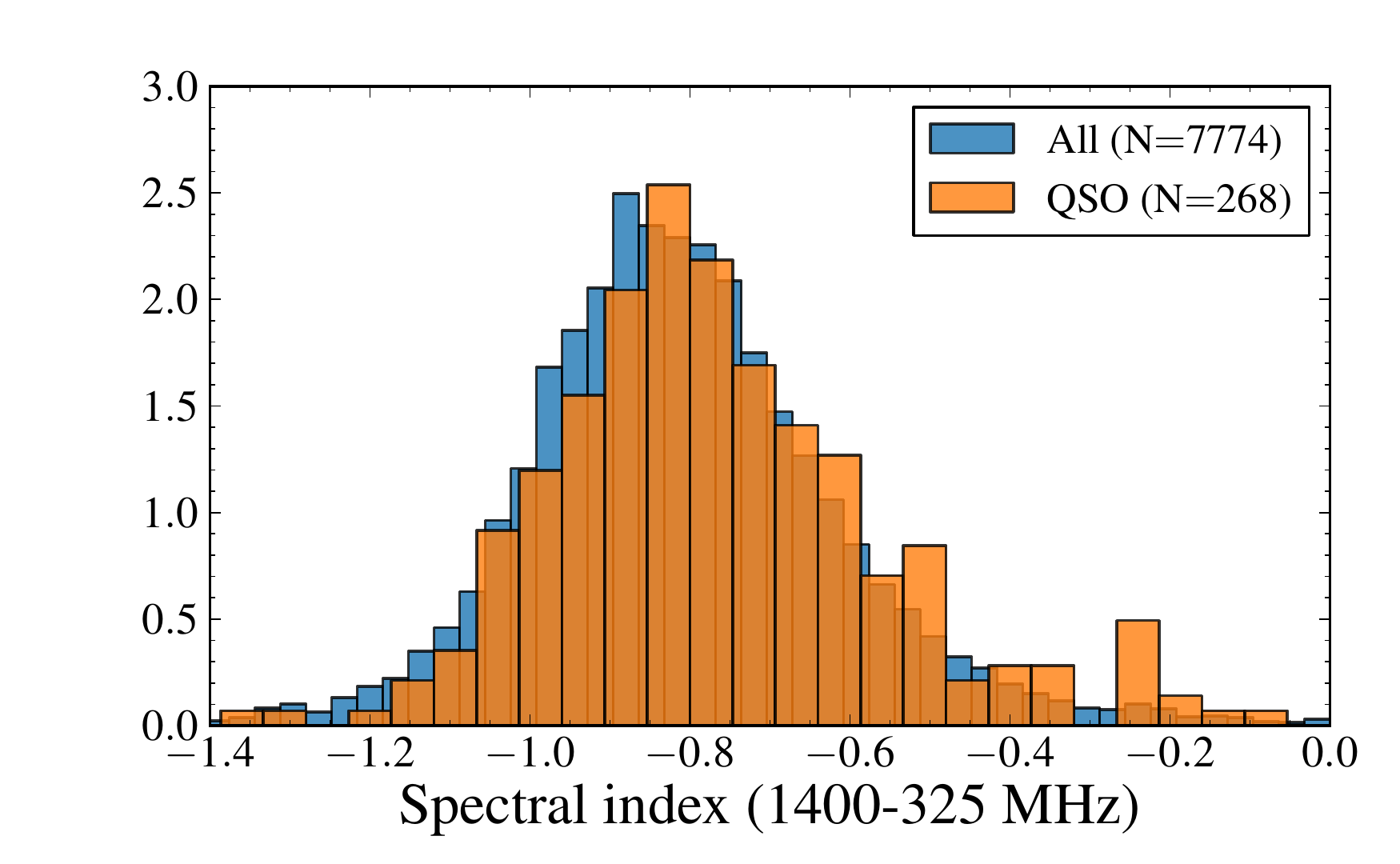}
\caption{Top left: The flux ratio of the two lobes. Top right: the ratio of the integrated flux to the peak flux (for each lobe). Formally, this ratio has to be greater or equal to unity, but for faint sources the FIRST detection algorithm can yield a peak flux that is higher than the integrated flux. Bottom left: the core flux over the sum of the flux of both lobes (we show only the sources with a detected core). Bottom right: the spectral index using the WENSS (325 MHz) and NVSS (1.4~GHz) flux. In the first three plots we show the doubeltjes sample before applying the quality cuts (Table~\ref{tab:cuts}); the cuts are indicated by the dashed line. For the bottom right panel, we show the sample after the quality cuts have been applied. All histograms are normalized (i.e., the integral over the binned parameter is equal to one).}\label{fig:prop}
\end{figure*}

\section{Redshift-independent properties of $\sim 10^4$ FR~IIs}\label{sec:prop}
In the previous section we obtained a well-defined sample of radio sources with an FR~II morphology by collecting pairs with similar flux. In Fig.~\ref{fig:prop} we present the basic properties of these doubeltjes: the flux ratio of the lobes, compactness ratio of both lobes, the core-to-lobe ratio, and the 1400-325~MHz spectral index. 

To measure the spectral index we used radio sources of the Westerbork Northern Sky Survey \citep[WENSS,][]{Rengelink97} that are within 20~arcsec of the centre of our FR~IIs. The images of WENSS have a  resolution of $54" \times 54"/\sin({\rm dec})$ FWHM, thus all of the FR~IIs in our sample are matched to a single WENSS source. This implies that the derived spectral index of sources with a detected core is slightly too steep (because the WENSS flux contains both the lobes and the core). The median spectral index of all sources is $\alpha=-0.85$; if we restrict to sources without a detected core, we find $\alpha=-0.82$. 
We also computed the spectral index using data from NVSS, which has a resolution similar to WENSS, finding a median spectral index of $\alpha=-0.81$. When we compute the spectral index using 74~MHz data from the VLA Low-frequency Sky Survey Redux \cite[VLSSr;][]{Lane14} we find $\alpha=-0.85$.

\begin{figure}
\centering
\includegraphics[trim=4mm 0mm 2mm 6mm, clip, width=0.5\textwidth]{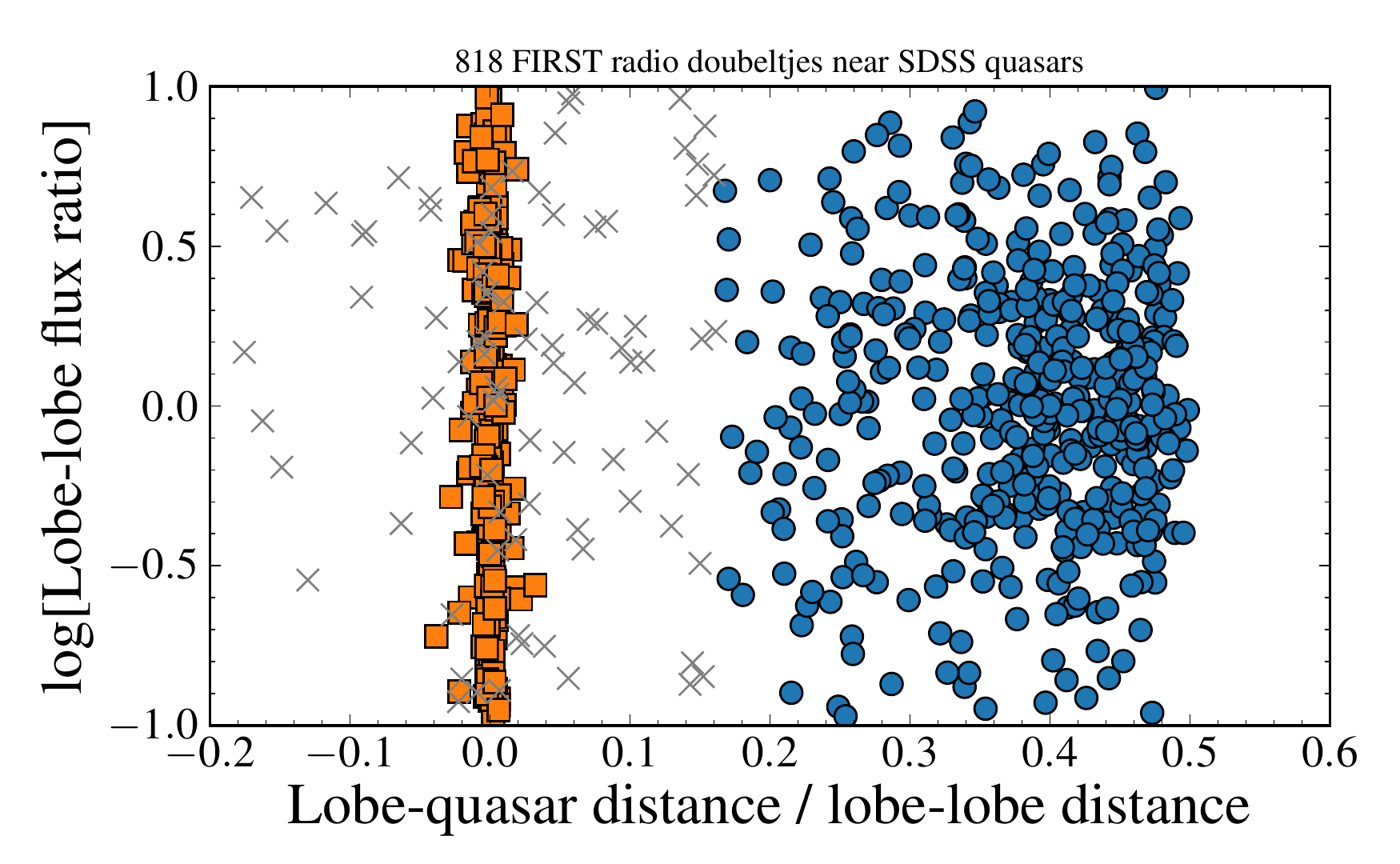}
\caption{The distance between the quasar and the lobe (Eq.~\ref{eq:lobedist}), divided by the lobe-lobe separation, versus the lobe-lobe flux ratio. A negative distance implies that the quasar is found outside the lobes. Our sample of FR~II quasars (Eq.~\ref{eq:dcut}) is labelled with blue circles. We see a clear overdensity of quasars near the centre of the lobes. Radio sources with a lobe-quasar distance less than 1~arcsec are labelled with orange squares. These are radio quasars that have been misidentified as doubles due to a nearby, but unrelated radio source. These misidentifications occur for 4\% of all quasars that are matched within 1~arcsec of a single FIRST source. }\label{fig:dist-sep}
\end{figure}

\subsection{$\sim10^2$ FR IIs matched to quasars}\label{sec:sdss_dub}
The next step is to match our collection of double-lobed radio sources to quasars. The sample of spectroscopically identified quasars from SDSS \citep{Richards02, schneider07} is most suited for this task since, by design, the sky coverage of FIRST almost fully overlaps with the SDSS footprint.  

We use the Seventh Data Release \citep[DR7;][]{Abazajian09} edition of the SDSS quasar catalogue \citep{Schneider10}, consisting of 105,783 quasars with $M_i<-22$. We also included the 87,810 quasars from DR9, the Baryon Oscillation Spectroscopic Survey \citep[BOSS;][]{Dawson13} quasar catalogue \citep{Ross12,Paris12}, which probes deeper than DR9 and includes more quasars at $z>2$. The majority of the SDSS quasars were selected for spectroscopic follow-up based on their optical colours, but information from other wavelengths was also used. We selected only SDSS DR7 quasars that were targeted based on their optical properties, leaving 77,319 sources. For the BOSS quasars we restricted the sample using the \verb UNIFORM>0  requirement, leaving 42,433 objects. For the DR7 quasars that have been re-observed by BOSS, we used the latest redshift determination (i.e., \verb Z_VI  from the BOSS catalog). Our final combined SDSS/BOSS quasar sample consists of 11,7174 unique sources. Since BOSS is part of SDSS-III \citep{Eisenstein11}, we hereafter refer to this sample simply as the SDSS quasar sample.

At the 5~arcsec resolution of the FIRST survey, most quasars are unresolved, double-lobed morphologies are rare \citep{deVries06,Kimball11}. To demonstrate this, we first match our quasar catalogue to the centers of our FR~IIs using a match radius of 30~arcsec (half of the maximum lobe-lobe separation). 
Since the centre of the FR~II is simply given by the geometrical centre of the two lobes, the quasar-lobe separation can be written as
\begin{equation}\label{eq:lobedist}
D_{\rm quasar-lobe} = d/2-s 
\end{equation}
with $d$ being the lobe-lobe distance and $s$ the separation between the quasar and the FR~II center. When $D_{\rm quasar-lobe}<0$, the quasar is located outside the circle that connects the two lobes. In Fig.~\ref{fig:dist-sep} we show $D_{\rm quasar-lobe}$  normalized by the lobe-lobe separation. We find that some quasars which are radio point sources are misidentified as FR~IIs by our algorithm. Of the 5863 FIRST sources that are within 1~arcsec of a SDSS quasar, 4\% have $D_{\rm quasar-lobe}<1$~arcsec. This confirms that the background due to matches between unrelated FIRST sources is low (Sec.~\ref{sec:background}).

Based on the observed distribution of lobe-quasar distances (Fig.~\ref{fig:dist-sep}), we adopt the following criterion to define the FR~II quasar sample 
\begin{equation}\label{eq:dcut}
s< d/3\quad.
\end{equation} 
This requirement yields 459 double-lobed radio quasars. 

If we shuffle the Declination of the SDSS quasars we find on average 9.9 matches, so the probability of a chance association of a quasar with a radio source pair using Eq.~\ref{eq:dcut} is 2.2\%.
When we make no quality cuts to the doubeltjes sample and we use all SDSS quasars, the number of matches is 1109, with an average of 1.7\% random associations. This catalogue of FR~II quasars, the largest to date, is presented in Appendix~\ref{sec:catalog}.

\begin{figure}
\centering
\includegraphics[trim=4mm 0mm 2mm 6mm, clip, width=0.5\textwidth]{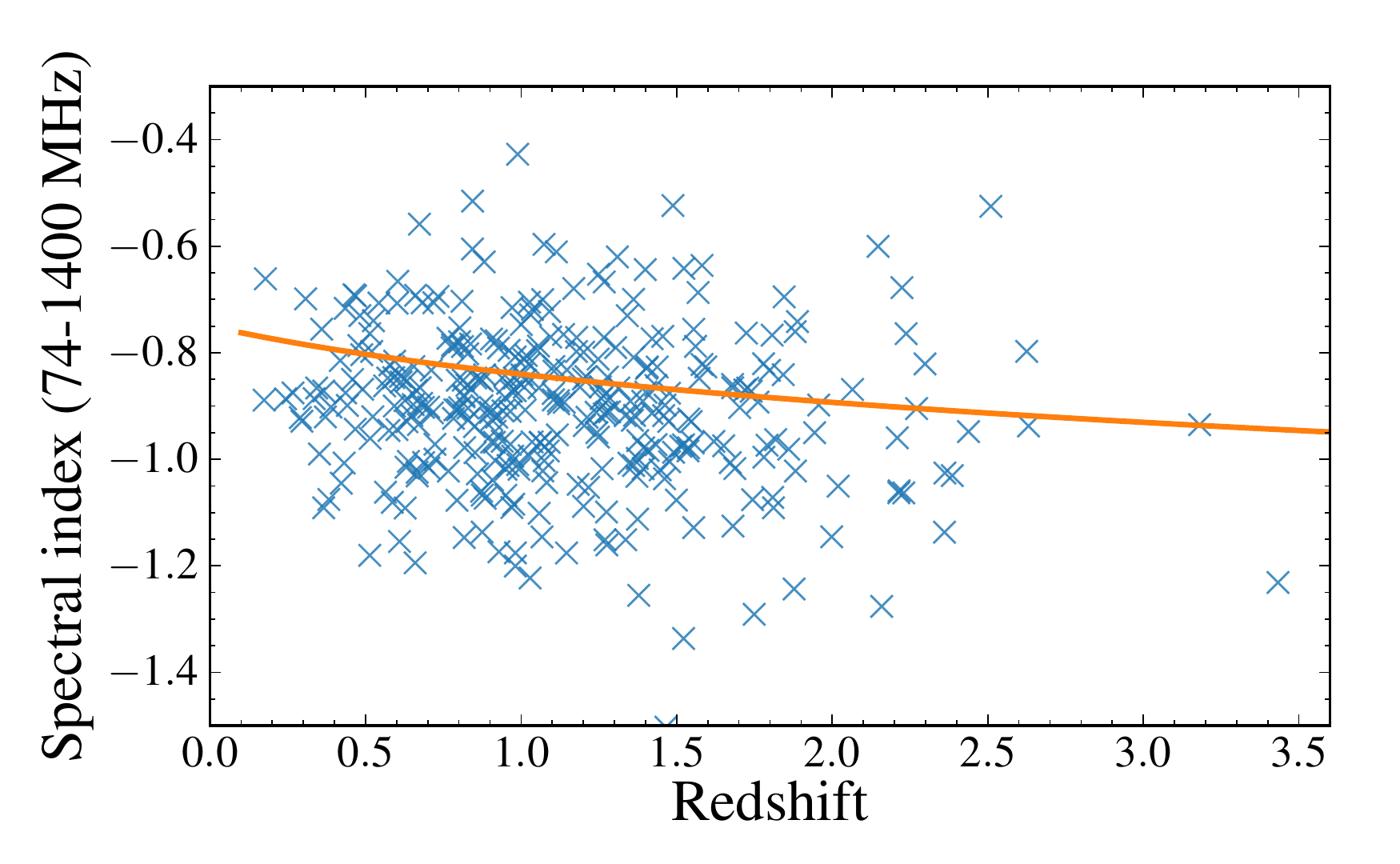}
\caption{Redshift and spectral index for the FR~II quasars that are detected in VLSSr. The curve shows the relation found by \citet{Ker12} using a dataset that is independent from ours.}\label{fig:alpha_evo}
\end{figure}

The median projected separation of the lobes of the SDSS quasars is 261~kpc; the largest and smallest separation are 55 and 497~kpc, respectively.  The highest redshift of the FR~II quasars is 3.4 and the median redshift is 1.06 (for the parent sample of quasars the maximum and median redshift are 5.46 and 1.83, respectively). Using the virial mass estimates of \citet{Shen11}, which are available for all SDSS quasars from DR7, we find a median black hole mass of $10^{9.2}\,M_\odot$ for the FR~II quasars \citep[for a recent study of the radio-loud fraction of SDSS quasars as a function of black hole mass, see][]{Kratzer14}. We find a weak correlation between redshift and the spectral index measured between 74~MHz and 1.4~GHz, $\alpha \propto (-0.11\pm 0.04) \times \log[1+z]$ (Fig.~\ref{fig:alpha_evo}).

To study the properties of the SDSS FR~II quasars it will prove valuable to have a second quasar sample selected at different wavelength. We therefore applied the elegant colour selection proposed by \citet{Stern12}, $W2>15$ and $W1-W2>0.8$ (Vega magnitudes), to data from the {\it Wide-field Infrared Survey Explorer} \citep[{\it WISE};][]{Wright10} All-Sky Release. This mid-IR requirement selects powerful AGN with high purity ($\approx 90\%$). We find 1546 {\it WISE} quasars at the centers of our FR~IIs (Eq.~\ref{eq:dcut}). Thanks to the lower optical depth at IR frequencies, the {\it WISE} sample contains more obscured AGN compared to optical/UV selection \citep[e.g.,][]{Lusso13}, but not all radio-selected AGN are retrieved \citep{Gurkan14}. 

Recalling that the main aim of this work is to test if the areal density of radio pairs can be obtained from the quasar luminosity function, the following two sections focus on two properties of the FR~II quasars that are needed to reach this goal: the efficiency factor for turning accretion power into lobe radio luminosity and the fraction of quasars that are found in our FR~II sample.

\begin{figure}
\centering
\includegraphics[trim=6mm 0mm 0mm 6mm, clip, width=0.5\textwidth]{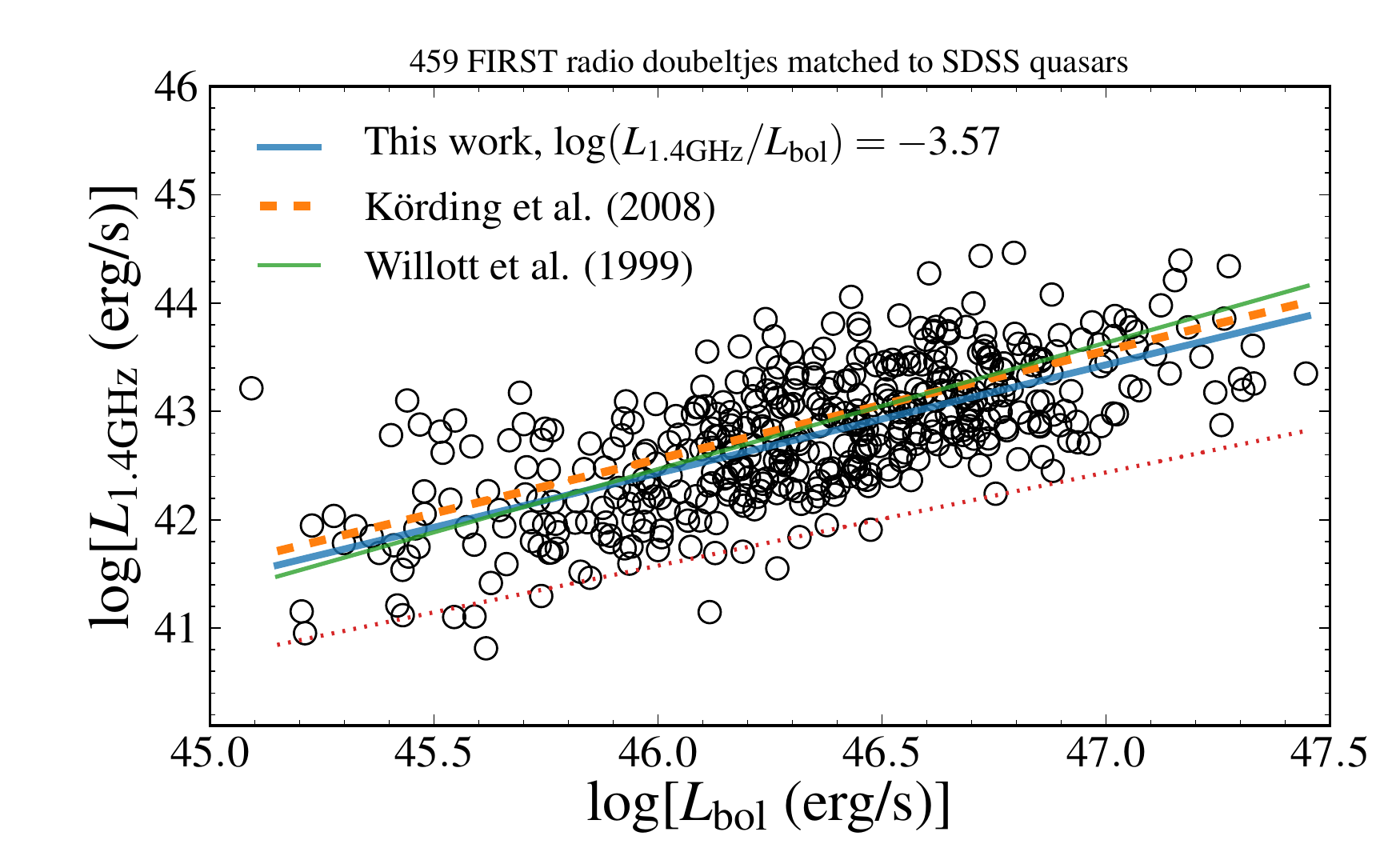}
\caption{We observe a linear correlation between the radio luminosity of the lobes ($\nu L_\nu$ at 1.4~GHz in the rest-frame) and the bolometric quasar luminosity (obtained from the \mbox{$i$-band} magnitude). The scatter of $L_{\rm 1.4GHz} / L_{\rm bol}$ is 0.47~dex (rms). Our normalization is consistent with the results of \citet{Koerding08}, who matched SDSS quasars to point sources at 74~MHz. Furthermore, the well-known minimum-energy relation between  jet power ($Q_j$) and lobe radio luminosity \citep{Willott99} combined with standard jet-disc coupling ($Q_j / L_{\rm bol} = 0.2$) is also consistent with our observations. The red dotted line shows the correlation that would be obtained if all quasars clustered at the flux limit of the radio survey. }\label{fig:Bol-Lradio}
\end{figure}

\subsubsection{Optical-radio correlation}\label{sec:optical-radio}
A linear coupling between the power of the accretion disc ($L_{\rm bol}$) and the power of the jet ($Q_j$) has been observed using radio emission from the compact jet core \citep{Falcke95I} and the extended lobes \citep{Rawlings91}. \citet{Willott99} used the 3CR  and 7CR radio samples to measure a correlation between the 151~MHz luminosity and the [OII] line luminosity of the central source. The two radio samples span over a decade in flux, and thus it was incontrovertibly shown that the disc luminosity predominantly correlates with radio luminosity, not redshift \citep{Willott99}. 

In this work, we use the \mbox{$i$-band} luminosity of the quasars as a proxy for disc power. The luminosity- and frequency-dependent bolometric correction of \citet*{Hopkins07} is used to convert the rest-frame optical luminosity of the quasars to their bolometric luminosity. The $K$-correction for the rest-frame radio luminosity is obtained using the mean spectral index of our sample ($\alpha=-0.85$). As expected based on the previous work summarized above, we find a linear relation between disc and lobe luminosity (Fig.~\ref{fig:Bol-Lradio}). The normalization is: 
\begin{equation}\label{eq:radeff}
\epsilon_r \equiv \log L_{\rm 1.4GHz}/L_{\rm bol} = -3.57\pm 0.47\quad.
\end{equation} 
This linear correlation is not induced by the finite flux limit of the surveys that we used. First of all, we observe a significant flux-flux correlation: the $p$-value of the Spearman rank correlation coefficient between the \mbox{$i$-band} flux and the 1.4~GHz flux is $10^{-15}$ (a Kendall's tau analysis yields the same significance level). Furthermore, the Spearman partial rank correlation \citep[e.g.,][]{Macklin82} was used to check the significance of the $L_{\rm bol}$-$L_{\rm 1.4 GHz}$ correlation, in the presence  of the $z$-$L_{\rm bol}$ correlation. We found $p=10^{-12}$ for this partial correlation analysis, confirming once more that the disc-lobe correlation is not induced by an underlying correlation between redshift and luminosity.

The depth of the FIRST data allows us to measure outliers to the disc-lobe correlation of about 1~dex (the median flux of the radio-selected FR~IIs is 26~mJy, while the median flux of the FR~II quasars is 80~mJy). As shown in \citet{vanVelzenFalcke13} the residuals to the disc-lobe correlation follow a Gaussian distribution and are dominated by environmental effects (i.e., the conversion of jet power to radio luminosity) rather than internal processes (i.e., fluctuating jet power at a fixed accretion rate). We stress once more that the observed correlation presented here applies only to double-lobed radio sources and is therefore not incompatible with the much broader radio-loudness distribution of SDSS quasars and single FIRST sources \citep[e.g.,][]{SingalPetrosian13}.

Our measurement of $\epsilon_r$ is consistent with minimal energy arguments for synchrotron emission as presented in \citet{Willott99}, $Q_j \propto L_{\rm 1.4GHz}^{6/7} f^{3/2}$. For a fudge factor that is typical for FR~II sources, $f=10$ \citep{Blundell00, Godfrey13}, our observations imply that about 20\% of the bolometric quasar luminosity ends up in the jet, $q_j \approx 0.2$. The empirical correlation between 74~MHz luminosity and $L_{\rm bol}$, as measured by \citet{Koerding08} using the DR5 version of the SDSS quasar catalogue and sources from VLSS \citep[][]{Cohen07}, is in excellent agreement with our result (Fig.~\ref{fig:Bol-Lradio}). 

\begin{figure*}
\includegraphics[trim=7mm 0mm 5mm 6mm, clip, width=0.48\textwidth]{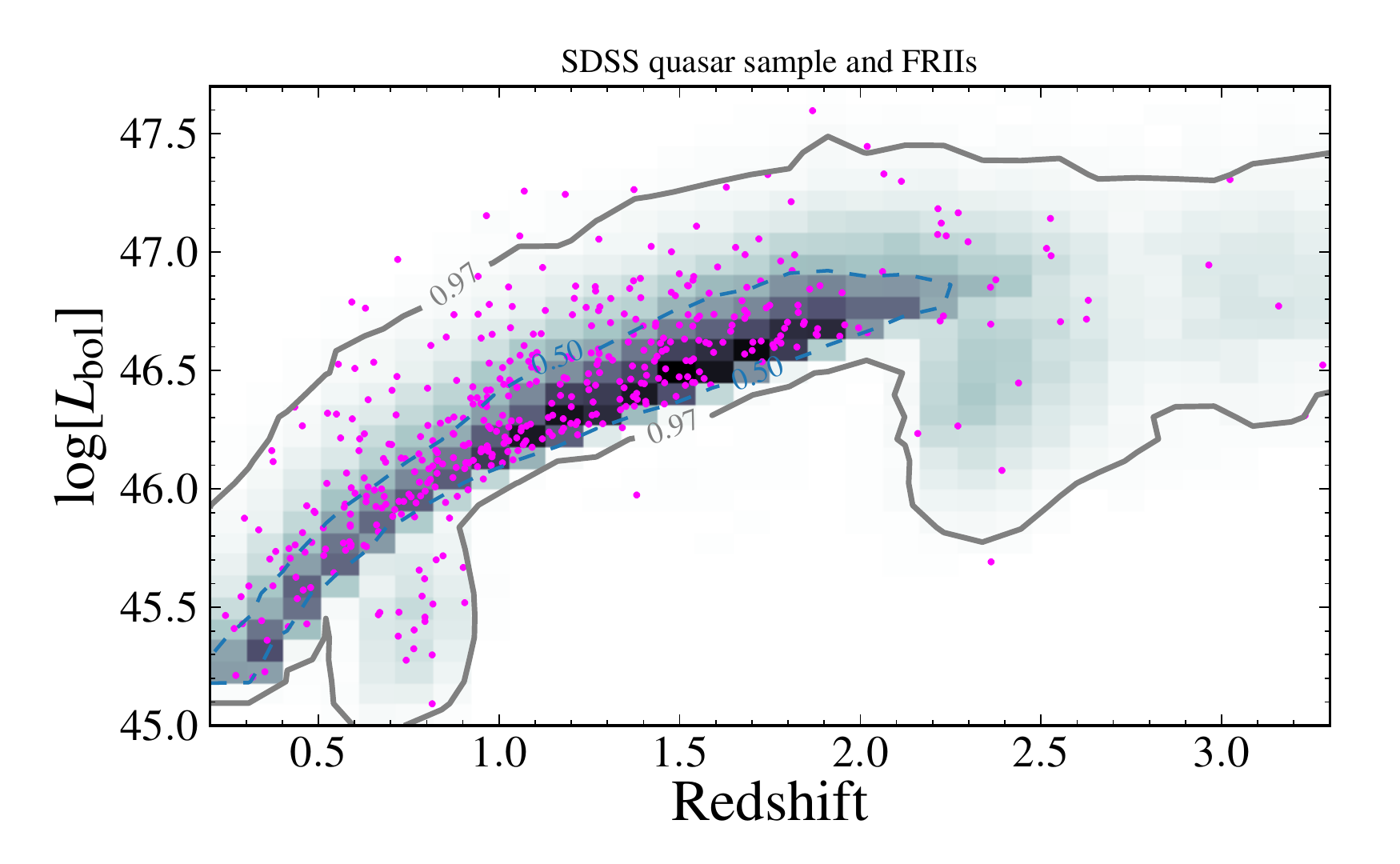} \quad
\includegraphics[trim=7mm 0mm 5mm 6mm, clip, width=0.48\textwidth]{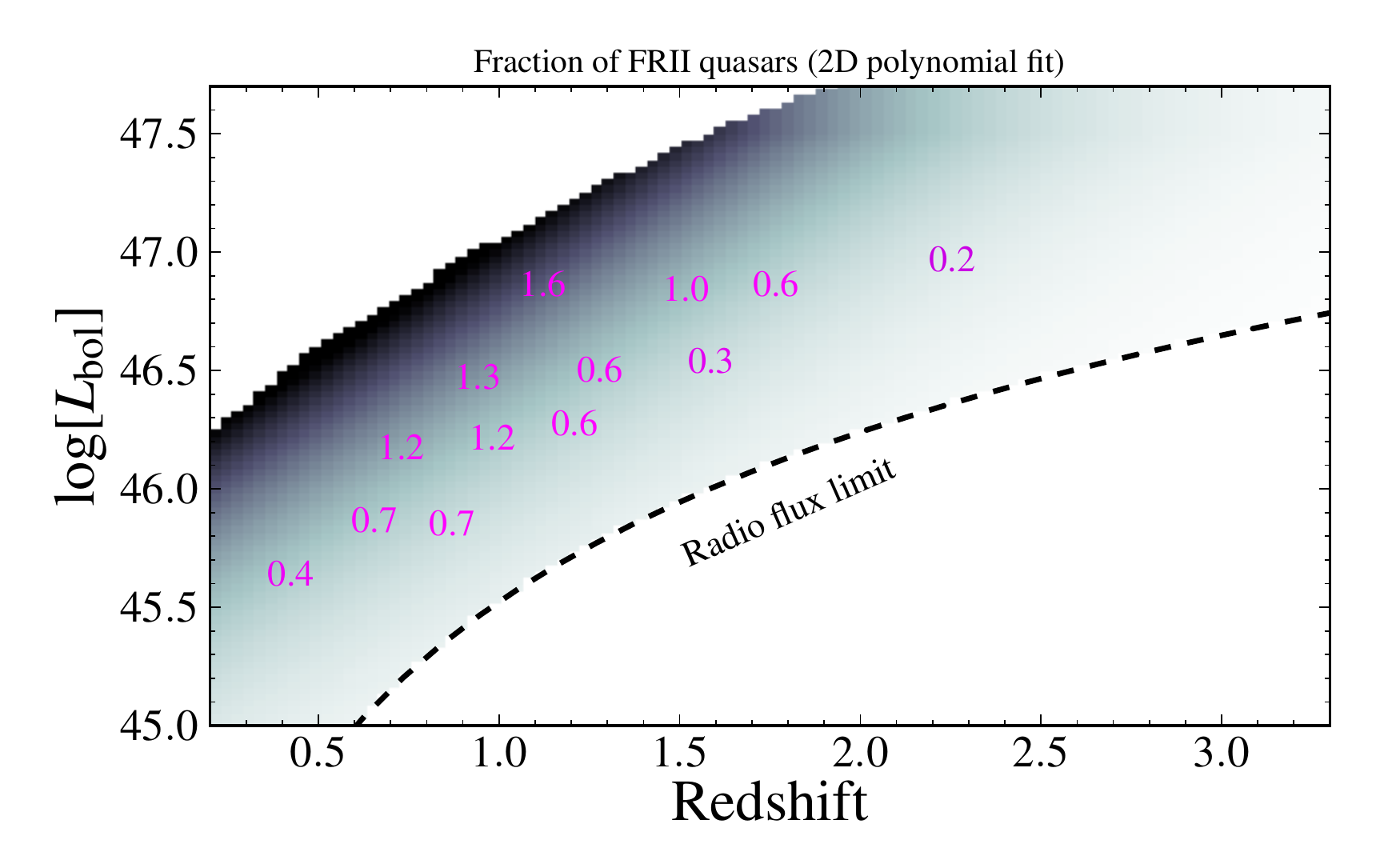} 
\includegraphics[trim=7mm 0mm 5mm 6mm, clip, width=0.48\textwidth]{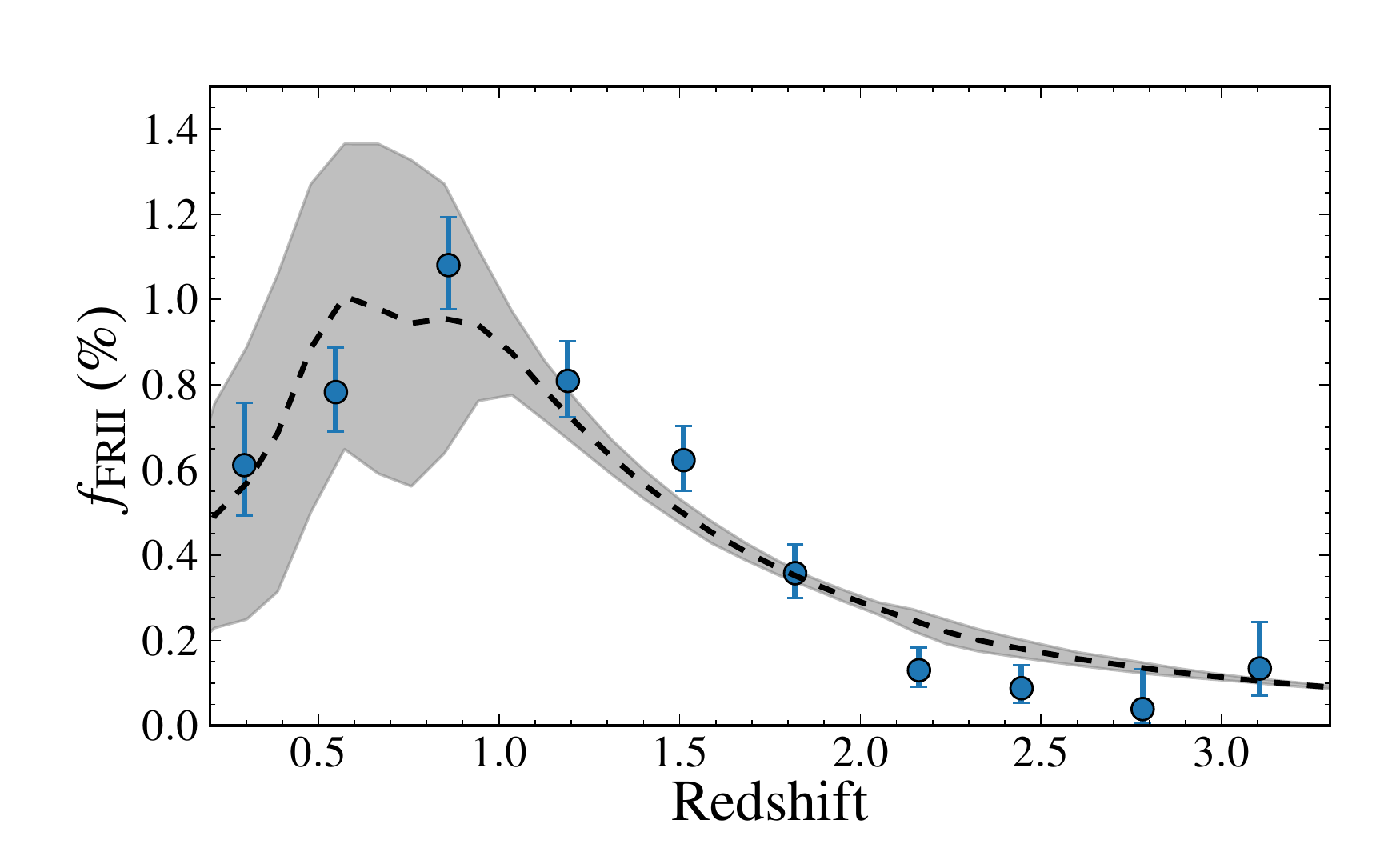} \quad
\includegraphics[trim=7mm 0mm 5mm 6mm, clip, width=0.48\textwidth]{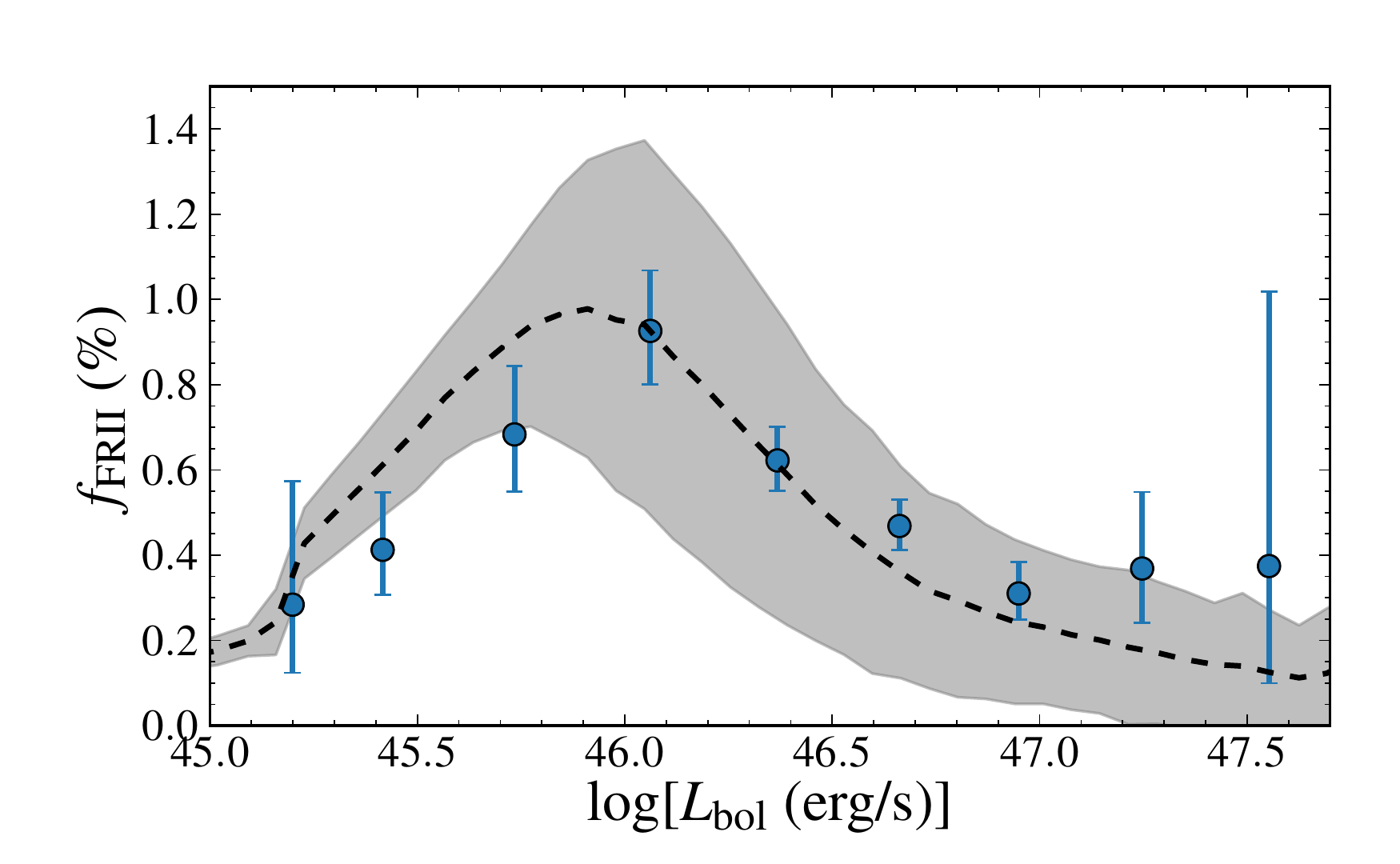}
\caption{Different representations of the fraction of SDSS quasars that are identified as FR~IIs (Sec.~\ref{sec:qcut}). We compute this fraction using only quasars with an estimated radio flux (from the optical-radio correlation, Fig.~\ref{fig:Bol-Lradio}) that is above our radio flux limit. Top left: the distribution of the redshift and bolometric luminosity of our quasar sample (greyscale) and detected FR~IIs (magenta dots). We see that the radio-loud quasars are clearly distributed differently than the parent population. Top right: we binned the detected FR~II quasars in luminosity and redshift, using variable width bins that contain at least 30 sources. For each bin, we compute $f_{\rm FRII}$ (percentages are shown by the magenta numbers). The greyscale shows the result of a 2-dimensional spline fit to guide the eye. We see evolution of $f_{\rm FRII}$ with both redshift (horizontal direction) and bolometric luminosity (vertical direction). The bottom two panels show the evolution of FR~IIs projected onto the redshift axis (left) and luminosity axis (right), and our analytical expression for $f_{\rm FRII}$ (Eq.~\ref{eq:RLfrac}); the grey area shows the rms of this prediction (due to binning in luminosity/redshift). }\label{fig:RLfrac}
\end{figure*}

\subsubsection{FR~II fraction}
The last property that we need to investigate is the fraction of quasars that are identified by our lobe-finding algorithm (Sec.~\ref{sec:init}) and pass the quality cuts (Sec.~\ref{sec:qcut}). This FR~II fraction ($f_{\rm FRII}$) is the product of our selection efficiency and the true fraction of quasars with double lobes.

To compute $f_{\rm FRII}$, we first restrict to the quasars that are inside the FIRST footprint. We also have to account for quasars that are not contained in our sample because their radio flux falls below our radio flux limit ($S_\nu=12$~mJy). Using the observed distribution of the optical-to-radio efficiency and the quasar redshift, we generate a hypothetical lobe luminosity for each quasar. We then remove the quasars with a simulated lobe flux that is below our flux limit. This requirement removes 30\% of the quasars in the parent sample; the majority of the quasars with an estimated radio flux below our threshold are at $z>2$. In the parent population of quasars with a radio flux that should be high enough to be detectable, only 0.57\% are detected, $\left<f_{\rm FRII}\right>=5.7 \times 10^{-3}$. In the left panel of Fig.~\ref{fig:RLfrac} we compare the luminosity and redshift of FR~II quasars to the parent population. We see that the FR~II quasars are typically found at higher luminosity and lower redshift than normal quasars.

The increase of $f_{\rm FRII}$ from low redshift to $z=1$ is partly due to a simple geometrical effect: at low redshift the angular separation of the lobes is larger than our upper limit of 1~arcmin. The number of FR~IIs that have been missed for this reason can be computed directly from the observed distribution of physical lobe-lobe separations. At $z=0.3$, the fraction of missing sources is 50\%, while at $z=0.6$ this is fraction is only 15\%. 

The fraction of quasars that are identified as doubeltjes by our algorithm can be described with the following expression:  
\begin{align}\label{eq:RLfrac}
\log \left[\frac{f_{\rm FRII}}{ 1-f_{\rm FRII}}\right] &= G(z) + c_0 + c_z \log[ (1+z)/2 ]  \nonumber \\ 
& +\log\left[1-\exp(-L/L_{\rm I/II})\right] \quad.
\end{align}
Here $G(z)$ is the geometrical factor that accounts for missing sources due to large angular separations at low redshift. The normalization of $f_{\rm FRII}$ at $z=1$ is given by $c_0$, while $c_z$ describes the redshift evolution. 

To account for the luminosity-dependent FR~I/II transition, we use an exponential function $\exp(-L/L_{\rm I/II})$.  At $z<0.5$ this transition is observed in the range $L_{\rm bol}=10^{43.7-46.2} \, {\rm erg} \, {\rm s}^{-1}$ \citep{Ledlow96}. Unfortunately, the dynamic range of the SDSS quasar catalogue is not sufficient to directly determine this break from the full sample: very few quasars with $L_{\rm bol}<10^{46}\,{\rm erg}\,{\rm s}^{-1}$ are found for $z>1$ (Fig.~\ref{fig:RLfrac}, top left panel).  We therefore restrict to $0.6<z<0.9$, a small redshift window with the highest dynamical range containing 93 FR~II quasars. Using an unbinned maximum likelihood method we find $L_{\rm I/II}=45.9$. Taking into account the degeneracy with redshift evolution ($c_z$ in Eq.~\ref{eq:RLfrac}), the 68\% confidence interval of $L_{\rm I/II}$ is [45.7, 46.2]. We shall adopt $L_{\rm I/II}=46$ as our fiducial choice, but we will also compute how the uncertainty on the location of the break influences our results. For the remaining two parameters that describe $f_{\rm FRII}(z,L)$ we find $\log c_0=-1.96 \pm 0.02$, and $c_z=-3.1 \pm 0.2 $. 

When we repeat the measurement of the FR~II fraction using only sources with a triple morphology, we find $\log c_0=-2.22 \pm 0.03$, $c_z=-3.2 \pm 0.3 $ (again for $L_{\rm I/II}=46$). For this subclass of FR~IIs, the observed disc-to-lobe efficiency is $\epsilon_r = -3.72\pm 0.42$.

\section{Predicting the FR~II density}\label{sec:predict}
In the previous sections we obtained a well-defined sample of $10^4$ double-lobed radio sources and,  using a subset of these doubeltjes, we measured a linear correlation between the rest-frame radio luminosity and the bolometric disc power. We shall now try to reproduce the areal density of the double-lobed radio sources under the assumption that all of these lobes are created by jets from quasars that follow the observed disc-lobe correlation (Fig.~\ref{fig:Bol-Lradio}).

We start by converting lobe flux ($S_{\nu}$) to 1.4~GHz rest-frame luminosity,
\begin{equation}
L_{\rm 1.4GHz}(z) = 4\pi d_L^2   S_{\nu}  (1+z)^{-1-\alpha} \quad .
\end{equation}
Here $d_L$ is the luminosity distance. We model the radio spectral energy distribution (SED) using the observed mean spectral index between 1.4~GHz and 325~MHz ($\alpha=-0.85$) to make the K-correction. Next, we convert this radio luminosity to the bolometric luminosity of the quasar,
\begin{equation}\label{eq:StoL}
L_{\rm bol}(S_\nu, z) = \frac{4\pi d_L^2   S_{\nu}  (1+z)^{-1-\alpha} 1.4~{\rm GHz}}{\epsilon_r} \quad, 
\end{equation}
using the disc-to-lobe efficiency as measured using our sample of 459 FR~II quasars (Fig.~\ref{fig:Bol-Lradio}). 

Since the environment that is probed by the jet can vary from galaxy to galaxy, two jets with identical kinetic power may not yield lobes with equal radio luminosity \citep[e.g.,][]{Hardcastle13}. When the number density decreases with luminosity, this intrinsic scatter of the bolometric-to-radio efficiency ($\epsilon_r$, Eq.~\ref{eq:radeff}) will lead to an overall increase of the predicted number of sources.  The intrinsic scatter of the bolometric-to-radio conversion, $\sigma(\epsilon_r)$, is unknown, but can be constrained to a relatively narrow range. Since the fluctuations in environment as probed by the two jets from a single black hole are at least as large as the fluctuations in environment between black holes, a lower limit for $\sigma(\epsilon_r)$  follows from the observed lobe-lobe flux ratio of the FR~II quasars,  $\sigma(\epsilon_r)>0.30$.
An upper limit for $\sigma(\epsilon_r)$ follows from the observed rms scatter in the optical-radio correlation, minus the observational uncertainty. The latter is dominated by the dispersion of the bolometric correction, which is estimated to be 0.15~dex for our optical quasar sample \citep{Hopkins07}. We thus obtain $\sigma(\epsilon_r) \leq (0.47^2 - 0.15^2)^{1/2} = 0.45$. 

The final ingredient that is required to predict the number of FR~IIs per square degree is the accretion density, i.e., the disc power per unit volume that is available to create jets. We will use the bolometric quasar luminosity function as determined by \citet*{Hopkins07}. These authors combined 28 different quasar samples from optical (both broad-band and emission lines), soft and hard X-ray, and near- and mid-IR bands in the redshift interval $z=[0, 6.5]$. From this wealth of data one can obtain a luminosity-dependent SED and a distribution of column densities, which allows the datasets from different wavelengths to be fitted to a single luminosity function. The best fit is obtained for a broken power law: 
\begin{align}\label{eq:phi}
\phi(L) = \frac{d N}{d \log L } = \frac{\phi_*}{(L/L_*)^{\gamma_1}+(L/L_*)^{\gamma_2}} \quad,
\end{align}
All four parameters of this power law can vary with redshift, $\phi_*(z)$, $L_*(z)$, $\gamma_1(z)$, $\gamma_2(z)$.
The luminosity density peaks at $z\approx 2$, where $j_{\rm bol}\sim 10^8\,L_\odot\,{\rm Mpc}^{-3}$ \citep{Hopkins07}.

\begin{figure*}
\begin{centering}
\includegraphics[trim=9mm 0mm 5mm 6mm, width=0.490\textwidth]{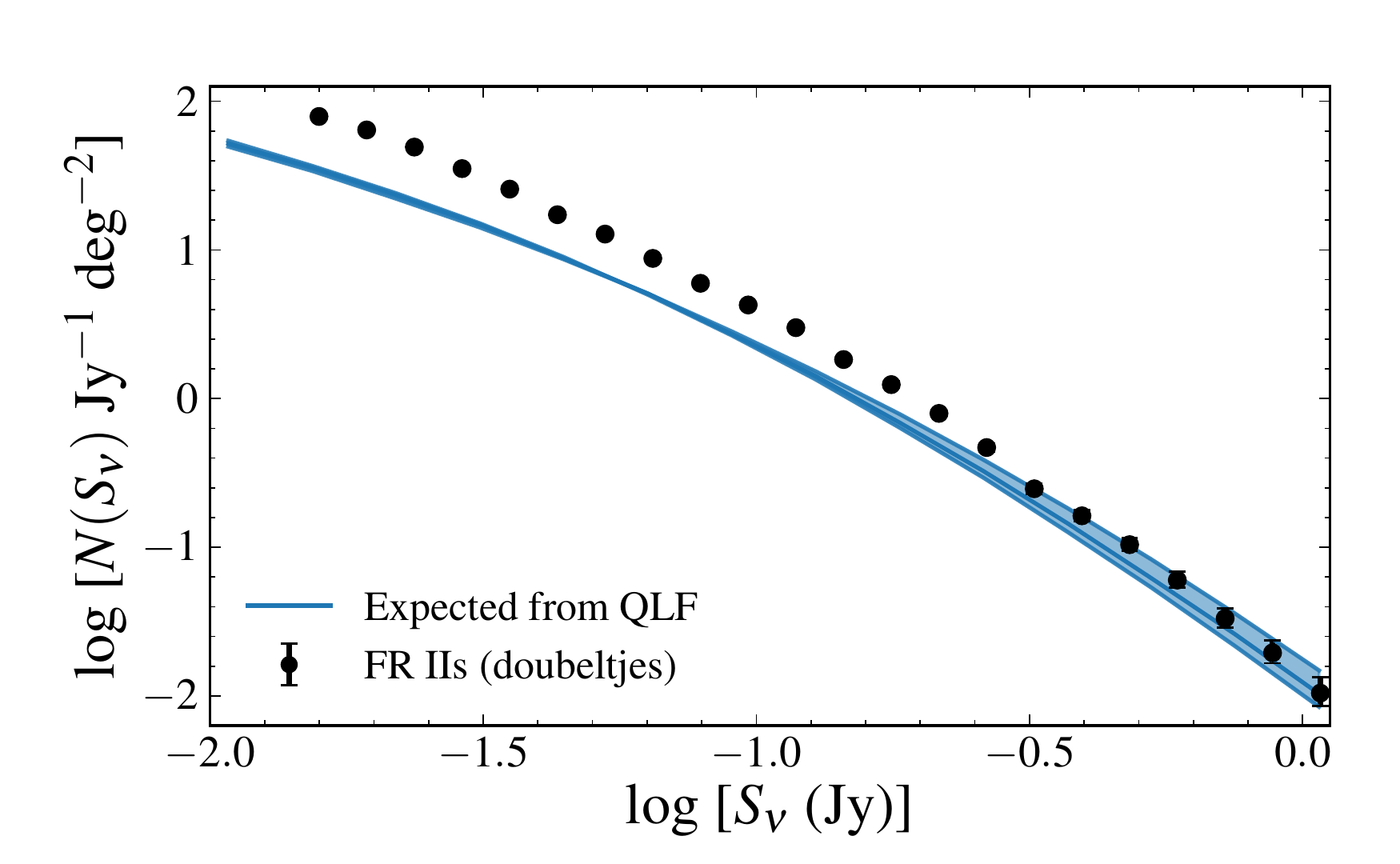}
\includegraphics[trim=17mm 0mm -3mm 6mm, clip, width=0.490\textwidth]{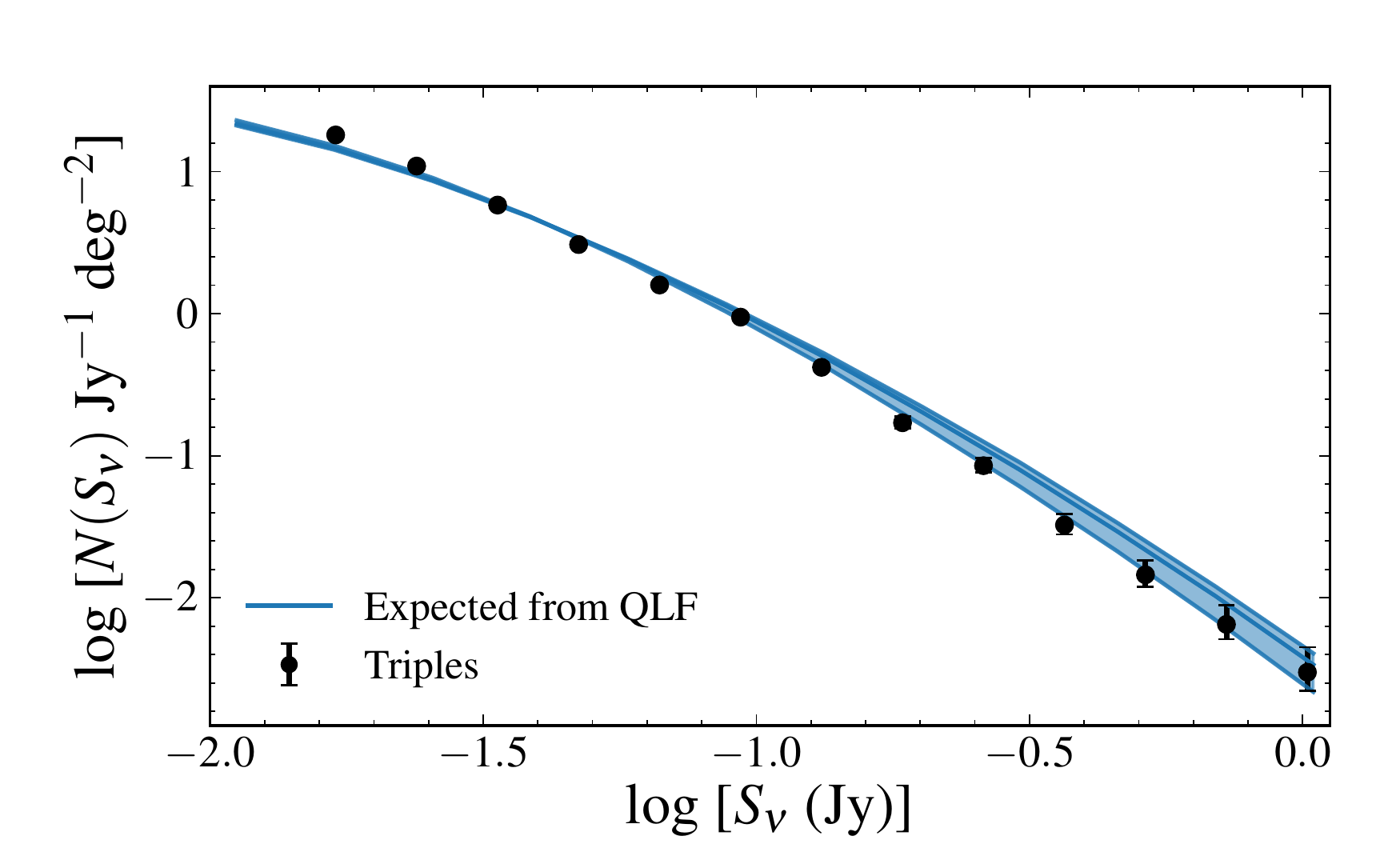}
\caption{The observed distribution of FR~II radio sources at 1.4~GHz and the predicted number using the quasar luminosity function, for our fiducial model (Sec.~\ref{sec:sum_dub}). The line thickness indicates the uncertainty due to the unknown scatter in the bolometric-to-radio efficiency. In the left panel we show all FR~IIs, while in the right panel we show the result using only sources with a triple morphology (i.e., sources with radio emission from the geometrical centre of the lobes). }\label{fig:logSlogN}
\end{centering}
\end{figure*}

To find the total number of FR~IIs on the sky, we simply integrate the quasar luminosity function over the comoving distance $d_C(z)$. We first give the analytic expression that is valid only for $\sigma(\epsilon_r)=0$:
\begin{align}
 N(>S_\nu) &= 4\pi D_H \int_{0}^{6} {d}z\,\frac{d_C^2}{E(z)}\, ~ \times  \nonumber \\ & \int_{}^{} {d}(\log L) \, f_{\rm FRII}(z, L) \phi(z, L) \quad, \label{eq:int}
\end{align} 
with $E(z)=\sqrt{\Omega_M(1+z)^3+\Omega_\Lambda}$ and $D_H$ the Hubble distance \citep[e.g.,][]{Hogg99}. For the upper limit of the integration over quasar luminosity we use the most luminous SDSS quasar observed within that redshift (note that the steep bright-end slope of the luminosity function, $\gamma_2\approx2$, implies that the final predicted density is almost independent of this upper limit). To take into account the scatter of the bolometric-to-radio conversion, we replace the integration over the quasar luminosity by a Monte Carlo simulation. At each redshift, we generate a sample of quasars with a bolometric luminosity drawn from the luminosity function for that redshift. We then assign a 1.4~GHz lobe luminosity (Eq.~\ref{eq:StoL}) to each quasar using a normal distribution centred at $\log[L_{\rm bol}] - \epsilon_r $. We confirm that for $\sigma(\epsilon_r)=0$, this Monte Carlo method yields the same result as the numerical integration (Eq.~\ref{eq:int}).

\subsection{Summary}\label{sec:sum_dub}
In summary, we derived a model that predicts the FR~II density on the radio sky by using the luminosity function of quasars and a linear relation between the accretion disc and lobe radio luminosity. This approach requires four parameters, all of these parameters are well constrained by observations:

\begin{itemize}

\item $\phi(z, L)$: the quasar bolometric luminosity function \citep{Hopkins07}, determined for $0<z<6$ by combining a large number of quasars from (rest-frame) infrared (IR; 8--15$\mu$m), optical ($\approx 1500$\AA), soft X-ray (0.5--2~keV), hard X-ray (2--20~keV), and emission line observations (H$\alpha$, [O~II], [O~III]). 

\item $f_{\rm FRII}(z, L)$: the fraction of quasars that we identify as double-lobed radio sources. This parameter is a function of disc luminosity and redshift. It is the product of the true radio-loud fraction and our selection efficiency. We determined this parameter using $2 \times 10^5$ quasars from SDSS (DR7 and DR9) as test cases. For our fiducial model, we use an analytical expression for $f_{\rm FRII}$ (Eq.~\ref{eq:RLfrac}), but we will also present the results obtained using a non-parametric approach (see Fig.~\ref{fig:logSlogN_sys}).

\item $\epsilon_r \equiv \log[L_{\rm 1.4~GHz} /L_{\rm bol}] =-3.57$: the conversion from disc luminosity to the luminosity of the radio lobes at 1.4~GHz (rest-frame). This linear relation was obtained using 459 FR~II SDSS quasars (Fig.~\ref{fig:Bol-Lradio}) and is in good agreement with earlier work \citep{Rawlings91, Falcke95II, Serjeant98, Willott99, Koerding08, Buttiglione10, Antognini12}. 

\item $\sigma(\epsilon_r)$: the intrinsic scatter in $\epsilon_r$, i.e., the spread of lobe radio luminosity for jets launched by black holes with identical bolometric disc luminosity. We can constrain this parameter to $\sigma(\epsilon_r)=[0.30, 0.45]$, as given by the observed lobe-lobe flux ratio and the observed rms of $\epsilon_r$. 
For our fiducial model we adopt $\sigma_r=0.35$.
\end{itemize}

In Fig.~\ref{fig:logSlogN} we show the result for the fiducial model. The predicted FR~II density is most accurate in range of radio flux that is well covered by the SDSS quasars: 30~mJy to 400~mJy. At the median radio luminosity of the FR~II quasars, 80~mJy, we observe an excess of $N_{\rm data}/N_{\rm QLF}=1.76$. For the radio sources with a triple morphology,  the prediction is within 10\% of the observed density ($N_{\rm data, triples}/N_{QLF}=0.9$). Before discussing this result in Section~\ref{sec:disc}, below we first present a detailed discussion of potential sources of systematic uncertainty. 

\subsection{Systematic uncertainty}\label{sec:sys}
Below we discuss sources of systematic uncertainty of our model for the FR~II density. The results are summarized in Fig.~\ref{fig:logSlogN_sys}. At the median radio flux of the FR~II quasars (80~mJy), we find that the systematic uncertainty is 10\%. For lower fluxes, however, the different models start to diverge and our prediction is less robust.


\subsubsection{Luminosity and redshift evolution}
\begin{figure*}
\centering
\includegraphics[trim=9mm 0mm 5mm 6mm,  clip, width=0.490\textwidth]{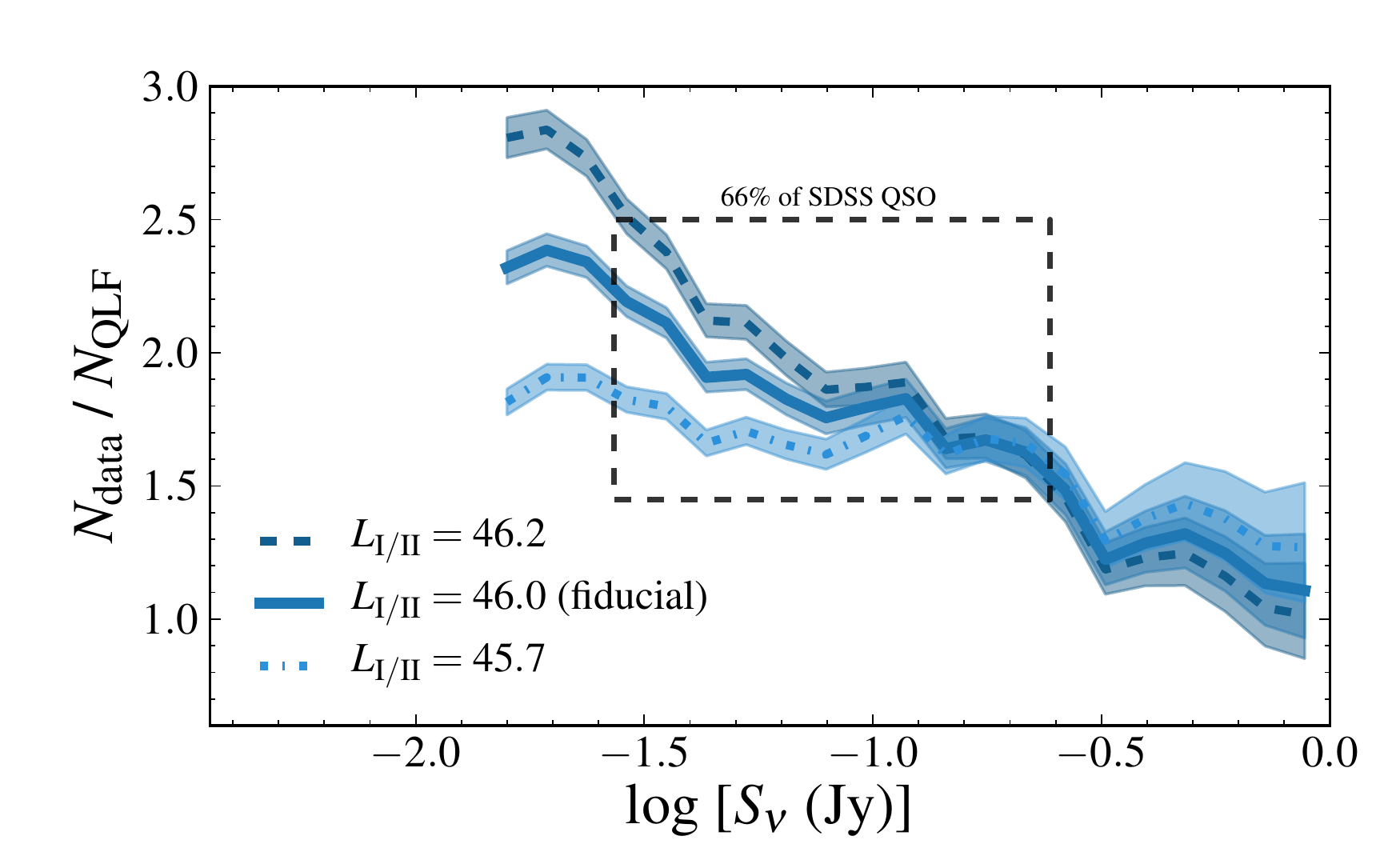}
\includegraphics[trim=17mm 0mm -3mm 6mm,  clip, width=0.490\textwidth]{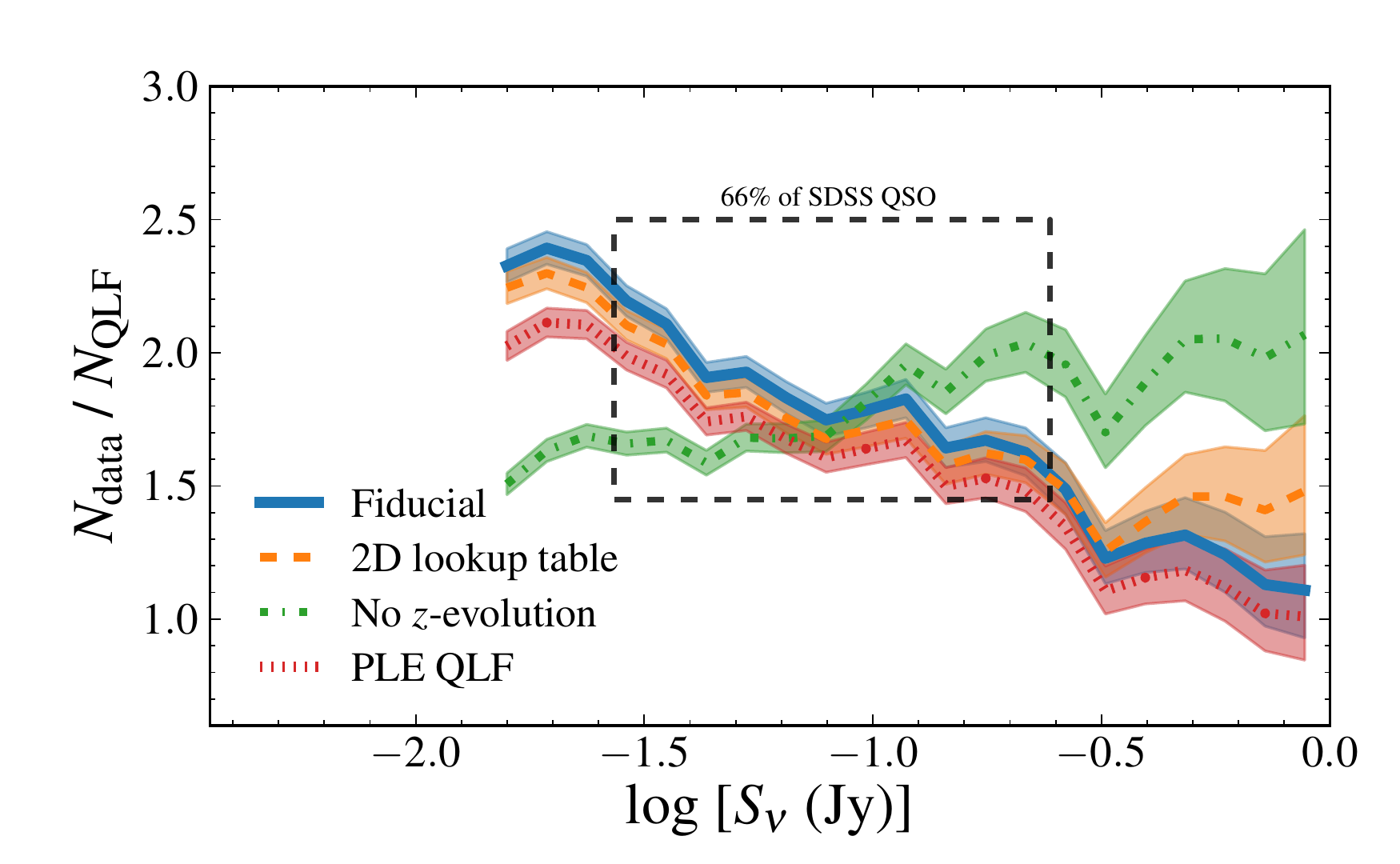}
\caption{Left: Difference between the number of observed FR~II radio sources and the predicted number from the quasar luminosity function. We show the result for the allowed range of $L_{\rm I/II}$ (see Eq.~\ref{eq:RLfrac}). Right: The difference between the model and the data for fiducial model and three slightly less sophisticated methods (Sec.~\ref{sec:sys}). The factor two difference between the prediction and the data is robust with the range of radio fluxes covered by our 459 FR~II quasars (indicated by the dashed lines). The shaded regions indicated the 1-$\sigma$ confidence interval for Poisson statistics of the detected number of sources.}\label{fig:logSlogN_sys}
\end{figure*}

The fraction of quasars that are identified as FR~II radio sources by our method ($f_{\rm FRII}$) is observed to evolve with redshift and luminosity (Fig.~\ref{fig:RLfrac}).  For the fiducial model this evolution is parametrized by Eq.~\ref{eq:RLfrac} with $L_{\rm I/II}=10^{46}\,{\rm erg}\,{\rm s}^{-1}$. We first investigate how the results changes for other break luminosities that are allowed by the data, $L_{\rm I/II}=10^{46.2}\,{\rm erg}\,{\rm s}^{-1}$ and $L_{\rm I/II}=10^{45.7}\,{\rm erg}\,{\rm s}^{-1}$. As shown in the left panel of Fig.~\ref{fig:logSlogN_sys}, a change in $L_{\rm I/II}$ is only important at low radio flux (at higher flux, the bolometric luminosity of most of the FR~IIs is well above the break). At 80~mJy, $N_{\rm data}/N_{\rm QLF}=[1.86,\,1.76,\,1.62]$ for $\log L_{\rm I /II}=[46.2,\,46,\,45.7]$, respectively.

We also considered evolution of the break luminosity with redshift: $L_{\rm I/II}(z) \propto (1+z)^{c_L}$. Our fit for $f_{\rm FRII}$ with this extra free parameter yields $c_0=-1.82\pm 0.02$, $c_z=-4.0 \pm 0.2$, and $c_L=2.3 \pm 0.5$. The resulting predicted density at 80~mJy is 3\% lower. 

To investigate the effect of redshift evolution we first repeat the calculation using an FR~II fraction that depends only on luminosity. Even though this description of the FR~II fraction is clearly inaccurate (see Fig.~\ref{fig:RLfrac}), the resulting density at 80~mJy is only 3\% higher than the fiducial model. As expected, this model diverges at lower and higher fluxes (Fig.~\ref{fig:logSlogN_sys}, right panel). We also considered a different functional form to describe the fraction of FR~II quasars:
\begin{equation}\label{eq:RLfrac_full}
\log f_{\rm FRII} = c_{0'} + c_{L'} (\log L_{\rm bol} -46) + c_{z'} \log[(1+z)/2] \quad,
\end{equation}
with $c_{0'}=-2.33\pm 0.02$, $c_{L'}=0.77\pm 0.03$, $c_{z'}=-4.4\pm0.2$. For this function we find a predicted density that is 13\% lower than the fiducial model. We also repeated the calculation for the predicted FR~II density using an evolving spectral index as measured by \citet{Ker12}, $\alpha(z) = -0.30 \log[1+z] - 0.75$. This curved spectrum slightly reduces the detectability of high-redshift FR~IIs, but yields no significant difference on the predicted density ($N_{\rm data}/N_{\rm QLF}=1.72$). 

Finally, we consider a non-parametric description for the fraction of quasars that are detected as FR~IIs. We binned the 459 FR~II quasars in redshift and luminosity, using 19 or 20 sources per bin, to create a `2D lookup table'. Using linear interpretation we extract $f_{\rm FRII}(z,L_{\rm bol})$ from this dataset. To be able to extrapolate outside the range of the table, we set $f_{\rm FRII}$ to zero at $z=4$ and $L_{\rm bol}=10^{45}\,{\rm erg}\,{\rm s}^{-1}$. Using this approach, we find $N_{\rm data}/N_{\rm QLF}=1.67$,  which is 5\% lower than our fiducial model. Since the linear extrapolation over-estimates the number of low-luminosity FR~IIs at $z\sim 1$, we should treat the density predicted by the lookup table as an upper limit.

\subsubsection{Parametrization of the quasar luminosity function}
To estimate the uncertainty due to the adopted parametrization of the quasar luminosity function, we repeat the prediction using `pure luminosity evolution' (PLE) for the quasar number density, i.e., a double power law with evolving normalization ($\phi_*$) but no change in the faint or bright end slopes. This functional form of the luminosity function is a worse description of the observed number density, $\Delta \chi^2 =1924 - 1007 = 917$, with $511-508=3$ degrees of freedom \citep{Hopkins07}. The resulting change to our fiducial model is an 8\% increase to the predicted number of radio sources (see Fig.~\ref{fig:logSlogN_sys}). When we use the luminosity function that \citet{Hopkins07} obtain after adding a scatter of 0.05~dex to the error estimates of each quasar sample, which reflects the variance in the sample-to-sample normalization (and thus underweights the most well-constrained samples), we obtain a predicted FR~II density that is 10\% higher.

\subsubsection{Sample incompleteness}\label{sec:incompleteness}
Our FR~II sample is not complete: we miss small ($d<18$~arcsec) and large FR~IIs ($d>1$~arcmin), corresponding to a projected size  at $z=1$ of 150 and 500~kpc, respectively. This incompleteness does not limit our conclusions because we measure the {\it fraction} of quasars that are identified by our radio-based selection method. To demonstrate the robustness of this approach we removed a random subset of doubeltjes from our sample, reducing its size by a factor of 2.  For this smaller sample we obtain $N_{\rm data}/N_{\rm QLF}=1.88$, which is consistent with excess of sources measured using the full doubeltjes sample. 

For $i<19.1$, the areal density of a complete optically-selected quasars sample is estimated to be 10.2~deg$^{-2}$ \citep[][]{VandenBerk05} or 12.4~deg$^{-2}$ (by integrating the optical quasar luminosity function). The areal density of the SDSS quasar sample used in this work is about 50\% lower, 6.7~deg$^{-2}$. Our conclusions are not affected by this incompleteness, because we only use the SDSS quasars as {\it examples} of unobscured AGN; completeness is obtained by using the luminosity function. For our analysis, a complete quasar sample is not required, but the sample should be large enough to accurately determine $f_{\rm FRII}$ and $\epsilon_r$. To confirm that our quasar sample is indeed large enough, we repeated our analysis using only quasars from SDSS DR7 (i.e., we exclude the targets from DR9 which reduces the number of quasars by a factor $\approx 2$), finding a change to our prediction of 3\% ($N_{\rm data}/N_{\rm QLF}= 1.82$). Even if we increase the incompleteness of our quasar sample further by using only a random subsets of 50\% of the SDSS DR7 quasars, the predicted density is not changed significantly (for different subsets we find $N_{\rm data}/N_{\rm QLF}=1.8\pm 0.1$). 

We can thus conclude that our radio sample and optical quasars sample are large enough to make a robust prediction of the FR~II areal density. The only remaining problem could be that our SDSS quasars are not a representative subset of the full population of quasars. This is discussed in the next section.

\subsubsection{Potential bias due to calibration with optical AGN}\label{sec:unification_bias}
We implicitly assumed that the SDSS quasars that were used to determine $\epsilon_r$ (Eq.~\ref{eq:radeff}) and $f_{\rm FRII}$ (Eq.~\ref{eq:RLfrac}) are a representative subsample of all quasars. Below we discuss some potential biases that this method may introduce. Again, none of these biases is found to influence our result by more than 10\%.

To test whether the FR~II quasars have a different disc SED compared to normal quasars, we can compare the median optical colours: $g-i=0.27\pm0.01$, $g-i=0.25\pm0.01$ for the FR~II quasars and the full quasar sample, respectively (this comparison was made using a subsample of quasars, drawn from the probability distribution of the FR~II quasar redshifts). Similar to \citet[][cf. their Fig.~7]{deVries06} we find a very small colour difference in the optical SED of double-lobed quasars compared to normal quasars, implying that both populations likely have the same bolometric correction. 

In the grand orientation-based unification scheme of radio-loud AGN \citep{Urry95}, the number counts of type-I and type-II AGN yield a maximum jet inclination of broad-line (type-I) AGN, $i_1<60^{\circ}$ \citep[e.g.,][]{Willott00,Baldi13,Wilkes13}. Since we used type-1 AGN to measure  $f_{\rm FRII}$, this fraction could be underestimated because we sampled a restricted range of jet inclinations.

\begin{figure}
\centering
\includegraphics[trim=6mm 0mm 0mm 6mm, width=0.49\textwidth]{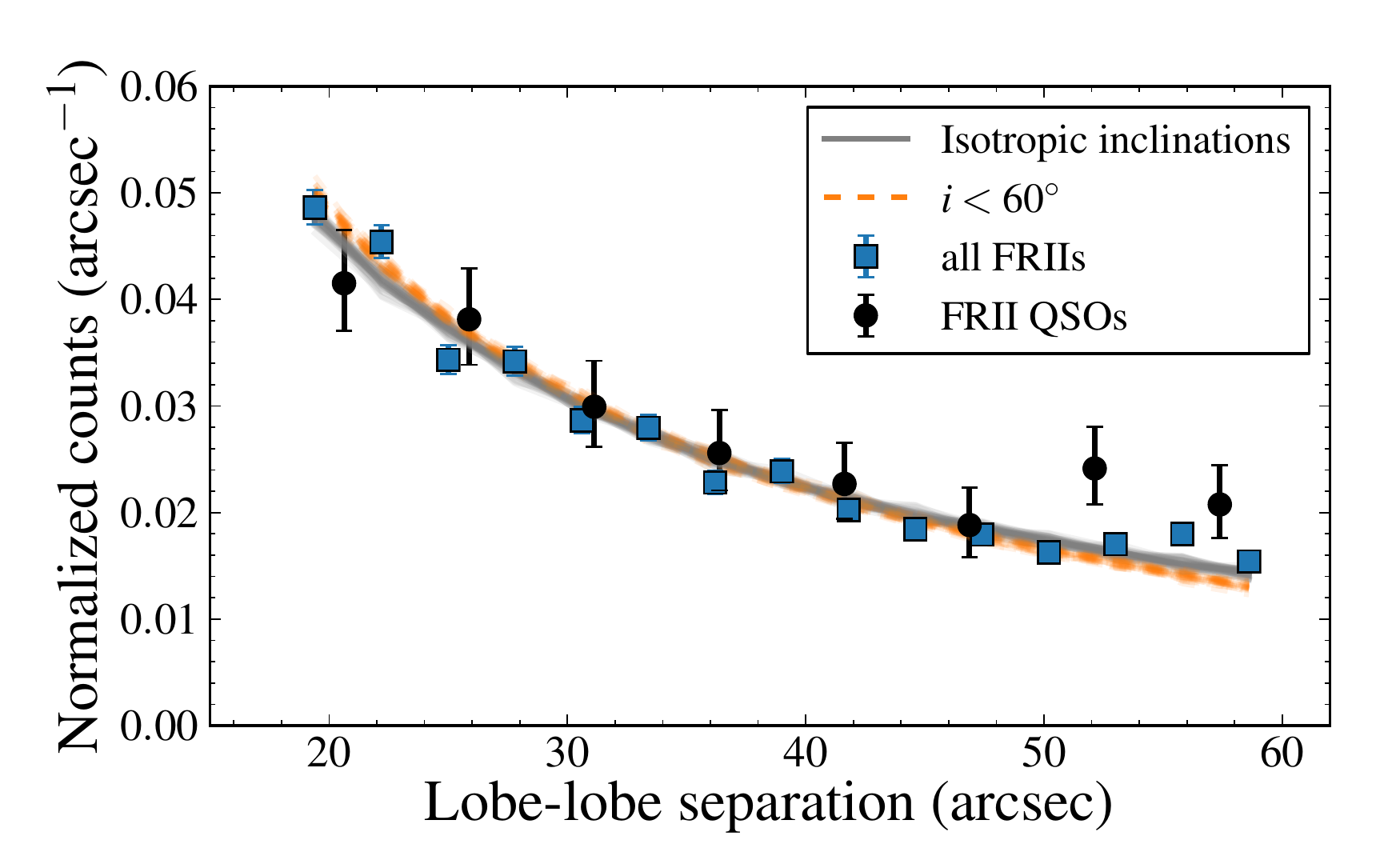}
\caption{The distribution of angular lobe-lobe separation for all FR~IIs and the FR~II quasars. We also show the simulated distribution, as obtained using a Monte Carlo simulation (Sec.~\ref{sec:unification_bias}), for an isotropic distribution of jet inclinations and a capped distribution of jet inclinations.}\label{fig:lobelobe-count}
\end{figure}

To quantify the effect of orientation bias, we ran a Monte Carlo simulation to predict the observed distribution of the lobe-lobe angular separation. We modelled the physical size ($l$) distribution using a power law, $P(l) \propto l^{p}$ (with $100~{\rm kpc}<l<1$~Mpc); the redshifts were drawn from the FR~II quasar redshift distribution. A power-law index of $p=-1$ was found to reproduce the observed distribution of angular distances (Fig.~\ref{fig:lobelobe-count}). Assuming the intrinsic size distribution of quasars and radio galaxies are similar (as predicted in the unified model), we find that for $i<60^{\circ}$, the number of FR~IIs in our sample (i.e., with $18"<d<60"$) is only 4\% lower compared to an isotropic distribution of jet inclinations. 

The effect of orientation bias may even be smaller than 4\% since at the high disc luminosity of our sample, the torus could be flattened due to dust sublimation \citep{Simpson05}. A decrease of the fraction of obscured AGN with luminosity \citep[e.g.,][]{Willott00, Ueda03, Grimes04, Assef13, Lusso13} is often taken as evidence for this receding torus model \citep{Lawrence91,FalckeGopalKrishna95}. Orientation bias may not be relevant at all if the axis that connects the two lobes is not perpendicular to a torus \citep{Singal93,Lawrence10}.  

\section{Results}\label{sec:conclusion}
The main results of this work are as follows.
\begin{itemize}
\item We obtained a sample of $10^4$ FR~II radio sources from the FIRST survey (Fig.~\ref{fig:sep-flux}). 
\item We identified 459 photometrically selected and spectroscopically confirmed SDSS (DR7 plus DR9) quasars that match to the centre of our double-lobed radio sources (Fig.~\ref{fig:dist-sep}).
\item Using the FR~II quasars, we observe a linear correlation between the bolometric disc luminosity and the lobe radio luminosity (Fig.~\ref{fig:Bol-Lradio}), see also \citet{vanVelzenFalcke13}.
\item Our disc-lobe correlation is in very good agreement with the results that \citet{Koerding06} obtained using VLSS (74~MHz) sources that match to SDSS quasars. Since the VLSS-SDSS matches are made regardless of morphology, we can conclude that sources that are (much) smaller than our FR~IIs follow a similar disc-lobe correlation. 
\item The fraction of quasars that show $10^2$ kpc-scale radio lobes decreases strongly with redshift. For $z>1$, $f_{\rm FRII} \propto (1+z)^{-4}$. 
Independently of the redshift evolution, the typical bolometric disc luminosity to obtain an FR~II morphology is observed at $\sim 10^{46}~{\rm erg}\,{\rm s}^{-1}$ (Fig.~\ref{fig:RLfrac}).
\item We observed no significant correlation between the project lobe-lobe separation and the radio luminosity or the disc luminosity (Fig.~\ref{fig:size}).  
\item The observed density of FR~II radio sources exceeds the number predicted from the bolometric quasar luminosity function by a factor $\approx 2$ (Fig.~\ref{fig:logSlogN}, left panel). At $S_\nu =80$~mJy, where this result is most accurate, the systematic uncertainty is estimated to be 10\% (Fig.~\ref{fig:logSlogN_sys}).
\item The observed density of FR~II sources with a triple radio morphology agrees with the number predicted from the bolometric quasar luminosity function (Fig.~\ref{fig:logSlogN}, right panel).
\end{itemize}

\section{Discussion}\label{sec:disc}
Below we first discuss potential explanations for the factor of two difference between the observed and predicted FR~II density. We consider three possibilities: radiatively inefficient accretion flows,  extreme obscuration, and the lifetime of the radio lobes. 
The fact that the predicted and observed density agree when we restrict to radio sources with an active core provides evidence for the last explanation.    

\subsection{Explaining the excess of radio lobes}
It is important to recall that our prediction for the FR~II density is obtained using an empirical estimate for the fraction of quasars that are detected as double-lobed sources by our algorithm ($f_{\rm FRII}$, Eq.~\ref{eq:RLfrac}). In the range of radio flux that is well sampled by the SDSS quasars ($30<S_\nu<400$~mJy); by construction, the predicted FR~II density should not exceed the observed density. 
The observed density, on the other hand, can exceed our prediction.  
The ratio of the predicted and observed FR~II density measures directly what fraction of FR~IIs is matched to jets from {\it known quasars}. Here, by known quasars we mean active black holes whose density has been accounted for in the bolometric luminosity function (Eq.~\ref{eq:phi}). We stress once more that this luminosity function was constructed using observations at mid-IR to hard X-ray frequencies with the aim to  ``represent all AGNs with intrinsic (obscuration-corrected) luminosities above the observational limits at each redshift'' \citep{Hopkins07}. Below we discuss three explanations for the excess of observed double-lobed radio sources.

\subsubsection{Radiatively inefficient accretion and FR~I radio galaxies}
Below an accretion rate of about 1\% of the Eddington luminosity, accreting stellar mass black holes in X-ray binaries are observed \citep{RemillardMcClintock06} to switch to a radiatively-inefficient accretion mode \citep{Narayan95,Yuan14}. A consistent feature of this state is a steady (compact) jet \citep*{Fender04}. There is considerable evidence that accretion discs of AGN also have two states, depending on the (Eddington-normalized) accretion rate \citep{Ho99,Ghisellini01,Falcke04,Koerding06,Plotkin12,Best12}. Radiatively inefficient AGN are likely to be missing in the bolometric quasar luminosity function that we used. Jets from these sources, also known as low-excitation radio galaxies \citep{Laing94}, may appear to be the solution for the observed excess of radio lobes. Since the excess is a factor of 2, at least half of the doubeltjes in our sample must be low-excitation radio galaxies (LERGs).  

The first argument against this scenario is that it would not naturally explain why the observed areal density of radio triples agrees with our predicted density. While high-excitation radio galaxies (HERGs) have a higher core-to-lobe ratio than LERGs, it would be a coincidence that nearly all low-excitation sources are removed when one selects triple morphologies based on FIRST data. In fact, the \citet{Best12} sample of HERGs and LERGs shows that the incidence of triple morphologies is similar for these two classes: of the 2442 HERGs that match to our doubeltjes, 27\% is identified as a triple by our algorithm, for the 81 HERGs that match to doubeltjes this fraction is 33\%.

The luminosity function of \citet{Best12} shows that HERGs dominate over LERGs for $L_{\rm 1.4 GHz}>10^{42} \,{\rm erg}\,{\rm s}^{-1} $, which corresponds to $z>0.5$ when $S_\nu = 80$~mJy. At low redshift, our selection efficiency is poor (as parametrized by $G(z)$ in Eq.~\ref{eq:RLfrac}), hence we expect that HERG dominate the source count of doubeltjes at $S_\nu > 80$~mJy. To make a rough estimate of the relatively numbers of high-excitation and low-excitation sources we integrated the LERG and HERG luminosity functions over redshift (cf. Eq.~\ref{eq:int}), using a single power law to extrapolate the luminosity functions for $L_{\rm 1.4 GHz}>10^{41} \,{\rm erg}\,{\rm s}^{-1}$. For $S>80$~mJy, the estimated HERG to LERG ratio is 3:1. We note that this method yields a lower limit to the HERG contribution, because the \citet{Best12} luminosity functions are valid only at $z\sim 0.1$, while the high-excitation source density is known to increase steeply from $z=0.1$ to $z=1$. 

While our method is optimized for identifying relatively compact lobe emission, we will also pick up some FR~I radio galaxies. The same argument used above for LERGs applies to FR~I radio galaxies; ($i$) our triple sample will also contain FR~Is and ($ii$) the typically lower luminosity of FR~Is implies they likely contribute little to the source count at $S_\nu \sim 10^2$~mJy. 

\subsubsection{Extreme obscuration}
The excess of radio-selected FR~IIs could be explained by a population of powerful active black holes that have been missed in high-frequency quasar surveys due to extreme obscuration. 
In this scenario, one is forced to conclude that the selection of radio sources with a triple morphology completely removes the obscured population (because for the triples, the predicted areal density agrees with the observed density, Fig.~\ref{fig:logSlogN}). This could be possible within the grand unification scheme of radio-loud AGN \citep{Urry95}, but only when the extreme obscuration is caused by the ``dusty torus''.  A possible scenario is the following. Due to Doppler boosting, the selection of sources with a detected core leads to a lower mean jet inclination. If the jet is oriented perpendicular to the obscuring torus,  restricting to radio triples leads to a less obscured view of the accretion disc. Below we briefly discuss the observations of obscured quasars that could be missing in the \citet{Hopkins07} bolometric luminosity function. 

Type-2 quasars have been found in the SDSS spectroscopic galaxy sample \citep{Zakamska03} and about 10\% of these are radio-loud \citep{Lal10}. Heavily obscured quasars can be selected by their mid- or far-IR colours due to reprocessing of the optical/UV disc emission by warm to cool dust \citep[for recent examples of this selection technique see][]{Roseboom13,Mateos13}. We note that the \citet{Hopkins07} luminosity function includes two mid-IR selected quasar samples, namely 8~$\mu$m {\it Spitzer} observations of the Bo\"otes field \citep{Brown06} and a 15$\mu$m sample compiled by \citet{Matute06}, but these are relatively small and contribute little to the fit for the parameters of the luminosity function. 

Estimating the percentage of obscured quasars that are missing in the 2--10 keV AGN samples \citep[e.g.,][]{Ueda03, Barger05,Silverman05} that were used in the \citet{Hopkins07} quasar luminosity function is non-trivial. Analysis of the cosmic X-ray background allows for the presence of Compton-thick AGN (with Hydrogen column densities $N_H>10^{24}\,{\rm cm}^{-2}$) in roughly equal numbers as less obscured AGN \citep{Ueda03}. However, for the high bolometric luminosities of the quasars in our sample ($L_{\rm bol}\sim 10^{46.5}\,{\rm erg}\,{\rm s}^{-1}$), the fraction of Compton-thick AGN is estimated to be less than 10\%  \citep*{Hasinger08,Gilli13}. If the fraction of Compton-thick AGN is indeed about 10\% at the highest disc power, the luminosity function used in this work is nearly complete for our purpose, i.e.,  the excess of radio-selected FR~IIs is not due to extreme, orientation-dependent obscuration of the accretion disc.

Observations of FR~II sources selected from the 3CRR survey show that a sizable fraction (10\% to 50\%) of radio galaxies are underluminous at 10--70$\mu{\rm m}$ compared to radio-loud quasars \citep{Shi05,Ogle06}. Most of these sources are currently classified as low-excitation radio galaxies \citep[e.g.,][]{Gurkan14}, but their lobes could have been created when the accretion rate was higher and the disc was radiating efficiently. So perhaps 10--50\% of galaxies with luminous radio lobes no longer host powerful active engines, which supports our final hypothesis for the excess of radio lobes: the lobe lifetime.

\subsubsection{Lobe lifetime}
To appreciate our final, and preferred, explanation for the excess of radio lobes, we have to consider the nature of the lobe radio emission: synchrotron radiation. When the accretion phase ends, the jet stops supplying power to the hotspots within $\tau_{\rm delay} \sim 10^6$~yr (the light travel time), but the synchrotron cooling time of the electrons can be an order of magnitude longer. The electron cooling time can be estimated for a given synchrotron-emitting region if the properties of the magnetic field in the region are known. The typical cooling time of FR~II lobes at a rest-frame frequency of 2~GHz is $\tau_{\rm sync} \sim 10^{7\pm 0.5}$~yr, depending on the details of the equipartition assumption \citep[e.g.,][]{Komissarov94,Blundell00}. 

The lobe fading time also depends on the evolution of the magnetic field energy and the lobe dynamics, which are non-trivial to calculate. As the lobes expand, the particle density decreases, leading to lower synchrotron luminosity. If the magnetic fields remains in equipartition with the electrons, the field strength will rapidly decrease as the lobes expand. 
The electrons in the lobes will also cool by inverse-Compton scattering of CMB electrons, which can significantly shorten the cooling time at $z>2$ \citep*{Mocz11}. 

If the lifetime of the radio lobes is equal to the duration of the quasar phase \citep[$\sim 10^{6-8}$~yr;][]{Martini04}, radio-selected lobes outnumber FR~II quasars by a factor two. If the accretion disc luminosity is episodic, the density of radio lobes compared to double-lobed quasars could increase further\footnote{Eddington-limited time variability of the accretion disc may also explain why the fraction of radio sources with an quasar counterpart is observed to increase with luminosity \citep{Willott00}.}. Radio sources with a nested morphology (e.g., two young hotspots inside older lobes) are often interpreted as a sign of AGN intermittency \citep[][]{Stanghellini90, Schoenmakers00, Saikia09, Filho11, Nandi14}. 

A few low-redshift ``relic lobes'' \citep{Komissarov94}  have been described in the literature \citep{Parma07,Dwarakanath09,Murgia11}. The estimates of the fading timescales are in the range $10^{6-7}$~yr \citep{Parma07}; longer lifetimes are found for lobes in galaxy clusters $10^{7-8}$~yr \citep{Murgia11}. One may be able to measure the importance of synchrotron cooling for the lobe fading timescale at $z \sim 1$ using observations of doubeltjes at low frequency ($\nu<1$~GHz). The synchrotron cooling time scales as $\nu^{-1/2}$, so the excess of radio lobes should be larger at lower frequency (as long as the fading timescale is longer than the light travel time to the lobes). Testing this idea requires a resolution of at least 20~arcsec, which is currently possible with LOFAR \citep[][]{vanHaarlem13} and the Giant Metrewave Radio Telescope (GMRT).

Finally, we recall that for sources with a triple morphology, the predicted areal density from the quasar luminosity function agrees with the observed density (Fig.~\ref{fig:logSlogN}). For each radio triple, we know that the jet is currently active and thus their areal density cannot be significantly enhanced by the lifetime of the radio lobes. This suggests that the factor two excess of radio-selected FR~IIs can be simply explained by a delayed response of the lobes to the shutdown of the jet.

\subsection{Redshift evolution of FR~II properties and jet feedback efficiency}
From $z=1$ to $z=3$, we observe a factor $\approx 10$ decrease in the fraction of quasars that are identified as FR~IIs (Fig.~\ref{fig:RLfrac}). This decrease is much stronger than the change of the angular diameter distance over this redshift interval (which only decreases the largest lobe-lobe separation that we can detect from 500 to 450~kpc, Fig.~\ref{fig:sep-redshift}). We stress that we measured $f_{\rm FRII}$ using only quasars with a predicted radio flux (from the radio-optical correlation) that is above our flux limit, i.e., the observed evolution is not a selection effect. 
We observed that the FR~II fraction increases with disc luminosity, so the decrease of $f_{\rm FRII}$ for $z>1$ is entirely due to redshift evolution (i.e., Malmquist bias would only lead to an increase $f_{\rm FRII}$ with redshift). 

\begin{figure}
\centering
\includegraphics[trim=4mm 0mm 0mm 4mm, clip, width=0.47\textwidth]{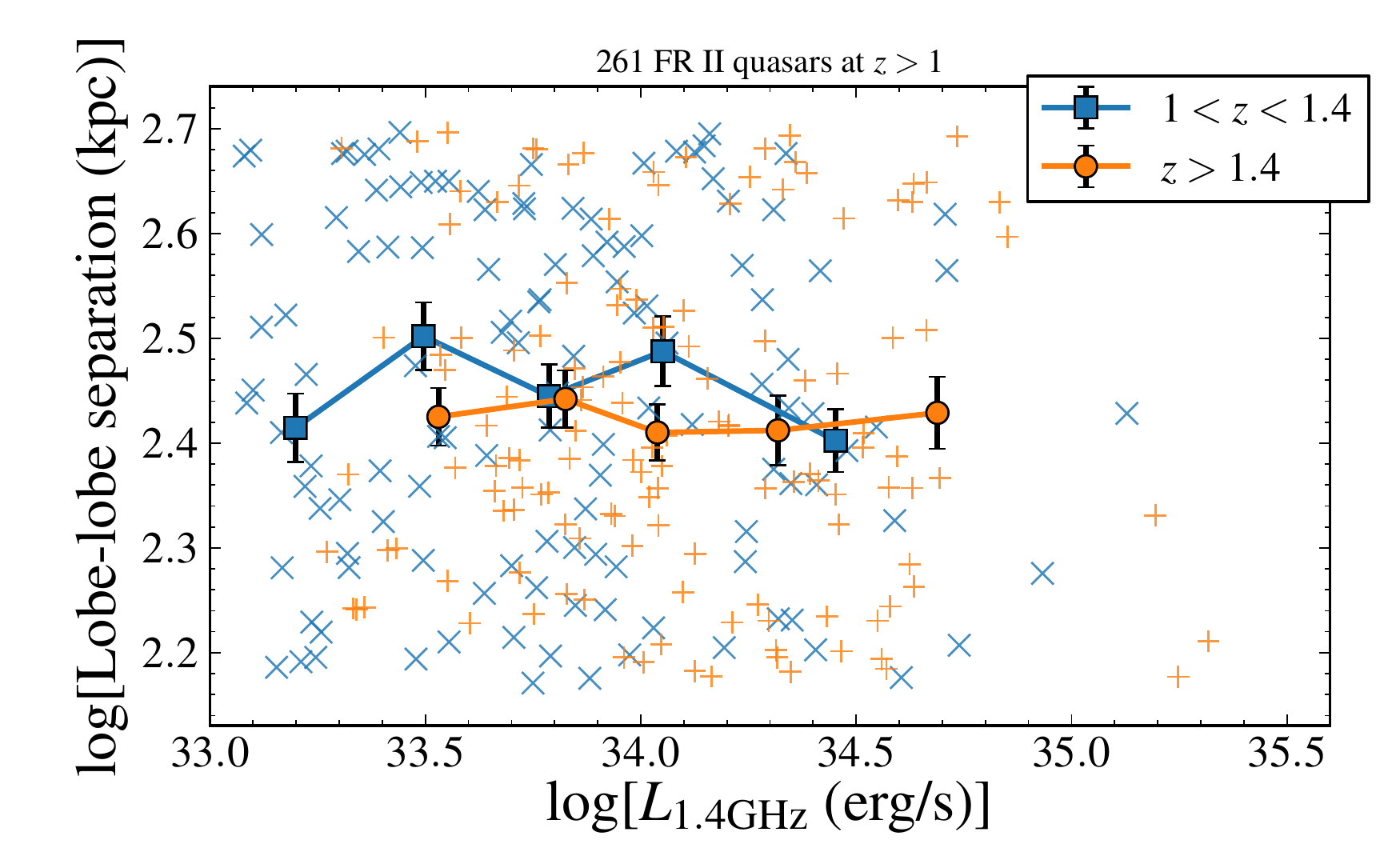}\caption{The mean projected physical distance between the lobes versus the lobe radio luminosity for two redshift intervals.}\label{fig:size}
\end{figure}

Two different processes could explain the observed redshift evolution: a change of external pressure which acts to disrupt the jet before reaching $10^2$~kpc or electron cooling due to inverse-Compton scattering of CMB photons. Evidence for the former hypothesis follows from matching the SDSS quasars to VLSSr (74~MHz) sources. The VLSSr catalogue is slightly more shallow ($F_{\rm 74MHz}>0.5$~Jy or 50~mJy at 1.4~GHz for $\alpha=-0.8$) than our sample of double-lobe sources from FIRST, yet the fraction of quasars with a VLSS match is nearly a factor 4 higher. From $z=1$ to $z=3$ the decrease of the fraction of quasars that match to a VLSSr source is only a factor five. This suggest that about half of the redshift evolution of the 1.4~GHz doubeltjes can be explained by size evolution: powerful radio sources are intrinsically smaller at higher redshift. We note that compact jets at very high redshift ($z \approx 10$) can be identified by their observed radio SED, which peaks at $\sim 10^2$~MHz \citep*{FalckeKoerdingNagar04}.

Earlier studies of FR~II sources have also reported evolution of the physical size of FR~IIs with (photometric) redshift \citep{Oort87,Kapahi89} ---see \citet{Blundell99} for a review. \citet{Neeser95} were the first to use (mainly) spectroscopic redshifts and complete radio samples to find $d\propto (1+z)^{-1.7\pm 0.5}$ (for $\Omega_M=1$, $\Omega_\Lambda =0$). We note that the median size of our 459 FR~II quasars is not observed to change with redshift (Fig.~\ref{fig:size}). This implies that size evolution at $z>1$ only affects sources smaller than $d=150$~kpc and thus support the idea that FR~II size evolution is determined by the interaction of the jet with gas close to the host galaxy, as suggested by \citet{Neeser95}. 
The lack of a correlation between cluster richness and physical size for 3CR sources at $z>0.4$ \citep{Harvanek02} is further evidence that size evolution is due to changes on scales below $10^2$~kpc.

Strong evidence for evolution of the radio-loud fraction with redshift and luminosity has been observed by \citet{Jiang07}. These authors matched SDSS quasars to all FIRST sources within 30~arcsec, yielding a sample that is dominated by unresolved radio quasars. \citet{Jiang07} found that the radio-loud fraction scales as $(1 + z)^{-2}$. Using the same parametrization as \citet{Jiang07}, Eq.~\ref{eq:RLfrac_full}, we find $(1 + z)^{-4}$. This steeper slope again confirms that fewer powerful jets can grow to $d>150$~kpc as the redshift increases. Jets at high redshift thus deposit more of the accreted energy into (or close to) their host galaxy compared to low-redshift sources; at high redshift, powerful radio jets appear to be more effective at supplying feedback to their host galaxy.

\subsection{The quasar fraction}\label{sec:quasar_fraction}
A traditional approach to study unification of quasars and radio galaxies is to use follow-up observations of a complete radio sample to measure what fraction of FR~II radio sources have a quasar at their center. Here the meaning of `quasar' can vary from study to study, depending on the selection method (e.g, the frequency or spectral resolution). In this work, we presented the ``true'' quasar fraction, i.e., as obtained from the bolometric luminosity function. For the full FR~II sample we measure a true quasar fraction of about 50\%, while for the triples this fraction is close to 100\%.  

Most relevant for a comparison to our results is the 7CRS. This survey contains 130 radio sources with $S_{151}>0.5$~Jy from the 7C survey \citep{McGilchrist90} with near-IR spectroscopic follow-up data that are 90\% complete. At $L_{\rm 1.4GHz}\sim 10^{43.5}\,{\rm erg}\,{\rm s}^{-1}$ and $z=1.5\pm 0.5$, the fraction of FR~IIs  that are observed to have broad lines in their spectrum is $34\pm6$\% At  $L_{\rm 1.4GHz}\sim 10^{42.5}\,{\rm erg}\,{\rm s}^{-1}$ and $z=0.5\pm 0.5$, this fraction is $11\pm4$\% \citep{Willott00}. Measured over all redshifts, the 68\% CL interval of the 7CRS quasar fraction is 20--37\% \citep{Grimes04}. 
This quasar fraction is consistent with our result if the near-IR selection of 7CRS has missed a factor $\approx 1.7$ of the full quasar population (due to obscuration). 

For the sake of comparing the SDSS quasars to the 7CRS quasars, we restrict our doubeltjes sample to $S_\nu>0.5~{\rm Jy} \times (1.4/0.15)^{-0.8} =84$~mJy, leaving 3748 sources. For the optically-selected SDSS quasars we require $m_i>19.1$ and we use the radio-optical flux correlation to estimate that 13\% of FR~II quasars with $S_\nu>84$~mJy will fall below this flux limit. We find 203 FR~II quasars, with a median radio luminosity of $L_{\rm 1.4GHz}\sim 10^{43.1}\,{\rm erg}\,{\rm s}^{-1}$, giving a quasar fraction of 5\%.  
This lower fraction compared to 7CRS is due to both the quasar selection method (optical versus near-IR) and the incompleteness of the photometrically selected SDSS quasar sample. When we match the SDSS quasars to all radio sources from the 7C survey with $S_{151}>0.5$~Jy, we again find a quasar fraction of 5\%. (We note that the lower radio frequency and smaller size of the 7CRS sources may also influence the quasar fraction.)

Repeating the exercise for the quasars selected by their mid-IR colours in {\it WISE} (see Section~\ref{sec:sdss_dub}), we find a quasar fraction of 23\%.  Correcting for the 78\% efficiency of the photometric selection \citep{Stern12}, the fraction is 30\%, which is consistent with the 7CRS result. By comparing this with our true quasar fraction, we estimate that mid-IR colour selection can retrieve about half of the full population of powerful radio-loud quasars. Finally, if we adopt a 50\% completeness for the SDSS sample (sec.~\ref{sec:incompleteness}), we estimate that quasar selection based on the UV excess \citep[][]{Richards02, Bovy11} can retrieve about one-fourth of the true quasar population. These two estimates for the obscured/missing fraction are consistent with those found by \citet[][cf. Eq. 4]{Hopkins07}.

Finally, we compare our estimate of the density of radio galaxies to the luminosity function of radio-loud AGN obtained by \citet{Best12} using the SDSS spectroscopic galaxy sample. At $L=10^{25.1}\,{\rm W}\,{\rm Hz}^{-1}$ ($L_{\rm 1.4 GHz}= 10^{42.2}\,{\rm erg} \,{\rm s}^{-1}$), the density of radio-selected AGN is $\log[\phi_{\rm BH12}]= -7.2_{-0.2}^{+0.1}$ (measured per Mpc$^{3}$ per logarithmic luminosity bin) with a typical redshift of $z\approx 0.3$ \citep{Best12}. Using our disc-lobe correlation, this radio luminosity yields a bolometric luminosity of $L_{\rm bol} = 10^{45.7}\,{\rm erg}\,{\rm s}^{-1}$, corresponding to a quasar density of $\log[\phi_{\rm H07}(z=0.3)] = 10^{-5.4}$ \citep{Hopkins07}. Estimating $f_{\rm FRII}$ to be 0.8\% at this luminosity (Fig.~\ref{fig:RLfrac}), we find an FR~II density of $\log[\phi_{\rm dub}] = -7.5$. Hence our estimate of the FR~II fraction combined with the quasar luminosity function agrees reasonably well with the \citet{Best12} radio-loud AGN luminosity function at low-redshift. We note that our density is expected to be lower than obtained from the \citet{Best12} luminosity function since we are not retrieving all radio-loud AGN due to our cut on the angular size.
\section{Conclusion: nature and evolution of radio galaxies}
We have obtained a very large sample of double-lobed radio sources and matched these to SDSS quasars, allowing for arguably the most comprehensive view of FR~II evolution to date.  

We found a strong redshift dependence: at $z>1$, fewer radio-loud quasars have large radio lobes. The areal density of our radio-selected FR~II sample exceeds the expected density based on the bolometric (obscuration-corrected) quasar luminosity function by a factor two. The fraction of FR~II quasars that have an active radio core (i.e., emission within 5~arcsec of the centre of the lobes) was also found to be a factor two higher. Indeed when we use only sources with a triple morphology, the predicted areal density agrees with the observed areal density. We therefore argue that the excess of radio-selected FR~IIs can be simply explained by the lifetime of the lobes, bringing us to the following conclusion on the nature of powerful FR~II radio galaxies at $z\sim 1$: {\it the majority of jets that power FR~II lobes originate from radiatively efficient accretion flows that obey a linear jet-disc coupling.} 
 
Our conclusion implies an almost complete unification of radio galaxies and quasars: nearly all FR~II lobes are powered (or have been powered) by jets from quasars. 

\subsection{Outlook}
We presented an automated method that faithfully selects powerful AGN using only radio data. When applied to the FIRST survey, our method yields the largest sample of FR~II sources to date. 
Surveys from the Jansky-VLA and SKA precursors (LOFAR, MeerKAT, ASKAP) will observe large parts of the radio sky to deeper flux limits, higher resolution, or lower frequencies than was technologically feasible at the time when FIRST was conducted. It will be exciting to apply our morphological selection method to these near-future surveys: we can expect to obtain a complete view of accreting black holes throughout the entire universe.

\section*{Acknowledgements}
\small
We would like to thank G. de Bruyn, D. Cseh, A. Gruzinov, P.\,F. Hopkins, J.\,H. Krolik, H. R\"ottering, P. Uttley, R.\,L. White, and N.\,L. Zakamska for useful discussions. We thank the first referee and we are grateful to the second referee, M. Hardcastle, for providing constructive comments.


The FIRST and VLSS surveys were obtained with the Very Large Area, which is operated by the National Radio Astronomy Observatory. NRAO is a facility of the National Science Foundation operated under cooperative agreement by Associated Universities, Inc.

Funding for the SDSS and SDSS-II has been provided by the Alfred P. Sloan Foundation, the Participating Institutions, the National Science Foundation, the U.S. Department of Energy, the National Aeronautics and Space Administration, the Japanese Monbukagakusho, the Max Planck Society, and the Higher Education Funding Council for England. The SDSS Web Site is http://www.sdss.org/. The SDSS is managed by the Astrophysical Research Consortium for the Participating Institutions. The Participating Institutions are the American Museum of Natural History, Astrophysical Institute Potsdam, University of Basel, University of Cambridge, Case Western Reserve University, University of Chicago, Drexel University, Fermilab, the Institute for Advanced Study, the Japan Participation Group, Johns Hopkins University, the Joint Institute for Nuclear Astrophysics, the Kavli Institute for Particle Astrophysics and Cosmology, the Korean Scientist Group, the Chinese Academy of Sciences (LAMOST), Los Alamos National Laboratory, the Max-Planck-Institute for Astronomy (MPIA), the Max-Planck-Institute for Astrophysics (MPA), New Mexico State University, Ohio State University, University of Pittsburgh, University of Portsmouth, Princeton University, the United States Naval Observatory, and the University of Washington. 

The WENSS project was a collaboration between the Netherlands Foundation for Research in Astronomy and the Leiden Observatory. The WENSS team consisted of Ger de Bruyn, Yuan Tang, Roeland Rengelink, George Miley, Huub R\"ottgering, Malcolm Bremer, Martin Bremer, Wim Brouw, Ernst Raimond and David Fullagar.

This publication makes use of data products from the Wide-field Infrared Survey Explorer, which is a joint project of the University of California, Los Angeles, and the Jet Propulsion Laboratory/California Institute of Technology, funded by the National Aeronautics and Space Administration.

This work was supported by an ERC Advanced Grant (no. 227610, PI: Falcke).

\bibliography{general_desk}


\appendix
\section{Catalogue \& example figures} \label{sec:catalog}
In Table~\ref{tab:cat}, we list the properties of our complete sample of 59,192 candidate FR II sources that remain after applying our flux limit ($S_\nu >12$~mJy, to ensure completeness), but no other cuts. In order to obtain the sample that was used for this paper, one can use the \verb quality_cut  flag (see Table~\ref{tab:cat}). We also give the properties of the FR~II quasars (e.g., redshift, bolometric luminosity). To maximize the size of the FR~II quasar catalogue we used all SDSS quasars from DR7 and DR9 (i.e., we made no cuts on the target selection flags, but only removed duplicates). We find 1108 matches, with an estimated background of 1.9\% random associations. To obtain the subsample of optically-selected FR~II quasars, one can use the \verb uniform    flag. 

In Fig.~\ref{fig:example} we show a random selection of sources from the catalogue. 
\onecolumn
\normalsize

\input{./figs/cutout/include_me.tex}

\begin{table*}
\centering
\begin{minipage}{0.93 \textwidth}
\begin{tabular}{l c l}
\hline \hline
\verb R.A. & deg & R.A., geometrical centre of lobes. \\
\verb Decl. & deg & Decl., geometrical centre of lobes. \\
\verb lobe_flux & Jy & Total lobe flux (integrated flux from FIRST, corrected for missing flux, see Sec.~\ref{sec:missingflux}). \\
\verb core_flux & Jy & Core flux (zero if no core is detected). \\
\verb lobe_ratio &   & Lobe-lobe flux ratio. \\
\verb separation & deg & Lobe-lobe separation. \\
\verb n_gauss &   & Total number of Gaussian components (i.e., at least two). \\
\verb quality_cut & bool & Flag: True if source passes quality cuts (Table~\ref{tab:cuts}). \\
\hline
\multicolumn{3}{c}{The following fields only apply for doubeltjes matched to SDSS quasars via Eq.~\ref{eq:dcut}.} \\ 
\hline
\verb sdss_ra & deg & Coordinates of quasars. \\
\verb sdss_dec & deg & Coordinates of quasars. \\
\verb redshift &   & Redshift \\
\verb i_mag & ABmag 
& SDSS $i$-band magnitude. \\
\verb log10_bol_lum & log erg\,s$^{-1}$ & Bolometric luminosity, using the Hopkins et al. (2007) bolometric corrections. \\
\verb DR &  & SDSS Data Release (i.e., 7 or 9) \\
\verb uniform & bool & Flag: True if QSO was targeted based on optical properties (either in DR7 or DR9). \\
\hline
\end{tabular}
\caption{Description of the doubeltjes catalogue. The catalogue is available as supplementary material (online).}\label{tab:cat}
\end{minipage}
\end{table*}

\end{document}

%% file: figs/cutout/include_me.tex
\begin{figure*}
 \centering 
  \includegraphics[width=43pt]{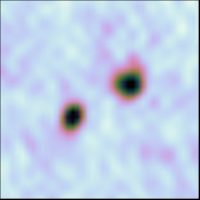} 
  \includegraphics[width=43pt]{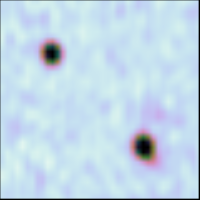} 
  \includegraphics[width=43pt]{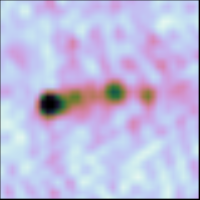} 
  \includegraphics[width=43pt]{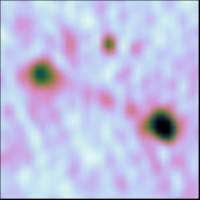} 
  \includegraphics[width=43pt]{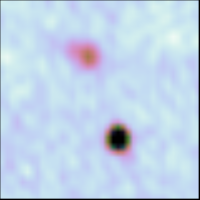} 
  \includegraphics[width=43pt]{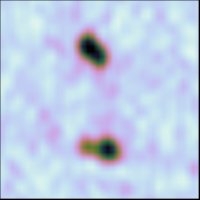} 
  \includegraphics[width=43pt]{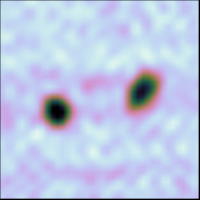} 
  \includegraphics[width=43pt]{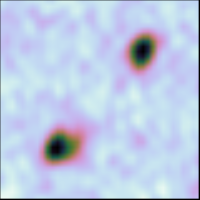} 
  \includegraphics[width=43pt]{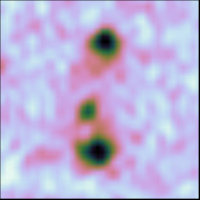} 
  \includegraphics[width=43pt]{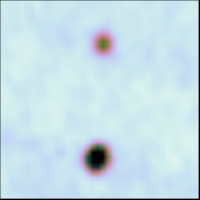} 
  \includegraphics[width=43pt]{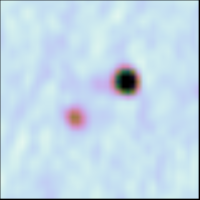} 
  \includegraphics[width=43pt]{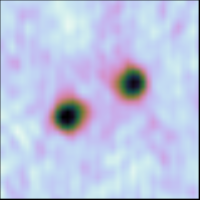} 
  \includegraphics[width=43pt]{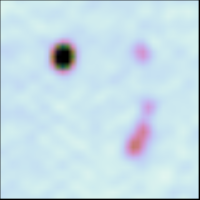} 
  \includegraphics[width=43pt]{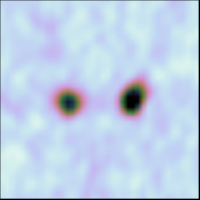} 
  \includegraphics[width=43pt]{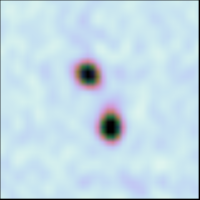} 
  \includegraphics[width=43pt]{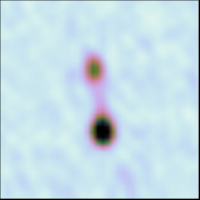} 
  \includegraphics[width=43pt]{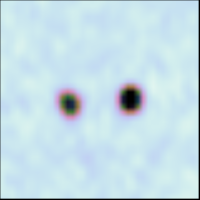} 
  \includegraphics[width=43pt]{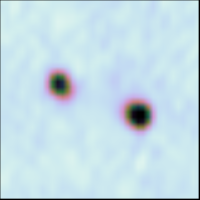} 
  \includegraphics[width=43pt]{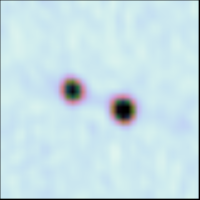} 
  \includegraphics[width=43pt]{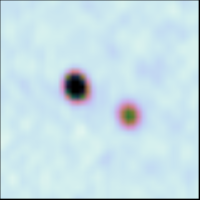} 
  \includegraphics[width=43pt]{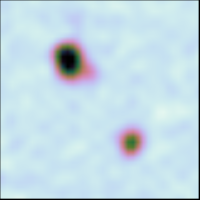} 
  \includegraphics[width=43pt]{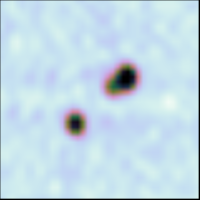} 
  \includegraphics[width=43pt]{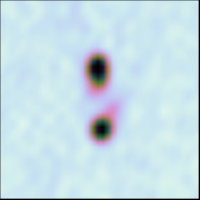} 
  \includegraphics[width=43pt]{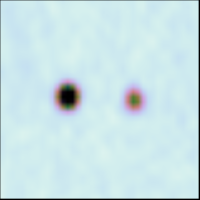} 
  \includegraphics[width=43pt]{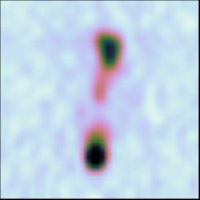} 
  \includegraphics[width=43pt]{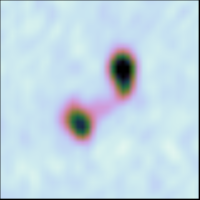} 
  \includegraphics[width=43pt]{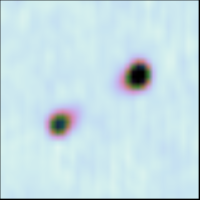} 
  \includegraphics[width=43pt]{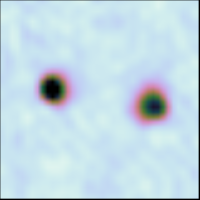} 
  \includegraphics[width=43pt]{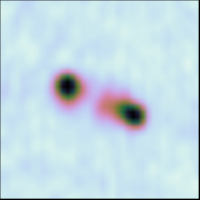} 
  \includegraphics[width=43pt]{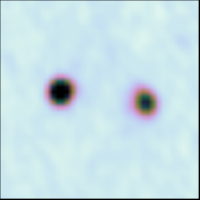} 
  \includegraphics[width=43pt]{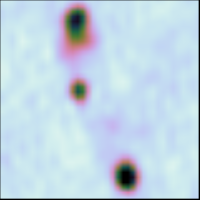} 
  \includegraphics[width=43pt]{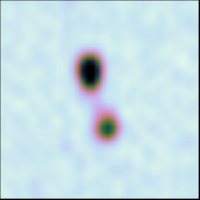} 
  \includegraphics[width=43pt]{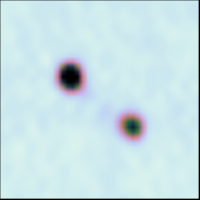} 
  \includegraphics[width=43pt]{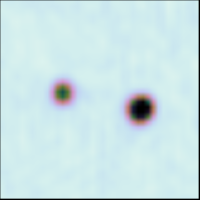} 
  \includegraphics[width=43pt]{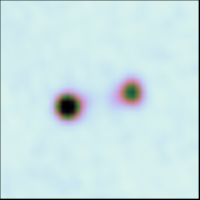} 
  \includegraphics[width=43pt]{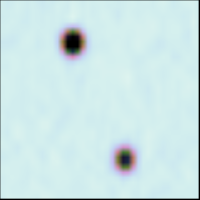} 
  \includegraphics[width=43pt]{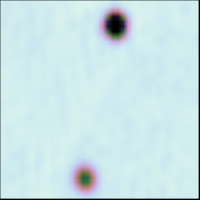} 
  \includegraphics[width=43pt]{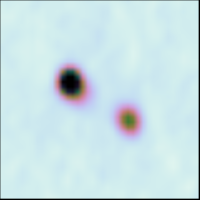} 
  \includegraphics[width=43pt]{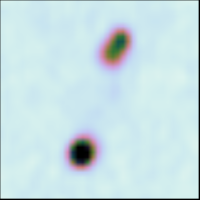} 
  \includegraphics[width=43pt]{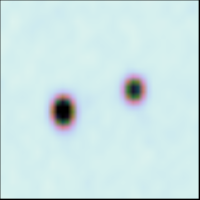} 
  \includegraphics[width=43pt]{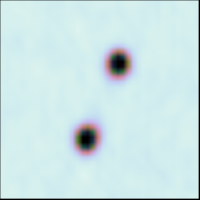} 
  \includegraphics[width=43pt]{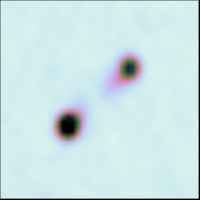} 
  \includegraphics[width=43pt]{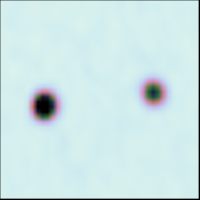} 
  \includegraphics[width=43pt]{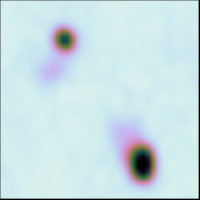} 
  \includegraphics[width=43pt]{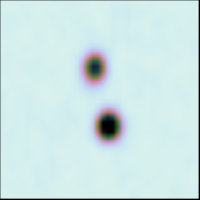} 
  \includegraphics[width=43pt]{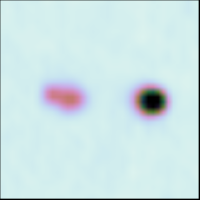} 
  \includegraphics[width=43pt]{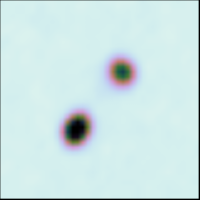} 
  \includegraphics[width=43pt]{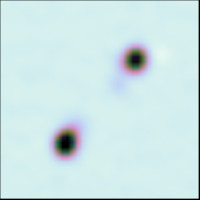} 
  \includegraphics[width=43pt]{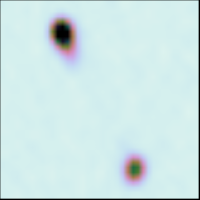} 
  \includegraphics[width=43pt]{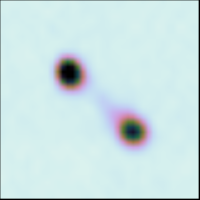} 
  \includegraphics[width=43pt]{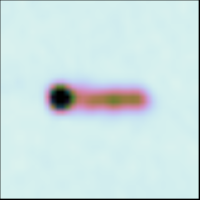} 
  \includegraphics[width=43pt]{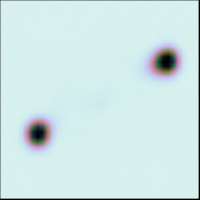} 
  \includegraphics[width=43pt]{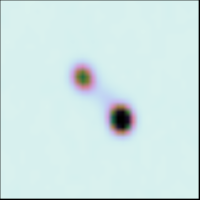} 
  \includegraphics[width=43pt]{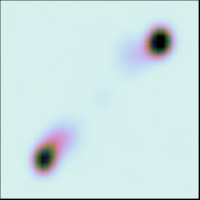}
  \includegraphics[width=43pt]{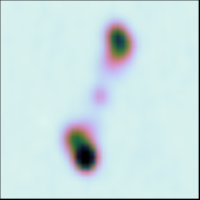} 
\caption{Random selection from the  $10^4$  FIRST doubeltjes that pass the quality cuts. Images are 1~arcmin wide and centered on the geometrical center of the lobes. Sources are selected using logarithmically-spaced flux-bins between 15~mJy and 222~mJy. }\label{fig:example}
\end{figure*}

%% file: dub.bbl
\begin{thebibliography}{149}
\expandafter\ifx\csname natexlab\endcsname\relax\def\natexlab#1{#1}\fi

\bibitem[{{Abazajian} {et~al}\mbox{.}(2009){Abazajian}, {Adelman-McCarthy},
  {Ag{\"u}eros}, {Allam}, {Allende Prieto}, {An}, {Anderson}, {Anderson},
  {Annis}, {Bahcall}, {Bailer-Jones}, {Barentine}, {Bassett}, {Becker},
  {Beers}, {Bell}, {Belokurov}, {Berlind}, {Berman}, {Bernardi}, {Bickerton},
  {Bizyaev}, {Blakeslee}, {Blanton}, {Bochanski}, {Boroski}, {Brewington},
  {Brinchmann}, {Brinkmann}, {Brunner}, {Budav{\'a}ri}, {Carey}, {Carliles},
  {Carr}, {Castander}, {Cinabro}, {Connolly}, {Csabai}, {Cunha}, {Czarapata},
  {Davenport}, {de Haas}, {Dilday}, {Doi}, {Eisenstein}, {Evans}, {Evans},
  {Fan}, {Friedman}, {Frieman}, {Fukugita}, {G{\"a}nsicke}, {Gates},
  {Gillespie}, {Gilmore}, {Gonzalez}, {Gonzalez}, {Grebel}, {Gunn},
  {Gy{\"o}ry}, {Hall}, {Harding}, {Harris}, {Harvanek}, {Hawley}, {Hayes},
  {Heckman}, {Hendry}, {Hennessy}, {Hindsley}, {Hoblitt}, {Hogan}, {Hogg},
  {Holtzman}, {Hyde}, {Ichikawa}, {Ichikawa}, {Im}, {Ivezi{\'c}}, {Jester},
  {Jiang}, {Johnson}, {Jorgensen}, {Juri{\'c}}, {Kent}, {Kessler}, {Kleinman},
  {Knapp}, {Konishi}, {Kron}, {Krzesinski}, {Kuropatkin}, {Lampeitl},
  {Lebedeva}, {Lee}, {Lee}, {Leger}, {L{\'e}pine}, {Li}, {Lima}, {Lin}, {Long},
  {Loomis}, {Loveday}, {Lupton}, {Magnier}, {Malanushenko}, {Malanushenko},
  {Mandelbaum}, {Margon}, {Marriner}, {Mart{\'{\i}}nez-Delgado}, {Matsubara},
  {McGehee}, {McKay}, {Meiksin}, {Morrison}, {Mullally}, {Munn}, {Murphy},
  {Nash}, {Nebot}, {Neilsen}, {Newberg}, {Newman}, {Nichol}, {Nicinski},
  {Nieto-Santisteban}, {Nitta}, {Okamura}, {Oravetz}, {Ostriker}, {Owen},
  {Padmanabhan}, {Pan}, {Park}, {Pauls}, {Peoples}, {Percival}, {Pier}, {Pope},
  {Pourbaix}, {Price}, {Purger}, {Quinn}, {Raddick}, {Fiorentin}, {Richards},
  {Richmond}, {Riess}, {Rix}, {Rockosi}, {Sako}, {Schlegel}, {Schneider},
  {Scholz}, {Schreiber}, {Schwope}, {Seljak}, {Sesar}, {Sheldon}, {Shimasaku},
  {Sibley}, {Simmons}, {Sivarani}, {Smith}, {Smith}, {Smol{\v c}i{\'c}},
  {Snedden}, {Stebbins}, {Steinmetz}, {Stoughton}, {Strauss}, {Subba Rao},
  {Suto}, {Szalay}, {Szapudi}, {Szkody}, {Tanaka}, {Tegmark}, {Teodoro},
  {Thakar}, {Tremonti}, {Tucker}, {Uomoto}, {Vanden Berk}, {Vandenberg},
  {Vidrih}, {Vogeley}, {Voges}, {Vogt}, {Wadadekar}, {Watters}, {Weinberg},
  {West}, {White}, {Wilhite}, {Wonders}, {Yanny}, {Yocum}, {York}, {Zehavi},
  {Zibetti}, \& {Zucker}}]{Abazajian09}
{Abazajian} K.~N. {et~al.}, 2009, \apjs, 182, 543

\bibitem[{{Alexander} \& {Hickox}(2012)}]{Alexander12}
{Alexander} D.~M., {Hickox} R.~C., 2012, \nar, 56, 93

\bibitem[{{Antognini} {et~al}\mbox{.}(2012){Antognini}, {Bird}, \&
  {Martini}}]{Antognini12}
{Antognini} J., {Bird} J., {Martini} P., 2012, \apj, 756, 116

\bibitem[{{Assef} {et~al}\mbox{.}(2013){Assef}, {Stern}, {Kochanek}, {Blain},
  {Brodwin}, {Brown}, {Donoso}, {Eisenhardt}, {Jannuzi}, {Jarrett}, {Stanford},
  {Tsai}, {Wu}, \& {Yan}}]{Assef13}
{Assef} R.~J. {et~al.}, 2013, \apj, 772, 26

\bibitem[{{Baldi} {et~al}\mbox{.}(2013){Baldi}, {Capetti}, {Buttiglione},
  {Chiaberge}, \& {Celotti}}]{Baldi13}
{Baldi} R.~D., {Capetti} A., {Buttiglione} S., {Chiaberge} M., {Celotti} A.,
  2013, \aap, 560, A81

\bibitem[{{Barger} {et~al}\mbox{.}(2005){Barger}, {Cowie}, {Mushotzky}, {Yang},
  {Wang}, {Steffen}, \& {Capak}}]{Barger05}
{Barger} A.~J., {Cowie} L.~L., {Mushotzky} R.~F., {Yang} Y., {Wang} W.-H.,
  {Steffen} A.~T., {Capak} P., 2005, \aj, 129, 578

\bibitem[{{Becker} {et~al}\mbox{.}(1995){Becker}, {White}, \&
  {Helfand}}]{Becker95}
{Becker} R.~H., {White} R.~L., {Helfand} D.~J., 1995, \apj, 450, 559

\bibitem[{{Best} \& {Heckman}(2012)}]{Best12}
{Best} P.~N., {Heckman} T.~M., 2012, \mnras, 421, 1569

\bibitem[{{Best} {et~al}\mbox{.}(2005){Best}, {Kauffmann}, {Heckman}, \&
  {Ivezi{\'c}}}]{Best05a}
{Best} P.~N., {Kauffmann} G., {Heckman} T.~M., {Ivezi{\'c}} {\v Z}., 2005,
  \mnras, 362, 9

\bibitem[{{Blandford} \& {Znajek}(1977)}]{BlandfordZnajek77}
{Blandford} R.~D., {Znajek} R.~L., 1977, \mnras, 179, 433

\bibitem[{{Blundell} \& {Rawlings}(2000)}]{Blundell00}
{Blundell} K.~M., {Rawlings} S., 2000, \aj, 119, 1111

\bibitem[{{Blundell} {et~al}\mbox{.}(1999){Blundell}, {Rawlings}, \&
  {Willott}}]{Blundell99}
{Blundell} K.~M., {Rawlings} S., {Willott} C.~J., 1999, \aj, 117, 677

\bibitem[{{Bovy} {et~al}\mbox{.}(2011){Bovy}, {Hennawi}, {Hogg}, {Myers},
  {Kirkpatrick}, {Schlegel}, {Ross}, {Sheldon}, {McGreer}, {Schneider}, \&
  {Weaver}}]{Bovy11}
{Bovy} J. {et~al.}, 2011, \apj, 729, 141

\bibitem[{{Bower} {et~al}\mbox{.}(2006){Bower}, {Benson}, {Malbon}, {Helly},
  {Frenk}, {Baugh}, {Cole}, \& {Lacey}}]{Bower06}
{Bower} R.~G., {Benson} A.~J., {Malbon} R., {Helly} J.~C., {Frenk} C.~S.,
  {Baugh} C.~M., {Cole} S., {Lacey} C.~G., 2006, \mnras, 370, 645

\bibitem[{{Brown} {et~al}\mbox{.}(2006){Brown}, {Brand}, {Dey}, {Jannuzi},
  {Cool}, {Le Floc'h}, {Kochanek}, {Armus}, {Bian}, {Higdon}, {Higdon},
  {Papovich}, {Rieke}, {Rieke}, {Smith}, {Soifer}, \& {Weedman}}]{Brown06}
{Brown} M.~J.~I. {et~al.}, 2006, \apj, 638, 88

\bibitem[{{Buttiglione} {et~al}\mbox{.}(2010){Buttiglione}, {Capetti},
  {Celotti}, {Axon}, {Chiaberge}, {Macchetto}, \& {Sparks}}]{Buttiglione10}
{Buttiglione} S., {Capetti} A., {Celotti} A., {Axon} D.~J., {Chiaberge} M.,
  {Macchetto} F.~D., {Sparks} W.~B., 2010, \aap, 509, A6

\bibitem[{{Cattaneo} {et~al}\mbox{.}(2009){Cattaneo}, {Faber}, {Binney},
  {Dekel}, {Kormendy}, {Mushotzky}, {Babul}, {Best}, {Br{\"u}ggen}, {Fabian},
  {Frenk}, {Khalatyan}, {Netzer}, {Mahdavi}, {Silk}, {Steinmetz}, \&
  {Wisotzki}}]{Cattaneo09}
{Cattaneo} A. {et~al.}, 2009, \nat, 460, 213

\bibitem[{{Churazov} {et~al}\mbox{.}(2005){Churazov}, {Sazonov}, {Sunyaev},
  {Forman}, {Jones}, \& {B{\"o}hringer}}]{Churazov05}
{Churazov} E., {Sazonov} S., {Sunyaev} R., {Forman} W., {Jones} C.,
  {B{\"o}hringer} H., 2005, \mnras, 363, L91

\bibitem[{{Cohen} {et~al}\mbox{.}(2007){Cohen}, {Lane}, {Cotton}, {Kassim},
  {Lazio}, {Perley}, {Condon}, \& {Erickson}}]{Cohen07}
{Cohen} A.~S., {Lane} W.~M., {Cotton} W.~D., {Kassim} N.~E., {Lazio} T.~J.~W.,
  {Perley} R.~A., {Condon} J.~J., {Erickson} W.~C., 2007, \aj, 134, 1245

\bibitem[{{Condon}(1984)}]{Condon84}
{Condon} J.~J., 1984, \apj, 287, 461

\bibitem[{{Condon} {et~al}\mbox{.}(1998){Condon}, {Cotton}, {Greisen}, {Yin},
  {Perley}, {Taylor}, \& {Broderick}}]{Condon98}
{Condon} J.~J., {Cotton} W.~D., {Greisen} E.~W., {Yin} Q.~F., {Perley} R.~A.,
  {Taylor} G.~B., {Broderick} J.~J., 1998, \aj, 115, 1693

\bibitem[{{Condon} {et~al}\mbox{.}(2013){Condon}, {Kellermann}, {Kimball},
  {Ivezi{\'c}}, \& {Perley}}]{Condon13}
{Condon} J.~J., {Kellermann} K.~I., {Kimball} A.~E., {Ivezi{\'c}} {\v Z}.,
  {Perley} R.~A., 2013, \apj, 768, 37

\bibitem[{{Condon} {et~al}\mbox{.}(1981){Condon}, {Odell}, {Puschell}, \&
  {Stein}}]{Condon81}
{Condon} J.~J., {Odell} S.~L., {Puschell} J.~J., {Stein} W.~A., 1981, \apj,
  246, 624

\bibitem[{{Croft} {et~al}\mbox{.}(2007){Croft}, {de Vries}, \&
  {Becker}}]{Croft07}
{Croft} S., {de Vries} W., {Becker} R.~H., 2007, \apjl, 667, L13

\bibitem[{{Croton} {et~al}\mbox{.}(2006){Croton}, {Springel}, {White}, {De
  Lucia}, {Frenk}, {Gao}, {Jenkins}, {Kauffmann}, {Navarro}, \&
  {Yoshida}}]{Croton06}
{Croton} D.~J. {et~al.}, 2006, \mnras, 365, 11

\bibitem[{{Dawson} {et~al}\mbox{.}(2013){Dawson}, {Schlegel}, {Ahn},
  {Anderson}, {Aubourg}, {Bailey}, {Barkhouser}, {Bautista}, {Beifiori},
  {Berlind}, {Bhardwaj}, {Bizyaev}, {Blake}, {Blanton}, {Blomqvist}, {Bolton},
  {Borde}, {Bovy}, {Brandt}, {Brewington}, {Brinkmann}, {Brown}, {Brownstein},
  {Bundy}, {Busca}, {Carithers}, {Carnero}, {Carr}, {Chen}, {Comparat},
  {Connolly}, {Cope}, {Croft}, {Cuesta}, {da Costa}, {Davenport}, {Delubac},
  {de Putter}, {Dhital}, {Ealet}, {Ebelke}, {Eisenstein}, {Escoffier}, {Fan},
  {Filiz Ak}, {Finley}, {Font-Ribera}, {G{\'e}nova-Santos}, {Gunn}, {Guo},
  {Haggard}, {Hall}, {Hamilton}, {Harris}, {Harris}, {Ho}, {Hogg}, {Holder},
  {Honscheid}, {Huehnerhoff}, {Jordan}, {Jordan}, {Kauffmann}, {Kazin},
  {Kirkby}, {Klaene}, {Kneib}, {Le Goff}, {Lee}, {Long}, {Loomis}, {Lundgren},
  {Lupton}, {Maia}, {Makler}, {Malanushenko}, {Malanushenko}, {Mandelbaum},
  {Manera}, {Maraston}, {Margala}, {Masters}, {McBride}, {McDonald}, {McGreer},
  {McMahon}, {Mena}, {Miralda-Escud{\'e}}, {Montero-Dorta}, {Montesano},
  {Muna}, {Myers}, {Naugle}, {Nichol}, {Noterdaeme}, {Nuza}, {Olmstead},
  {Oravetz}, {Oravetz}, {Owen}, {Padmanabhan}, {Palanque-Delabrouille}, {Pan},
  {Parejko}, {P{\^a}ris}, {Percival}, {P{\'e}rez-Fournon},
  {P{\'e}rez-R{\`a}fols}, {Petitjean}, {Pfaffenberger}, {Pforr}, {Pieri},
  {Prada}, {Price-Whelan}, {Raddick}, {Rebolo}, {Rich}, {Richards}, {Rockosi},
  {Roe}, {Ross}, {Ross}, {Rossi}, {Rubi{\~n}o-Martin}, {Samushia},
  {S{\'a}nchez}, {Sayres}, {Schmidt}, {Schneider}, {Sc{\'o}ccola}, {Seo},
  {Shelden}, {Sheldon}, {Shen}, {Shu}, {Slosar}, {Smee}, {Snedden}, {Stauffer},
  {Steele}, {Strauss}, {Streblyanska}, {Suzuki}, {Swanson}, {Tal}, {Tanaka},
  {Thomas}, {Tinker}, {Tojeiro}, {Tremonti}, {Vargas Maga{\~n}a}, {Verde},
  {Viel}, {Wake}, {Watson}, {Weaver}, {Weinberg}, {Weiner}, {West}, {White},
  {Wood-Vasey}, {Yeche}, {Zehavi}, {Zhao}, \& {Zheng}}]{Dawson13}
{Dawson} K.~S. {et~al.}, 2013, \aj, 145, 10

\bibitem[{{de Vries} {et~al}\mbox{.}(2006){de Vries}, {Becker}, \&
  {White}}]{deVries06}
{de Vries} W.~H., {Becker} R.~H., {White} R.~L., 2006, \aj, 131, 666

\bibitem[{{Dunlop} \& {Peacock}(1990)}]{Dunlop90}
{Dunlop} J.~S., {Peacock} J.~A., 1990, \mnras, 247, 19

\bibitem[{{Dwarakanath} \& {Kale}(2009)}]{Dwarakanath09}
{Dwarakanath} K.~S., {Kale} R., 2009, \apjl, 698, L163

\bibitem[{{Eisenstein} {et~al}\mbox{.}(2011){Eisenstein}, {Weinberg}, {Agol},
  {Aihara}, {Allende Prieto}, {Anderson}, {Arns}, {Aubourg}, {Bailey},
  {Balbinot}, \& et~al.}]{Eisenstein11}
{Eisenstein} D.~J. {et~al.}, 2011, \aj, 142, 72

\bibitem[{{Fabian}(2012)}]{Fabian12}
{Fabian} A.~C., 2012, \araa, 50, 455

\bibitem[{{Falcke} \& {Biermann}(1995)}]{Falcke95I}
{Falcke} H., {Biermann} P.~L., 1995, \aap, 293, 665

\bibitem[{{Falcke} {et~al}\mbox{.}(1999){Falcke}, {Bower}, {Lobanov},
  {Krichbaum}, {Patnaik}, {Aller}, {Aller}, {Ter{\"a}sranta}, {Wright}, \&
  {Sandell}}]{Falcke99}
{Falcke} H. {et~al.}, 1999, \apjl, 514, L17

\bibitem[{{Falcke} {et~al}\mbox{.}(1995{\natexlab{a}}){Falcke},
  {Gopal-Krishna}, \& {Biermann}}]{FalckeGopalKrishna95}
{Falcke} H., {Gopal-Krishna}, {Biermann} P.~L., 1995{\natexlab{a}}, \aap, 298,
  395

\bibitem[{{Falcke} {et~al}\mbox{.}(2004{\natexlab{a}}){Falcke}, {K{\"o}rding},
  \& {Markoff}}]{Falcke04}
{Falcke} H., {K{\"o}rding} E., {Markoff} S., 2004{\natexlab{a}}, \aap, 414, 895

\bibitem[{{Falcke} {et~al}\mbox{.}(2004{\natexlab{b}}){Falcke}, {K{\"o}rding},
  \& {Nagar}}]{FalckeKoerdingNagar04}
{Falcke} H., {K{\"o}rding} E., {Nagar} N.~M., 2004{\natexlab{b}}, \nar, 48,
  1157

\bibitem[{{Falcke} {et~al}\mbox{.}(1995{\natexlab{b}}){Falcke}, {Malkan}, \&
  {Biermann}}]{Falcke95II}
{Falcke} H., {Malkan} M.~A., {Biermann} P.~L., 1995{\natexlab{b}}, \aap, 298,
  375

\bibitem[{{Falcke} {et~al}\mbox{.}(1996){Falcke}, {Sherwood}, \&
  {Patnaik}}]{Falcke96}
{Falcke} H., {Sherwood} W., {Patnaik} A.~R., 1996, \apj, 471, 106

\bibitem[{{Falle}(1991)}]{Falle91}
{Falle} S.~A.~E.~G., 1991, \mnras, 250, 581

\bibitem[{{Fanaroff} \& {Riley}(1974)}]{Fanaroff74}
{Fanaroff} B.~L., {Riley} J.~M., 1974, \mnras, 167, 31P

\bibitem[{{Fender} {et~al}\mbox{.}(2004){Fender}, {Belloni}, \&
  {Gallo}}]{Fender04}
{Fender} R.~P., {Belloni} T.~M., {Gallo} E., 2004, \mnras, 355, 1105

\bibitem[{{Filho} {et~al}\mbox{.}(2011){Filho}, {Brinchmann}, {Lobo}, \&
  {Ant{\'o}n}}]{Filho11}
{Filho} M.~E., {Brinchmann} J., {Lobo} C., {Ant{\'o}n} S., 2011, \aap, 536, A35

\bibitem[{{Gendre} {et~al}\mbox{.}(2010){Gendre}, {Best}, \& {Wall}}]{Gendre10}
{Gendre} M.~A., {Best} P.~N., {Wall} J.~V., 2010, \mnras, 404, 1719

\bibitem[{{Gendre} \& {Wall}(2008)}]{Gendre08}
{Gendre} M.~A., {Wall} J.~V., 2008, \mnras, 390, 819

\bibitem[{{Ghisellini} \& {Celotti}(2001)}]{Ghisellini01}
{Ghisellini} G., {Celotti} A., 2001, \aap, 379, L1

\bibitem[{{Gilli}(2013)}]{Gilli13}
{Gilli} R., 2013, ArXiv e-prints

\bibitem[{{Godfrey} \& {Shabala}(2013)}]{Godfrey13}
{Godfrey} L.~E.~H., {Shabala} S.~S., 2013, \apj, 767, 12

\bibitem[{{Grimes} {et~al}\mbox{.}(2004){Grimes}, {Rawlings}, \&
  {Willott}}]{Grimes04}
{Grimes} J.~A., {Rawlings} S., {Willott} C.~J., 2004, \mnras, 349, 503

\bibitem[{{G{\"u}rkan} {et~al}\mbox{.}(2014){G{\"u}rkan}, {Hardcastle}, \&
  {Jarvis}}]{Gurkan14}
{G{\"u}rkan} G., {Hardcastle} M.~J., {Jarvis} M.~J., 2014, \mnras, 438, 1149

\bibitem[{{Hardcastle} \& {Krause}(2013)}]{Hardcastle13}
{Hardcastle} M.~J., {Krause} M.~G.~H., 2013, \mnras, 430, 174

\bibitem[{{Harvanek} \& {Stocke}(2002)}]{Harvanek02}
{Harvanek} M., {Stocke} J.~T., 2002, \aj, 124, 1239

\bibitem[{{Hasinger}(2008)}]{Hasinger08}
{Hasinger} G., 2008, \aap, 490, 905

\bibitem[{{Ho}(1999)}]{Ho99}
{Ho} L.~C., 1999, \apj, 516, 672

\bibitem[{{Hodge} {et~al}\mbox{.}(2009){Hodge}, {Zeimann}, {Becker}, \&
  {White}}]{Hodge09}
{Hodge} J.~A., {Zeimann} G.~R., {Becker} R.~H., {White} R.~L., 2009, \aj, 138,
  900

\bibitem[{{Hogg}(1999)}]{Hogg99}
{Hogg} D.~W., 1999, astro-ph/9905116

\bibitem[{{Hopkins} {et~al}\mbox{.}(2006){Hopkins}, {Hernquist}, {Cox}, {Di
  Matteo}, {Robertson}, \& {Springel}}]{Hopkins06}
{Hopkins} P.~F., {Hernquist} L., {Cox} T.~J., {Di Matteo} T., {Robertson} B.,
  {Springel} V., 2006, \apjs, 163, 1

\bibitem[{{Hopkins} {et~al}\mbox{.}(2007){Hopkins}, {Richards}, \&
  {Hernquist}}]{Hopkins07}
{Hopkins} P.~F., {Richards} G.~T., {Hernquist} L., 2007, \apj, 654, 731

\bibitem[{{Ivezi{\'c}} {et~al}\mbox{.}(2002){Ivezi{\'c}}, {Menou}, {Knapp},
  {Strauss}, {Lupton}, {Vanden Berk}, {Richards}, {Tremonti}, {Weinstein},
  {Anderson}, {Bahcall}, {Becker}, {Bernardi}, {Blanton}, {Eisenstein}, {Fan},
  {Finkbeiner}, {Finlator}, {Frieman}, {Gunn}, {Hall}, {Kim}, {Kinkhabwala},
  {Narayanan}, {Rockosi}, {Schlegel}, {Schneider}, {Strateva}, {SubbaRao},
  {Thakar}, {Voges}, {White}, {Yanny}, {Brinkmann}, {Doi}, {Fukugita},
  {Hennessy}, {Munn}, {Nichol}, \& {York}}]{Ivezic02}
{Ivezi{\'c}} {\v Z}. {et~al.}, 2002, \aj, 124, 2364

\bibitem[{{Jiang} {et~al}\mbox{.}(2007){Jiang}, {Fan}, {Ivezi{\'c}},
  {Richards}, {Schneider}, {Strauss}, \& {Kelly}}]{Jiang07}
{Jiang} L., {Fan} X., {Ivezi{\'c}} {\v Z}., {Richards} G.~T., {Schneider}
  D.~P., {Strauss} M.~A., {Kelly} B.~C., 2007, \apj, 656, 680

\bibitem[{{Kaiser} \& {Alexander}(1997)}]{Kaiser97}
{Kaiser} C.~R., {Alexander} P., 1997, \mnras, 286, 215

\bibitem[{{Kaiser} \& {Best}(2007)}]{Kaiser07}
{Kaiser} C.~R., {Best} P.~N., 2007, \mnras, 381, 1548

\bibitem[{{Kapahi}(1989)}]{Kapahi89}
{Kapahi} V.~K., 1989, \aj, 97, 1

\bibitem[{{Kapi{\'n}ska} {et~al}\mbox{.}(2012){Kapi{\'n}ska}, {Uttley}, \&
  {Kaiser}}]{Kapinska12}
{Kapi{\'n}ska} A.~D., {Uttley} P., {Kaiser} C.~R., 2012, \mnras, 424, 2028

\bibitem[{{Kauffmann} \& {Haehnelt}(2000)}]{Kauffmann00}
{Kauffmann} G., {Haehnelt} M., 2000, \mnras, 311, 576

\bibitem[{{Kellermann} {et~al}\mbox{.}(1989){Kellermann}, {Sramek}, {Schmidt},
  {Shaffer}, \& {Green}}]{Kellermann89}
{Kellermann} K.~I., {Sramek} R., {Schmidt} M., {Shaffer} D.~B., {Green} R.,
  1989, \aj, 98, 1195

\bibitem[{{Ker} {et~al}\mbox{.}(2012){Ker}, {Best}, {Rigby}, {R{\"o}ttgering},
  \& {Gendre}}]{Ker12}
{Ker} L.~M., {Best} P.~N., {Rigby} E.~E., {R{\"o}ttgering} H.~J.~A., {Gendre}
  M.~A., 2012, \mnras, 420, 2644

\bibitem[{{Kimball} \& {Ivezi{\'c}}(2008)}]{Kimball08}
{Kimball} A.~E., {Ivezi{\'c}} {\v Z}., 2008, \aj, 136, 684

\bibitem[{{Kimball} {et~al}\mbox{.}(2011){Kimball}, {Ivezi{\'c}}, {Wiita}, \&
  {Schneider}}]{Kimball11}
{Kimball} A.~E., {Ivezi{\'c}} {\v Z}., {Wiita} P.~J., {Schneider} D.~P., 2011,
  \aj, 141, 182

\bibitem[{{Komissarov} \& {Gubanov}(1994)}]{Komissarov94}
{Komissarov} S.~S., {Gubanov} A.~G., 1994, \aap, 285, 27

\bibitem[{{K{\"o}rding} {et~al}\mbox{.}(2006){K{\"o}rding}, {Jester}, \&
  {Fender}}]{Koerding06}
{K{\"o}rding} E.~G., {Jester} S., {Fender} R., 2006, \mnras, 372, 1366

\bibitem[{{K{\"o}rding} {et~al}\mbox{.}(2008){K{\"o}rding}, {Jester}, \&
  {Fender}}]{Koerding08}
{K{\"o}rding} E.~G., {Jester} S., {Fender} R., 2008, \mnras, 383, 277

\bibitem[{{Kratzer}(2014)}]{Kratzer14}
{Kratzer} R.~M., 2014, PhD thesis, Drexel University

\bibitem[{{Laing} {et~al}\mbox{.}(1994){Laing}, {Jenkins}, {Wall}, \&
  {Unger}}]{Laing94}
{Laing} R.~A., {Jenkins} C.~R., {Wall} J.~V., {Unger} S.~W., 1994, in
  Astronomical Society of the Pacific Conference Series, Vol.~54, The Physics
  of Active Galaxies, {Bicknell} G.~V., {Dopita} M.~A., {Quinn} P.~J., eds., p.
  201

\bibitem[{{Laing} {et~al}\mbox{.}(1983){Laing}, {Riley}, \&
  {Longair}}]{Laing83}
{Laing} R.~A., {Riley} J.~M., {Longair} M.~S., 1983, \mnras, 204, 151

\bibitem[{{Lal} \& {Ho}(2010)}]{Lal10}
{Lal} D.~V., {Ho} L.~C., 2010, \aj, 139, 1089

\bibitem[{{Lane} {et~al}\mbox{.}(2014){Lane}, {Cotton}, {van Velzen}, {Clarke},
  {Kassim}, {Helmboldt}, {Lazio}, \& {Cohen}}]{Lane14}
{Lane} W.~M., {Cotton} W.~D., {van Velzen} S., {Clarke} T.~E., {Kassim} N.~E.,
  {Helmboldt} J.~F., {Lazio} T.~J.~W., {Cohen} A.~S., 2014, \mnras, 440, 327

\bibitem[{{Lawrence}(1991)}]{Lawrence91}
{Lawrence} A., 1991, \mnras, 252, 586

\bibitem[{{Lawrence} \& {Elvis}(2010)}]{Lawrence10}
{Lawrence} A., {Elvis} M., 2010, \apj, 714, 561

\bibitem[{{Ledlow} \& {Owen}(1996)}]{Ledlow96}
{Ledlow} M.~J., {Owen} F.~N., 1996, \aj, 112, 9

\bibitem[{{Lusso} {et~al}\mbox{.}(2013){Lusso}, {Hennawi}, {Comastri},
  {Zamorani}, {Richards}, {Vignali}, {Treister}, {Schawinski}, {Salvato}, \&
  {Gilli}}]{Lusso13}
{Lusso} E. {et~al.}, 2013, \apj, 777, 86

\bibitem[{{MacDonald} {et~al}\mbox{.}(1968){MacDonald}, {Kenderdine}, \&
  {Neville}}]{MacDonald68}
{MacDonald} G.~H., {Kenderdine} S., {Neville} A.~C., 1968, \mnras, 138, 259

\bibitem[{{Mackay}(1971)}]{Mackay71}
{Mackay} C.~D., 1971, \mnras, 154, 209

\bibitem[{{Macklin}(1982)}]{Macklin82}
{Macklin} J.~T., 1982, \mnras, 199, 1119

\bibitem[{{Martini}(2004)}]{Martini04}
{Martini} P., 2004, Coevolution of Black Holes and Galaxies, 169

\bibitem[{{Massardi} {et~al}\mbox{.}(2010){Massardi}, {Bonaldi}, {Negrello},
  {Ricciardi}, {Raccanelli}, \& {de Zotti}}]{Massardi10}
{Massardi} M., {Bonaldi} A., {Negrello} M., {Ricciardi} S., {Raccanelli} A.,
  {de Zotti} G., 2010, \mnras, 404, 532

\bibitem[{{Mateos} {et~al}\mbox{.}(2013){Mateos}, {Alonso-Herrero}, {Carrera},
  {Blain}, {Severgnini}, {Caccianiga}, \& {Ruiz}}]{Mateos13}
{Mateos} S., {Alonso-Herrero} A., {Carrera} F.~J., {Blain} A., {Severgnini} P.,
  {Caccianiga} A., {Ruiz} A., 2013, \mnras, 434, 941

\bibitem[{{Matute} {et~al}\mbox{.}(2006){Matute}, {La Franca}, {Pozzi},
  {Gruppioni}, {Lari}, \& {Zamorani}}]{Matute06}
{Matute} I., {La Franca} F., {Pozzi} F., {Gruppioni} C., {Lari} C., {Zamorani}
  G., 2006, \aap, 451, 443

\bibitem[{{McGilchrist} {et~al}\mbox{.}(1990){McGilchrist}, {Baldwin}, {Riley},
  {Titterington}, {Waldram}, \& {Warner}}]{McGilchrist90}
{McGilchrist} M.~M., {Baldwin} J.~E., {Riley} J.~M., {Titterington} D.~J.,
  {Waldram} E.~M., {Warner} P.~J., 1990, \mnras, 246, 110

\bibitem[{{McHardy} {et~al}\mbox{.}(2006){McHardy}, {Koerding}, {Knigge},
  {Uttley}, \& {Fender}}]{McHardy06}
{McHardy} I.~M., {Koerding} E., {Knigge} C., {Uttley} P., {Fender} R.~P., 2006,
  \nat, 444, 730

\bibitem[{{McNamara} \& {Nulsen}(2012)}]{McNamara12}
{McNamara} B.~R., {Nulsen} P.~E.~J., 2012, New Journal of Physics, 14, 055023

\bibitem[{{Mocz} {et~al}\mbox{.}(2011){Mocz}, {Fabian}, \& {Blundell}}]{Mocz11}
{Mocz} P., {Fabian} A.~C., {Blundell} K.~M., 2011, \mnras, 413, 1107

\bibitem[{{Mocz} {et~al}\mbox{.}(2013){Mocz}, {Fabian}, \& {Blundell}}]{Mocz13}
{Mocz} P., {Fabian} A.~C., {Blundell} K.~M., 2013, \mnras, 432, 3381

\bibitem[{{Morganti} {et~al}\mbox{.}(2013){Morganti}, {Fogasy}, {Paragi},
  {Oosterloo}, \& {Orienti}}]{Morganti13}
{Morganti} R., {Fogasy} J., {Paragi} Z., {Oosterloo} T., {Orienti} M., 2013,
  Science, 341, 1082

\bibitem[{{Mullin} {et~al}\mbox{.}(2008){Mullin}, {Riley}, \&
  {Hardcastle}}]{Mullin08}
{Mullin} L.~M., {Riley} J.~M., {Hardcastle} M.~J., 2008, \mnras, 390, 595

\bibitem[{{Murgia} {et~al}\mbox{.}(2011){Murgia}, {Parma}, {Mack}, {de Ruiter},
  {Fanti}, {Govoni}, {Tarchi}, {Giacintucci}, \& {Markevitch}}]{Murgia11}
{Murgia} M. {et~al.}, 2011, \aap, 526, A148

\bibitem[{{Nandi} {et~al}\mbox{.}(2014){Nandi}, {Roy}, {Saikia}, {Singh},
  {Chandola}, {Baes}, {Joshi}, {Gentile}, \& {Patgiri}}]{Nandi14}
{Nandi} S. {et~al.}, 2014, \apj, 789, 16

\bibitem[{{Narayan} \& {Yi}(1995)}]{Narayan95}
{Narayan} R., {Yi} I., 1995, \apj, 452, 710

\bibitem[{{Neeser} {et~al}\mbox{.}(1995){Neeser}, {Eales}, {Law-Green},
  {Leahy}, \& {Rawlings}}]{Neeser95}
{Neeser} M.~J., {Eales} S.~A., {Law-Green} J.~D., {Leahy} J.~P., {Rawlings} S.,
  1995, \apj, 451, 76

\bibitem[{{Nipoti} {et~al}\mbox{.}(2005){Nipoti}, {Blundell}, \&
  {Binney}}]{Nipoti05}
{Nipoti} C., {Blundell} K.~M., {Binney} J., 2005, \mnras, 361, 633

\bibitem[{{Ogle} {et~al}\mbox{.}(2006){Ogle}, {Whysong}, \&
  {Antonucci}}]{Ogle06}
{Ogle} P., {Whysong} D., {Antonucci} R., 2006, \apj, 647, 161

\bibitem[{{Oort} {et~al}\mbox{.}(1987){Oort}, {Katgert}, \&
  {Windhorst}}]{Oort87}
{Oort} M.~J.~A., {Katgert} P., {Windhorst} R.~A., 1987, \nat, 328, 500

\bibitem[{{Owen} {et~al}\mbox{.}(1993){Owen}, {White}, \& {Ge}}]{Owen93}
{Owen} F.~N., {White} R.~A., {Ge} J., 1993, \apjs, 87, 135

\bibitem[{{P{\^a}ris} {et~al}\mbox{.}(2012){P{\^a}ris}, {Petitjean}, {Aubourg},
  {Bailey}, {Ross}, {Myers}, {Strauss}, {Anderson}, {Arnau}, {Bautista},
  {Bizyaev}, {Bolton}, {Bovy}, {Brandt}, {Brewington}, {Browstein}, {Busca},
  {Capellupo}, {Carithers}, {Croft}, {Dawson}, {Delubac}, {Ebelke},
  {Eisenstein}, {Engelke}, {Fan}, {Filiz Ak}, {Finley}, {Font-Ribera}, {Ge},
  {Gibson}, {Hall}, {Hamann}, {Hennawi}, {Ho}, {Hogg}, {Ivezi{\'c}}, {Jiang},
  {Kimball}, {Kirkby}, {Kirkpatrick}, {Lee}, {Le Goff}, {Lundgren}, {MacLeod},
  {Malanushenko}, {Malanushenko}, {Maraston}, {McGreer}, {McMahon},
  {Miralda-Escud{\'e}}, {Muna}, {Noterdaeme}, {Oravetz},
  {Palanque-Delabrouille}, {Pan}, {Perez-Fournon}, {Pieri}, {Richards},
  {Rollinde}, {Sheldon}, {Schlegel}, {Schneider}, {Slosar}, {Shelden}, {Shen},
  {Simmons}, {Snedden}, {Suzuki}, {Tinker}, {Viel}, {Weaver}, {Weinberg},
  {White}, {Wood-Vasey}, \& {Y{\`e}che}}]{Paris12}
{P{\^a}ris} I. {et~al.}, 2012, \aap, 548, A66

\bibitem[{{Parma} {et~al}\mbox{.}(2007){Parma}, {Murgia}, {de Ruiter}, {Fanti},
  {Mack}, \& {Govoni}}]{Parma07}
{Parma} P., {Murgia} M., {de Ruiter} H.~R., {Fanti} R., {Mack} K.-H., {Govoni}
  F., 2007, \aap, 470, 875

\bibitem[{{Plotkin} {et~al}\mbox{.}(2012){Plotkin}, {Anderson}, {Brandt},
  {Markoff}, {Shemmer}, \& {Wu}}]{Plotkin12}
{Plotkin} R.~M., {Anderson} S.~F., {Brandt} W.~N., {Markoff} S., {Shemmer} O.,
  {Wu} J., 2012, \apjl, 745, L27

\bibitem[{{Proctor}(2011)}]{Proctor11}
{Proctor} D.~D., 2011, \apjs, 194, 31

\bibitem[{{Rawlings} \& {Saunders}(1991)}]{Rawlings91}
{Rawlings} S., {Saunders} R., 1991, \nat, 349, 138

\bibitem[{{Remillard} \& {McClintock}(2006)}]{RemillardMcClintock06}
{Remillard} R.~A., {McClintock} J.~E., 2006, \araa, 44, 49

\bibitem[{{Rengelink} {et~al}\mbox{.}(1997){Rengelink}, {Tang}, {de Bruyn},
  {Miley}, {Bremer}, {Roettgering}, \& {Bremer}}]{Rengelink97}
{Rengelink} R.~B., {Tang} Y., {de Bruyn} A.~G., {Miley} G.~K., {Bremer} M.~N.,
  {Roettgering} H.~J.~A., {Bremer} M.~A.~R., 1997, \aaps, 124, 259

\bibitem[{{Richards} {et~al}\mbox{.}(2002){Richards}, {Fan}, {Newberg},
  {Strauss}, {Vanden Berk}, {Schneider}, {Yanny}, {Boucher}, {Burles},
  {Frieman}, {Gunn}, {Hall}, {Ivezi{\'c}}, {Kent}, {Loveday}, {Lupton},
  {Rockosi}, {Schlegel}, {Stoughton}, {SubbaRao}, \& {York}}]{Richards02}
{Richards} G.~T. {et~al.}, 2002, \aj, 123, 2945

\bibitem[{{Roseboom} {et~al}\mbox{.}(2013){Roseboom}, {Lawrence}, {Elvis},
  {Petty}, {Shen}, \& {Hao}}]{Roseboom13}
{Roseboom} I.~G., {Lawrence} A., {Elvis} M., {Petty} S., {Shen} Y., {Hao} H.,
  2013, \mnras, 429, 1494

\bibitem[{{Ross} {et~al}\mbox{.}(2012){Ross}, {Myers}, {Sheldon}, {Y{\`e}che},
  {Strauss}, {Bovy}, {Kirkpatrick}, {Richards}, {Aubourg}, {Blanton}, {Brandt},
  {Carithers}, {Croft}, {da Silva}, {Dawson}, {Eisenstein}, {Hennawi}, {Ho},
  {Hogg}, {Lee}, {Lundgren}, {McMahon}, {Miralda-Escud{\'e}},
  {Palanque-Delabrouille}, {P{\^a}ris}, {Petitjean}, {Pieri}, {Rich}, {Roe},
  {Schiminovich}, {Schlegel}, {Schneider}, {Slosar}, {Suzuki}, {Tinker},
  {Weinberg}, {Weyant}, {White}, \& {Wood-Vasey}}]{Ross12}
{Ross} N.~P. {et~al.}, 2012, \apjs, 199, 3

\bibitem[{{Saikia} \& {Jamrozy}(2009)}]{Saikia09}
{Saikia} D.~J., {Jamrozy} M., 2009, Bulletin of the Astronomical Society of
  India, 37, 63

\bibitem[{{Schellart}(2013)}]{Schellart13}
{Schellart} P., 2013, {K3Match: Point Matching in 3D space}. Ascl:1307.003

\bibitem[{{Scheuer}(1974)}]{Scheuer74}
{Scheuer} P.~A.~G., 1974, \mnras, 166, 513

\bibitem[{{Schneider} {et~al}\mbox{.}(2007){Schneider}, {Hall}, {Richards},
  {Strauss}, {Vanden Berk}, {Anderson}, {Brandt}, {Fan}, {Jester}, {Gray},
  {Gunn}, {SubbaRao}, {Thakar}, {Stoughton}, {Szalay}, {Yanny}, {York},
  {Bahcall}, {Barentine}, {Blanton}, {Brewington}, {Brinkmann}, {Brunner},
  {Castander}, {Csabai}, {Frieman}, {Fukugita}, {Harvanek}, {Hogg},
  {Ivezi{\'c}}, {Kent}, {Kleinman}, {Knapp}, {Kron}, {Krzesi{\'n}ski}, {Long},
  {Lupton}, {Nitta}, {Pier}, {Saxe}, {Shen}, {Snedden}, {Weinberg}, \&
  {Wu}}]{schneider07}
{Schneider} D.~P. {et~al.}, 2007, \aj, 134, 102

\bibitem[{{Schneider} {et~al}\mbox{.}(2010){Schneider}, {Richards}, {Hall},
  {Strauss}, {Anderson}, {Boroson}, {Ross}, {Shen}, {Brandt}, {Fan}, {Inada},
  {Jester}, {Knapp}, {Krawczyk}, {Thakar}, {Vanden Berk}, {Voges}, {Yanny},
  {York}, {Bahcall}, {Bizyaev}, {Blanton}, {Brewington}, {Brinkmann},
  {Eisenstein}, {Frieman}, {Fukugita}, {Gray}, {Gunn}, {Hibon}, {Ivezi{\'c}},
  {Kent}, {Kron}, {Lee}, {Lupton}, {Malanushenko}, {Malanushenko}, {Oravetz},
  {Pan}, {Pier}, {Price}, {Saxe}, {Schlegel}, {Simmons}, {Snedden}, {SubbaRao},
  {Szalay}, \& {Weinberg}}]{Schneider10}
{Schneider} D.~P. {et~al.}, 2010, \aj, 139, 2360

\bibitem[{{Schoenmakers} {et~al}\mbox{.}(2001){Schoenmakers}, {de Bruyn},
  {R{\"o}ttgering}, \& {van der Laan}}]{Schoenmakers01}
{Schoenmakers} A.~P., {de Bruyn} A.~G., {R{\"o}ttgering} H.~J.~A., {van der
  Laan} H., 2001, \aap, 374, 861

\bibitem[{{Schoenmakers} {et~al}\mbox{.}(2000){Schoenmakers}, {de Bruyn},
  {R{\"o}ttgering}, {van der Laan}, \& {Kaiser}}]{Schoenmakers00}
{Schoenmakers} A.~P., {de Bruyn} A.~G., {R{\"o}ttgering} H.~J.~A., {van der
  Laan} H., {Kaiser} C.~R., 2000, \mnras, 315, 371

\bibitem[{{Serjeant} {et~al}\mbox{.}(1998){Serjeant}, {Rawlings}, {Lacy},
  {Maddox}, {Baker}, {Clements}, \& {Lilje}}]{Serjeant98}
{Serjeant} S., {Rawlings} S., {Lacy} M., {Maddox} S.~J., {Baker} J.~C.,
  {Clements} D., {Lilje} P.~B., 1998, \mnras, 294, 494

\bibitem[{{Shen} {et~al}\mbox{.}(2011){Shen}, {Richards}, {Strauss}, {Hall},
  {Schneider}, {Snedden}, {Bizyaev}, {Brewington}, {Malanushenko},
  {Malanushenko}, {Oravetz}, {Pan}, \& {Simmons}}]{Shen11}
{Shen} Y. {et~al.}, 2011, \apjs, 194, 45

\bibitem[{{Shi} {et~al}\mbox{.}(2005){Shi}, {Rieke}, {Hines}, {Neugebauer},
  {Blaylock}, {Rigby}, {Egami}, {Gordon}, \& {Alonso-Herrero}}]{Shi05}
{Shi} Y. {et~al.}, 2005, \apj, 629, 88

\bibitem[{{Silverman} {et~al}\mbox{.}(2005){Silverman}, {Green}, {Barkhouse},
  {Cameron}, {Foltz}, {Jannuzi}, {Kim}, {Kim}, {Mossman}, {Tananbaum},
  {Wilkes}, {Smith}, {Smith}, \& {Smith}}]{Silverman05}
{Silverman} J.~D. {et~al.}, 2005, \apj, 624, 630

\bibitem[{{Simpson}(2005)}]{Simpson05}
{Simpson} C., 2005, \mnras, 360, 565

\bibitem[{{Singal}(1993)}]{Singal93}
{Singal} A.~K., 1993, \mnras, 263, 139

\bibitem[{{Singal} \& {Rajpurohit}(2014)}]{Singal14}
{Singal} A.~K., {Rajpurohit} K., 2014, \apss, 353, 233

\bibitem[{{Singal} {et~al}\mbox{.}(2013){Singal}, {Petrosian}, {Stawarz}, \&
  {Lawrence}}]{SingalPetrosian13}
{Singal} J., {Petrosian} V., {Stawarz} {\L}., {Lawrence} A., 2013, \apj, 764,
  43

\bibitem[{{Stanghellini} {et~al}\mbox{.}(1990){Stanghellini}, {Baum}, {O'Dea},
  \& {Morris}}]{Stanghellini90}
{Stanghellini} C., {Baum} S.~A., {O'Dea} C.~P., {Morris} G.~B., 1990, \aap,
  233, 379

\bibitem[{{Stern} {et~al}\mbox{.}(2012){Stern}, {Assef}, {Benford}, {Blain},
  {Cutri}, {Dey}, {Eisenhardt}, {Griffith}, {Jarrett}, {Lake}, {Masci},
  {Petty}, {Stanford}, {Tsai}, {Wright}, {Yan}, {Harrison}, \&
  {Madsen}}]{Stern12}
{Stern} D. {et~al.}, 2012, \apj, 753, 30

\bibitem[{{Stocke} {et~al}\mbox{.}(1992){Stocke}, {Morris}, {Weymann}, \&
  {Foltz}}]{Stocke92}
{Stocke} J.~T., {Morris} S.~L., {Weymann} R.~J., {Foltz} C.~B., 1992, \apj,
  396, 487

\bibitem[{{Strittmatter} {et~al}\mbox{.}(1980){Strittmatter}, {Hill},
  {Pauliny-Toth}, {Steppe}, \& {Witzel}}]{Strittmatter80}
{Strittmatter} P.~A., {Hill} P., {Pauliny-Toth} I.~I.~K., {Steppe} H., {Witzel}
  A., 1980, \aap, 88, L12

\bibitem[{{Ueda} {et~al}\mbox{.}(2003){Ueda}, {Akiyama}, {Ohta}, \&
  {Miyaji}}]{Ueda03}
{Ueda} Y., {Akiyama} M., {Ohta} K., {Miyaji} T., 2003, \apj, 598, 886

\bibitem[{{Urry} \& {Padovani}(1995)}]{Urry95}
{Urry} C.~M., {Padovani} P., 1995, \pasp, 107, 803

\bibitem[{{van Haarlem} {et~al}\mbox{.}(2013){van Haarlem}, {Wise}, {Gunst},
  {Heald}, {McKean}, {Hessels}, {de Bruyn}, {Nijboer}, {Swinbank}, {Fallows},
  {Brentjens}, {Nelles}, {Beck}, {Falcke}, {Fender}, {H{\"o}randel},
  {Koopmans}, {Mann}, {Miley}, {R{\"o}ttgering}, {Stappers}, {Wijers},
  {Zaroubi}, {van den Akker}, {Alexov}, {Anderson}, {Anderson}, {van Ardenne},
  {Arts}, {Asgekar}, {Avruch}, {Batejat}, {B{\"a}hren}, {Bell}, {Bell}, {van
  Bemmel}, {Bennema}, {Bentum}, {Bernardi}, {Best}, {B{\^\i}rzan}, {Bonafede},
  {Boonstra}, {Braun}, {Bregman}, {Breitling}, {van de Brink}, {Broderick},
  {Broekema}, {Brouw}, {Br{\"u}ggen}, {Butcher}, {van Cappellen}, {Ciardi},
  {Coenen}, {Conway}, {Coolen}, {Corstanje}, {Damstra}, {Davies}, {Deller},
  {Dettmar}, {van Diepen}, {Dijkstra}, {Donker}, {Doorduin}, {Dromer}, {Drost},
  {van Duin}, {Eisl{\"o}ffel}, {van Enst}, {Ferrari}, {Frieswijk}, {Gankema},
  {Garrett}, {de Gasparin}, {Gerbers}, {de Geus}, {Grie{\ss}meier}, {Grit},
  {Gruppen}, {Hamaker}, {Hassall}, {Hoeft}, {Holties}, {Horneffer}, {van der
  Horst}, {van Houwelingen}, {Huijgen}, {Iacobelli}, {Intema}, {Jackson},
  {Jelic}, {de Jong}, {Kant}, {Karastergiou}, {Koers}, {Kollen}, {Kondratiev},
  {Kooistra}, {Koopman}, {Koster}, {Kuniyoshi}, {Kramer}, {Kuper},
  {Lambropoulos}, {Law}, {van Leeuwen}, {Lemaitre}, {Loose}, {Maat}, {Macario},
  {Markoff}, {Masters}, {McFadden}, {McKay-Bukowski}, {Meijering}, {Meulman},
  {Mevius}, {Millenaar}, {Miller-Jones}, {Mohan}, {Mol}, {Morawietz},
  {Morganti}, {Mulcahy}, {Mulder}, {Munk}, {Nieuwenhuis}, {van Nieuwpoort},
  {Noordam}, {Norden}, {Noutsos}, {Offringa}, {Olofsson}, {Omar}, {Orr{\'u}},
  {Overeem}, {Paas}, {Pandey-Pommier}, {Pandey}, {Pizzo}, {Polatidis},
  {Rafferty}, {Rawlings}, {Reich}, {de Reijer}, {Reitsma}, {Renting},
  {Riemers}, {Rol}, {Romein}, {Roosjen}, {Ruiter}, {Scaife}, {van der Schaaf},
  {Scheers}, {Schellart}, {Schoenmakers}, {Schoonderbeek}, {Serylak},
  {Shulevski}, {Sluman}, {Smirnov}, {Sobey}, {Spreeuw}, {Steinmetz}, {Sterks},
  {Stiepel}, {Stuurwold}, {Tagger}, {Tang}, {Tasse}, {Thomas}, {Thoudam},
  {Toribio}, {van der Tol}, {Usov}, {van Veelen}, {van der Veen}, {ter Veen},
  {Verbiest}, {Vermeulen}, {Vermaas}, {Vocks}, {Vogt}, {de Vos}, {van der Wal},
  {van Weeren}, {Weggemans}, {Weltevrede}, {White}, {Wijnholds}, {Wilhelmsson},
  {Wucknitz}, {Yatawatta}, {Zarka}, {Zensus}, \& {van Zwieten}}]{vanHaarlem13}
{van Haarlem} M.~P. {et~al.}, 2013, \aap, 556, A2

\bibitem[{{van Velzen} \& {Falcke}(2013)}]{vanVelzenFalcke13}
{van Velzen} S., {Falcke} H., 2013, \aap, 557, L7

\bibitem[{{van Velzen} {et~al}\mbox{.}(2012){van Velzen}, {Falcke},
  {Schellart}, {Nierstenh{\"o}fer}, \& {Kampert}}]{vanVelzen12}
{van Velzen} S., {Falcke} H., {Schellart} P., {Nierstenh{\"o}fer} N., {Kampert}
  K.-H., 2012, \aap, 544, A18

\bibitem[{{Vanden Berk} {et~al}\mbox{.}(2005){Vanden Berk}, {Schneider},
  {Richards}, {Hall}, {Strauss}, {Brunner}, {Fan}, {Baldry}, {York}, {Gunn},
  {Nichol}, {Meiksin}, \& {Brinkmann}}]{VandenBerk05}
{Vanden Berk} D.~E. {et~al.}, 2005, \aj, 129, 2047

\bibitem[{{White} {et~al}\mbox{.}(2000){White}, {Becker}, {Gregg},
  {Laurent-Muehleisen}, {Brotherton}, {Impey}, {Petry}, {Foltz}, {Chaffee},
  {Richards}, {Oegerle}, {Helfand}, {McMahon}, \& {Cabanela}}]{White00}
{White} R.~L. {et~al.}, 2000, \apjs, 126, 133

\bibitem[{{White} {et~al}\mbox{.}(1997){White}, {Becker}, {Helfand}, \&
  {Gregg}}]{White97}
{White} R.~L., {Becker} R.~H., {Helfand} D.~J., {Gregg} M.~D., 1997, \apj, 475,
  479

\bibitem[{{Wilkes} {et~al}\mbox{.}(2013){Wilkes}, {Kuraszkiewicz}, {Haas},
  {Barthel}, {Leipski}, {Willner}, {Worrall}, {Birkinshaw}, {Antonucci},
  {Ashby}, {Chini}, {Fazio}, {Lawrence}, {Ogle}, \& {Schulz}}]{Wilkes13}
{Wilkes} B.~J. {et~al.}, 2013, \apj, 773, 15

\bibitem[{{Willott} {et~al}\mbox{.}(1999){Willott}, {Rawlings}, {Blundell}, \&
  {Lacy}}]{Willott99}
{Willott} C.~J., {Rawlings} S., {Blundell} K.~M., {Lacy} M., 1999, \mnras, 309,
  1017

\bibitem[{{Willott} {et~al}\mbox{.}(2000){Willott}, {Rawlings}, {Blundell}, \&
  {Lacy}}]{Willott00}
{Willott} C.~J., {Rawlings} S., {Blundell} K.~M., {Lacy} M., 2000, \mnras, 316,
  449

\bibitem[{{Willott} {et~al}\mbox{.}(2001){Willott}, {Rawlings}, {Blundell},
  {Lacy}, \& {Eales}}]{Willott01}
{Willott} C.~J., {Rawlings} S., {Blundell} K.~M., {Lacy} M., {Eales} S.~A.,
  2001, \mnras, 322, 536

\bibitem[{{Wright} {et~al}\mbox{.}(2010){Wright}, {Eisenhardt}, {Mainzer},
  {Ressler}, {Cutri}, {Jarrett}, {Kirkpatrick}, {Padgett}, {McMillan},
  {Skrutskie}, {Stanford}, {Cohen}, {Walker}, {Mather}, {Leisawitz}, {Gautier},
  {McLean}, {Benford}, {Lonsdale}, {Blain}, {Mendez}, {Irace}, {Duval}, {Liu},
  {Royer}, {Heinrichsen}, {Howard}, {Shannon}, {Kendall}, {Walsh}, {Larsen},
  {Cardon}, {Schick}, {Schwalm}, {Abid}, {Fabinsky}, {Naes}, \&
  {Tsai}}]{Wright10}
{Wright} E.~L. {et~al.}, 2010, \aj, 140, 1868

\bibitem[{{Xu} {et~al}\mbox{.}(1999){Xu}, {Livio}, \& {Baum}}]{Xu99}
{Xu} C., {Livio} M., {Baum} S., 1999, \aj, 118, 1169

\bibitem[{{York} {et~al}\mbox{.}(2000){York}, {Adelman}, {Anderson},
  {Anderson}, {Annis}, {Bahcall}, {Bakken}, {Barkhouser}, {Bastian}, {Berman},
  {Boroski}, {Bracker}, {Briegel}, {Briggs}, {Brinkmann}, {Brunner}, {Burles},
  {Carey}, {Carr}, {Castander}, {Chen}, {Colestock}, {Connolly}, {Crocker},
  {Csabai}, {Czarapata}, {Davis}, {Doi}, {Dombeck}, {Eisenstein}, {Ellman},
  {Elms}, {Evans}, {Fan}, {Federwitz}, {Fiscelli}, {Friedman}, {Frieman},
  {Fukugita}, {Gillespie}, {Gunn}, {Gurbani}, {de Haas}, {Haldeman}, {Harris},
  {Hayes}, {Heckman}, {Hennessy}, {Hindsley}, {Holm}, {Holmgren}, {Huang},
  {Hull}, {Husby}, {Ichikawa}, {Ichikawa}, {Ivezi{\'c}}, {Kent}, {Kim},
  {Kinney}, {Klaene}, {Kleinman}, {Kleinman}, {Knapp}, {Korienek}, {Kron},
  {Kunszt}, {Lamb}, {Lee}, {Leger}, {Limmongkol}, {Lindenmeyer}, {Long},
  {Loomis}, {Loveday}, {Lucinio}, {Lupton}, {MacKinnon}, {Mannery}, {Mantsch},
  {Margon}, {McGehee}, {McKay}, {Meiksin}, {Merelli}, {Monet}, {Munn},
  {Narayanan}, {Nash}, {Neilsen}, {Neswold}, {Newberg}, {Nichol}, {Nicinski},
  {Nonino}, {Okada}, {Okamura}, {Ostriker}, {Owen}, {Pauls}, {Peoples},
  {Peterson}, {Petravick}, {Pier}, {Pope}, {Pordes}, {Prosapio},
  {Rechenmacher}, {Quinn}, {Richards}, {Richmond}, {Rivetta}, {Rockosi},
  {Ruthmansdorfer}, {Sandford}, {Schlegel}, {Schneider}, {Sekiguchi}, {Sergey},
  {Shimasaku}, {Siegmund}, {Smee}, {Smith}, {Snedden}, {Stone}, {Stoughton},
  {Strauss}, {Stubbs}, {SubbaRao}, {Szalay}, {Szapudi}, {Szokoly}, {Thakar},
  {Tremonti}, {Tucker}, {Uomoto}, {Vanden Berk}, {Vogeley}, {Waddell}, {Wang},
  {Watanabe}, {Weinberg}, {Yanny}, \& {Yasuda}}]{york02}
{York} D.~G. {et~al.}, 2000, \aj, 120, 1579

\bibitem[{{Yuan} \& {Narayan}(2014)}]{Yuan14}
{Yuan} F., {Narayan} R., 2014, ArXiv e-prints

\bibitem[{{Zakamska} \& {Greene}(2014)}]{Zakamska14}
{Zakamska} N.~L., {Greene} J.~E., 2014, \mnras, 442, 784

\bibitem[{{Zakamska} {et~al}\mbox{.}(2003){Zakamska}, {Strauss}, {Krolik},
  {Collinge}, {Hall}, {Hao}, {Heckman}, {Ivezi{\'c}}, {Richards}, {Schlegel},
  {Schneider}, {Strateva}, {Vanden Berk}, {Anderson}, \&
  {Brinkmann}}]{Zakamska03}
{Zakamska} N.~L. {et~al.}, 2003, \aj, 126, 2125

\end{thebibliography}
